\documentclass[11pt,draftclsnofoot,onecolumn]{IEEEtran}
\usepackage{amssymb}    

\usepackage{cite}

\ifCLASSINFOpdf
   \usepackage[pdftex]{graphicx}
\else
\fi
\usepackage[cmex10]{amsmath}
\usepackage{multirow} 

\usepackage{url}

\usepackage[pdftex]{hyperref}

\newtheorem{lemma}{Lemma}[section]
\newtheorem{remark}{Remark}[section]
\newtheorem{conjecture}{Conjecture}[section]
\newtheorem{corollary}{Corollary}[section]
\newtheorem{proposition}{Proposition}[section]
\newtheorem{theorem}{Theorem}[section]
\newtheorem{definition}{Definition}[section]

\DeclareMathOperator*{\opliminf}{liminf^{(P)}}
\newcommand{\dpliminf}[2]{\opliminf_{#1\to#2\phantom{(P)}}} 

\newcommand{\lt}{\left}
\newcommand{\rt}{\right}

\newcommand{\E}{\mathcal{E}}
\newcommand{\M}{\mathcal{M}}
\newcommand{\Ps}{\mathcal{P}}
\newcommand{\Pst}{\mathcal{P}^{*}}
\newcommand{\St}{\mathcal{S}}
\newcommand{\U}{\mathcal{U}}
\newcommand{\V}{\mathcal{V}}
\newcommand{\X}{\mathcal{X}}
\newcommand{\Y}{\mathcal{Y}}
\newcommand{\Z}{\mathcal{Z}}

\newcommand{\bc}{\overline{B}}

\newcommand{\C}{\overline{C}}

\newcommand{\uC}{\underline{C}}

\newcommand{\tc}{\tilde{c}}
\newcommand{\tC}{\tilde{C}}
\newcommand{\tu}{\tilde{u}}

\newcommand{\tP}{\tilde{P}}
\newcommand{\tR}{\tilde{R}}

\newcommand{\hp}{\hat{p}}
\newcommand{\hS}{\hat{S}}
\newcommand{\hu}{\hat{u}}
\newcommand{\hU}{\hat{U}}
\newcommand{\hX}{\hat{X}}
\newcommand{\hY}{\hat{Y}}

\newcommand{\1}[1]{\boldsymbol{1}_{\lt\{#1\rt\}}}
\newcommand{\be}{\boldsymbol{e}}
\newcommand{\infr}{\underline{\boldsymbol{I}}}
\newcommand{\bW}{\boldsymbol{W}}
\newcommand{\bU}{\boldsymbol{U}}
\newcommand{\bv}{\boldsymbol{v}}
\newcommand{\bV}{\boldsymbol{V}}
\newcommand{\bx}{\boldsymbol{x}}
\newcommand{\by}{\boldsymbol{y}}
\newcommand{\bY}{\boldsymbol{Y}}

\newcommand{\Ex}[1]{\mathbb{E}\!\left[\,#1\,\right]}
\newcommand{\ud}{\,\mathrm{d}}
\newcommand{\cn}{\,|\,}

\newcommand{\bA}{\boldsymbol{A}}
\newcommand{\bB}{\boldsymbol{B}}
\newcommand{\bF}{\boldsymbol{F}}
\newcommand{\bI}{\boldsymbol{I}}
\newcommand{\bJ}{\boldsymbol{J}}
\newcommand{\bM}{\boldsymbol{M}}
\newcommand{\bP}{\boldsymbol{P}}
\newcommand{\bSig}{\boldsymbol{\Sigma}}

\newcommand{\hnu}{\hat{\nu}}

\newcommand{\hm}{\hat{m}}

\newcommand{\oeta}{\bar{\eta}}
\newcommand{\omu}{\bar{\mu}}
\newcommand{\omus}{\bar{\mu}^{*}}
\newcommand{\op}{\bar{p}}
\newcommand{\osd}[1]{#1_{1}^{\infty}}

\newcommand{\teta}{\tilde{\eta}}

\newcommand{\tpi}{\tilde{\pi}}

\newcommand{\R}{\mathbb{R}}
\newcommand{\z}{\mathbb{Z}}

\begin{document}
%
\title{Capacity Analysis of Discrete Energy Harvesting Channels}
%
%
%

\author{Wei~Mao and~Babak~Hassibi,~\IEEEmembership{Member,~IEEE}
\thanks{Portions of this work were presented at the 2013 and 2015 IEEE International Symposiums on Information Theory\cite{Mao-Hassibi-EnHarv,Mao-Hassibi-EnHarvLinBdCoding}, and the 2014 IEEE Information Theory Workshop\cite{Mao-Hassibi-CausCSIEnHav}. Wei Mao was with the Department of Electrical Engineering,
California Institute of Technology, Pasadena, CA 91125 USA. He is now with with the Department of Electrical Engineering, University of California, Los Angeles, CA 90095 USA (e-mail: weimao@ucla.edu). Babak Hassibi is with the Department of Electrical Engineering,
California Institute of Technology, Pasadena, CA 91125 USA (email: hassibi@caltech.edu).}}

\maketitle

\begin{abstract}
We study the channel capacity of a general discrete energy harvesting channel with a finite battery. Contrary to traditional communication systems, the transmitter of such a channel is powered by a device that harvests energy from a random exogenous energy source and has a finite-sized battery. As a consequence, at each transmission opportunity the system can only transmit a symbol whose energy is no more than the energy currently available. This new type of power supply introduces an unprecedented input constraint for the channel, which is simultaneously random, instantaneous, and influenced by the full history of the inputs and the energy harvesting process. Furthermore, naturally, in such a channel the energy information is observed causally at the transmitter. Both of these characteristics pose great challenges for the analysis of the channel capacity. In this work we use techniques developed for channels with side information and finite state channels, to obtain lower and upper bounds on the capacity of energy harvesting channels. In particular, in a general case with Markov energy harvesting processes we use stationarity and ergodicity theory to compute and optimize the achievable rates for the channels, and derive series of computable capacity upper and lower bounds.
\end{abstract}


%
\IEEEpeerreviewmaketitle


\section{Introduction}
%
%
%
%
\IEEEPARstart{I}{n} many future wireless systems, such as low-power wireless sensor networks, one may encounter transmitters that harvest and store energy for transmission. Such communication systems were first introduced by Ulukus et al. \cite{Ozel-Yang-Ulukus,Yang-Ulukus} and have received a lot of recent interest. When the battery is unlimited, \cite{Ozel-Ulukus-AWGN} shows that the entire capacity of an additive white gaussian noise (AWGN) channel can be achieved. When there is no battery, in the continuous setting \cite{Ozel-Ulukus-0battery} provides an analysis of the AWGN channel capacity, but there are gaps in the proof. For the discrete setting, however, the treatment with zero-battery is rather elementary (cf. Section~\ref{subsec:numerical-ach-rates}). The intermediate case, i.e., the case with a finite nonzero battery, was first considered in \cite{Tutuncuoglu-Yener-EnHarvPolicy}, where the optimum offline transmission policy for an energy harvesting node is obtained. However, in general, determining the channel capacity in such a case remains open. For the simplest case of a binary energy harvesting transmitter with a unit-sized battery connected to a noiseless channel, under the assumption that the transmitter only uses the causal battery state information (which is called scenario~1 in the current paper, see Section~\ref{subsec:three-models}) \cite{Tutuncuoglu-EnHarvTimingCh} derives a capacity formula involving an auxiliary random variable and obtains its upper and lower bounds. Also under scenario~1, \cite{Ozel-EH-CSIR-Discrete} further assumes that the receiver also has the energy information and studies the discrete setting with an i.i.d. energy harvesting process. Assuming some recent results on finite state channels (see \cite{Permuter-FSC-FB,Chen-Berger-FSC-FB}) can be generalized to finite state channels with \emph{input constraints}, \cite{Ozel-EH-CSIR-Discrete} suggests the possibility of a single-letter capacity formula under some extra assumptions. For the continuous setting with i.i.d. energy harvesting, \cite{Dong-Ozgur-EH-AWGN-Bounds} and \cite{Shaviv-Minh-Ozgur} explore the AWGN channel and provide upper and lower bounds that have a constant gap. In addition, for general energy harvesting channels with i.i.d. energy harvesting \cite{Shaviv-Minh-Ozgur} obtains a multi-letter mutual information capacity formula, and also shows that the capacity does not depend on the initial battery level. \cite{Shaviv-Ozgur-Permuter} explores some special cases with feedback and shows that in these cases feedback can increase capacity.

In this work we study the capacity of a discrete energy harvesting channel with a finite battery in its full generality. We study both transmitter-side energy information scenarios that have appeared in the literature (i.e., causal battery information v.s. causal harvested energy information), with a general energy utilization model and a general energy cost function. In all the (finite-battery) literature above the energy harvesting process is assumed to be i.i.d., whereas in this paper we derive capacity formulas for arbitrary energy harvesting processes. In the special case when this process is finite-order Markov (which is not necessarily stationary), we obtain computable upper and lower bounds. As we will see, the difficulty of the finite-battery energy harvesting channel is mainly caused by 1) the random instantaneous input constraint, which is influenced by both the input and the energy harvesting process and evolves with time, and 2) the causal energy information that is available to the transmitter only. In what follows we briefly outline our approaches to tackle this capacity problem. Since energy harvesting channels have both channel side information and input constraints, we first use results from channels with causal transmitter-side information (CSIT) to convert each of them to a certain equivalent channel without side information or constraint, but with an enlarged alphabet and a more complicated channel transition probability. We then express the capacity of this channel in terms of a multi-letter formula using the Verd\'u-Han general framework \cite{Verdu-Han-Capacity}. As such formulas are not easy to evaluate in general, we impose some restrictions on the input of the equivalent channel to obtain a certain surrogate channel model, whose capacity provides a lower bound on the original channel capacity. For this surrogate channel we study the required stationarity and ergodicity conditions and use the Shannon-McMillan-Breiman theorem to obtain some achievable rates, which serve as capacity lower bounds for the energy harvesting channel. These rates can be computed and optimized using the generalized Blahut-Arimoto algorithm\cite{Vontobel-GBAA}. For the capacity upper bounds, we assume that the energy information is also known at the receiver, and use Gallager's methods for finite state channels\cite{Gallager-IT_Reliable} to obtain an upper bound in terms of maximized block mutual information for every block length. These bounds have high computational complexity as they are derived from the equivalent channel, so we use results from feedback channels\cite{Tatikonda-Mitter-Feedback} to rewrite them in terms of maximized directed information on the original channel, which have much less complexity. It turns out that in this form the upper bounds for scenario~1 allow for a linear complexity dynamic programming recursion, whereas those for scenario~2 can also be relaxed to obtain a similar recursion. Apart from the upper bounds, we also obtain a capacity lower bound in scenario~2 for i.i.d. energy harvesting processes, in terms of maximized block mutual information. This bound can serve as a simpler alternative achievability proof for the multi-letter capacity formula in \cite{Shaviv-Minh-Ozgur}. In addition to these main results, using the same methods we also analyze a certain finite state channel model that is closely related to energy harvesting channels.

The rest of the paper is organized as follows. First, in the rest of this section, we introduce our major notations. Section~\ref{sec:sys-model} describes the channel models for two different energy information scenarios, as well as a related finite state channel based model, and transforms them to their respective equivalent channels. In Section~\ref{sec:channel-capacity} we express the channel capacities using the Verd\'u-Han formula. In the next section, Section~\ref{sec:ach-rates}, we impose some restrictions on the equivalent channels, derive the required stationarity and ergodicity conditions, and use the Shannon-McMillan-Breiman theorem to compute some achievable rates. The capacity upper bounds are derived in Section~\ref{sec:capacity-bounds}, together with a lower bound for scenario~2 with i.i.d. energy harvesting. Section~\ref{sec:simplified-upper-bounds} then simplifies and relaxes these high-complexity upper bounds. In Section~\ref{sec:numerical} some numerical examples are given for the computation of the achievable rates and various capacity bounds. Section~\ref{sec:conclusions} concludes the paper. The appendices are devoted to the stationarity and ergodicity theory for our channels, which are necessary for the results in Section~\ref{sec:ach-rates}.

\subsection{Notation}

In the main text of this paper we use the following notational conventions:
\begin{itemize}
\item For random variables:
	\begin{itemize}
	\item capital letters denote the random variables, e.g., $X_{n}$, $Y_{n}$.
	\item corresponding lowercase letters denote the realizations, e.g., $x_{n}$, $y$.
	\item corresponding script letters denote the alphabets, e.g., $\X$, $\Y$.
	\end{itemize}
\end{itemize}
\begin{itemize}
\item A vector $(z_{m},z_{m+1},\cdots,z_{n})$ is usually denoted by $z_{m}^{n}$, whereas $z^{n}\triangleq z_{1}^{n}$. When $n<m$, $z_{m}^{n}$ denotes the empty set. In addition, sometimes we use the abbreviation $e^{-r} \triangleq e_{-r+1}^{0}$ for $r\geq0$.
\item Bold lowercase letters also denote vectors, e.g., $\be_{i}$, $\bv_{k}$.
\item Bold capital letters denote certain infinite collections, e.g., $\bW$, $\bU$.
\item $\1{\cdot}$ denotes the indicator function:
\[ \boldsymbol{1}_{A}(x) = \Bigg\{ \begin{array}{cl}
1 & \text{if } x \in A \\
0 & \text{o.w.}
\end{array}. \]
When $A$ is the solution set of an equation $f(x)=0$, we simply denote the function by $\1{f(x)=0}$.
\item $\{\cdot\}_{n=n_{1}}^{n_{2}}$ denotes a sequence of symbols, indexed by $n$. For example, $\{E_{n}\}_{n=1}^{\infty}$ denotes the random process $E_{1},E_{2},\ldots,E_{n},\ldots$ To be concise we sometimes drop the sub-/super- scripts and just write $\{E_{n}\}$ when the context is clear.
\end{itemize}

\section{System Models}\label{sec:sys-model}

 \begin{figure}[t]
   \centering
   \includegraphics{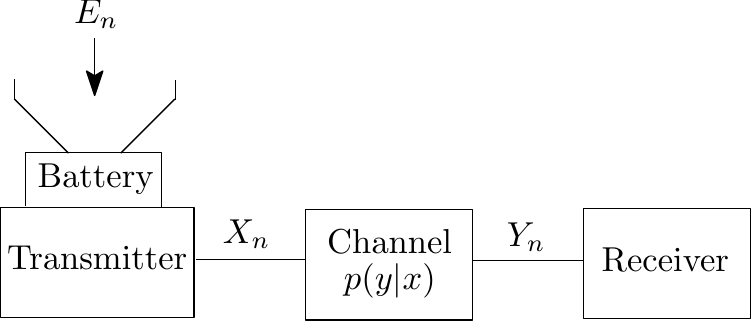}
   \caption{Energy harvesting system model}
   \label{fig:EH-model}
 \end{figure}
 
We consider a communication system powered by some energy harvesting mechanism with a battery, as depicted in Fig.~\ref{fig:EH-model}. At each transmission cycle $n$, the system first harvests some amount of energy, $E_{n}$, from the environment, and combines it with $B_{n}$, the energy stored in the battery after the last transmission, to transmit a symbol $X_{n}\in\X$. $X_{n}$ consumes some amount of energy $\gamma(X_{n})$, which cannot exceed the total available energy $S_{n}$ for the current cycle. The remainder, not exceeding the battery capacity $\bc$, is saved in the battery for future transmissions. The symbol $X_{n}$ is sent over the channel $p(y|x)$ and at the receiver a symbol $Y_{n}\in\Y$ is received. The alphabets $\X$ and $\Y$ are assumed to be finite with $\X\subset\mathbb{R}$ or $\mathbb{C}$, and the channel is discrete memoryless.

To be precise, the energy constraint on the system can be written as
\begin{equation}\label{eq:EH-en-constr}
\lt\{\begin{array}{ccl}
S_{n} &=& S(B_{n},E_{n}) \\
\gamma(X_{n}) &\leq& S_{n} \\
B_{n+1} &=& \min\left\{\, S_{n}-\gamma(X_{n}),\ \bc\, \right\}
\end{array}\rt.,
\end{equation}
where the total available energy $S_{n}$ is expressed as a function $S$ of the battery energy $B_{n}$ and the harvested energy $E_{n}$. The form of $S(\cdot)$ depends on how the system combines and utilizes $B_{n}$ and $E_{n}$. For example, if $E_{n}$ is immediately available for transmission, then simply
\begin{equation}\label{eq:EH-S-comb-1}
S(B_{n},E_{n}) = B_{n}+E_{n}.
\end{equation}
However, if the system can only use $E_{n}$ to charge the battery and draws energy solely from the battery for transmission, then
\begin{equation}\label{eq:EH-S-comb-2}
S(B_{n},E_{n}) = \min\lt\{\,B_{n}+E_{n},\,\bc\,\rt\}.
\end{equation}
This energy model can also take account of more real world influences. For example, if the battery is inefficient at charging and also has leakage, characterized by the ratios $\eta<1$ and $\beta<1$, respectively, then the model \eqref{eq:EH-S-comb-2} becomes
\[ S(B_{n},E_{n}) = \min\lt\{\,\beta B_{n}+\eta E_{n},\,\bc\,\rt\}. \] 
In view of the expression for $B_{n+1}$ in \eqref{eq:EH-en-constr}, for $n\geq 1$ sometimes we also write
\begin{equation}\label{eq:EH-S-ev}
S_{n+1} = S(X_{n},S_{n},E_{n+1}).
\end{equation}

The energy cost function $\gamma(\cdot)$, in general, can be any non-negative function on the alphabet $\X$. However, in this work, we require $\X$ to always include a zero symbol $0$ and that transmitting a zero does not consume any energy, i.e.,
\begin{equation}\label{eq:EH-0-cost-0}
\gamma(0) = 0.
\end{equation}
In addition, $\gamma$ is usually endowed with some physical meaning. For example, we often use the quadratic cost function to denote the instantaneous power:
\begin{equation}\label{eq:EH-cost-quadratic}
\gamma(x) = |x|^{2}.
\end{equation}

For readers' convenience the notations for the energy harvesting channel are summarized in Table~\ref{table:EH-notations}. Assume the initial energy $B_{1}$ stored in the battery is a random variable and the sequence of harvested energy $\{E_{n}\}_{n=1}^{\infty}$ is a random process independent of $B_{1}$. To simplify the problem, we only consider a \emph{finite discrete} system. Specifically, we assume $\bc <\infty$, and that all the energy quantities involved are quantized with the same interval size, i.e., all $E_{n}$, $B_{n}$, $S_{n}$, $\gamma(X_{n})$ and $\bc$ are integral multiples of some common unit of energy $\Delta_{E}$. Hence without loss of generality we can assume all these quantities are integers. Moreover, we further assume that the alphabet of $E_{n}$ is a bounded set $\E_{H}$ of non-negative integers, so that $B_{n}$ and $S_{n}$ can also only take values in finite integral sets $\E_{B}$ and $\St$, respectively.

\begin{table}[!t]
\caption{Energy harvesting channel notations}
\label{table:EH-notations}
\centering
\begin{tabular}{c|l|c}
\hline
Symbol & Definition & Alphabet\\
\hline
$E_{n}$ & Energy harvested between $(n-1)$-th and $n$-th transmission & $\E_{H}$ \\
$B_{n}$ & Energy stored in the battery after $(n-1)$-th transmission & $\E_{B}$ \\
$S_{n}$ & Energy available for $n$-th transmission & $\St$ \\
$X_{n}$ & Symbol transmitted at time $n$ & $\X$ \\
$Y_{n}$ & Symbol received at time $n$ & $\Y$ \\
$\bc$ & Battery capacity limit & - \\
$\gamma$ & Energy cost function & - \\
\hline
\end{tabular}
\end{table}

Because of the energy constraint \eqref{eq:EH-en-constr}, the operation of energy harvesting channels is much more complex than an ordinary DMC. During each transmission the transmitter is not free to choose any letter in $\X$; instead, at time $n$ it can only send a symbol $X_{n}$ that does not demand more than the current available energy $S_{n}$. Since $S_{n}$ determines how much energy the system can spend for the current transmission, we also call it the \emph{energy state} at time $n$. From the functional dependence of $S_{n}$ on $B_{n}$ and $E_{n}$, we see that $\{S_{n}\}$ is a random process with memory. Such a type of input constraints is unprecedented in traditional communication systems, which poses a major challenge for the analysis of the channels.

%

\subsection{Three Channel Models}\label{subsec:three-models}

For such energy harvesting channels, we study the following two scenarios with regard to the availability of energy information at transmitter.
\begin{itemize}
\item \emph{Scenario 1}: before the $n$-th transmission only the energy states $\{S_{i}\}_{i=1}^{n}$ are observed at the transmitter.
\item \emph{Scenario 2}: the transmitter knows the initial battery level $B_{1}$ and observes the harvested energy $\{E_{i}\}_{i=1}^{n}$ before the $n$-th transmission.
\end{itemize}
In both scenarios the receiver has no energy information. For convenience, in the following we refer to the channel models under these two scenarios as \emph{EH-SC1} and \emph{EH-SC2}, respectively. In a sense the second scenario is more general than the first, since we can recover the energy information for EH-SC1 from EH-SC2: by \eqref{eq:EH-en-constr}, with $X^{n}$ the transmitter can deduce $S^{n}$ from $E^{n}$ and $B_{1}$ (but not vice versa).

The energy information is a certain form of channel side information causally known at the transmitter, which is reminiscent of channels with causal CSIT \cite{Shannon-CSIT,Caire-Shamai-CSI}. The difference is that, in this new setting the energy states affect the input alphabets, instead of the channel transition probabilities. To assist the analysis of  such type of channels, we introduce a closely related, but simpler channel model: a certain finite state channel with causal CSIT and state-dependent input constraints, which is referred to as FSC-X.

\begin{definition}\label{def:Gallager-FSC}
A finite state channel\footnote{Defined by Gallager \cite{Gallager-IT_Reliable}; also see Appendix~\ref{subsec:Markov-Finite-State-channels} for more discussion. Compared to the original definition in \cite{Gallager-IT_Reliable}, we increase the indices of the states by 1 to better accommodate our channel model.} (FSC) is a channel with finite input, output, and state alphabets $\X$, $\Y$, and $\St$. The corresponding symbols at time $n$ are denoted by $X_{n}$, $Y_{n}$, and $S_{n}$, respectively, and the channel transitions is governed by a condtional probability $p(y_{n}s_{n+1}\,|\,x_{n}s_{n})$ which satisfies
\begin{equation}\label{eq:FSC-cond-prob}
p(y_{n}s_{n+1}\,|\,x^{n}s^{n}y^{n-1}) = p(y_{n}s_{n+1}\,|\,x_{n}s_{n})
\end{equation}
and which is time-invariant (i.e., independent of $n$).
\end{definition}

\begin{figure}[!t]
\centering
\includegraphics{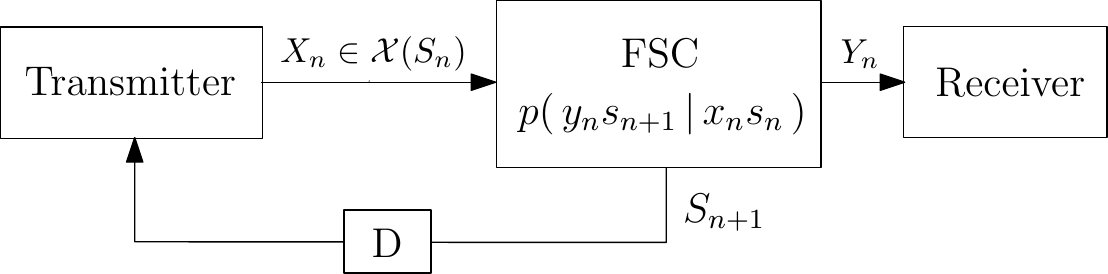}
\caption{FSC-X: FSC with input constraint and Causal CSIT}
\label{fig:FSC-model}
\end{figure}

\begin{definition}
A channel \emph{FSC-X} is an FSC with causal transmitter side CSI whose input is constrained by the current state. Precisely, at each time $n$, the state $S_{n}$ is fed to the encoder, which limits the input symbol $X_{n}$ to a subset $\X(S_{n})\subseteq\X$. The channel model is illustrated in Fig.~\ref{fig:FSC-model}.
\end{definition}

The connection of FSC-X to the energy harvesting channels is given by the following proposition.

\begin{proposition}\label{proposition:EH-SC1-FSC-X}
When the energy harvesting process $\{E_{n}\}$ is i.i.d., the channel EH-SC1 becomes a special case of FSC-X, whose states are exactly the energy states $\{S_{n}\}$.
\end{proposition}

\begin{IEEEproof}
By the DMC property and \eqref{eq:EH-S-ev}, for EH-SC1
\begin{align}
p(y_{n}s_{n+1}\,|\,x^{n}s^{n}y^{n-1}) &= p(y_{n}\cn x_{n})p(s_{n+1}\cn x^{n}y^{n}s^{n}) \nonumber\\
& = p(y_{n}\,|\,x_{n})\Pr\lt(S(x_{n},s_{n},E_{n+1}) = s_{n+1}\cn  x^{n}y^{n}s^{n}\rt) \nonumber\\
&\stackrel{(a)}{=} p(y_{n}|x_{n})\sum_{e_{n+1}}p(e_{n+1})\cdot \1{S(x_{n},s_{n},e_{n+1}) = s_{n+1}} \nonumber\\
& \stackrel{(b)}= p(y_{n}s_{n+1}\,|\,x_{n}s_{n}) \label{eq:EH-SC1-FSC-cond-prob},
\end{align}
where $(a)$ holds by the i.i.d. property of $\{E_{n}\}$, $(b)$ is established by expanding $p(y_{n}s_{n+1}\,|\,x_{n}s_{n})$ similarly. The expression of $p(y_{n}s_{n+1}\,|\,x_{n}s_{n})$ is independent of $n$, and so an FSC is defined.  Furthermore, by \eqref{eq:EH-en-constr} the input is constrained by $X_{n}\in\X(S_{n})$, where $\forall s\in\St$,
\begin{equation}\label{eq:EH-SC1-alphabet}
\X(s) \triangleq \{x\in\X: \gamma(x)\leq s\} .
\end{equation}
Since $S_{n}$ is causally known at the transmitter, the model fits exactly into the definition of FSC-X.
\end{IEEEproof}

Note that all the channels EH-SC1, EH-SC2, and FSC-X are subject to random input constraints, so an ordinary channel encoding scheme cannot function properly. In fact, if a message is mapped to any fixed input vector $x^{N}$, then chances are that some symbol $x_{n}$ does not satisfy the input constraint at the time of transmission, since the constraint value at that time might be incompatible with $x_{n}$. Hence for these channels we define new encoding schemes analogous to \cite{Shannon-CSIT,Caire-Shamai-CSI}, taking the CSIT into account to resolve this issue. Denote the set of messages to be transmitted as $\M = \{1,2,\cdots,M\}$.

\begin{definition}
A block code $f^{(N)}$ of length $N$ for EH-SC1 is defined by a sequence of $N$ encoding functions
$\{f_{n}:\M\times\St^{n}\to\X\}_{n=1}^{N}$,
such that $\forall m\in\M$ and $\forall s^{N}\in\St^{N}$,
i) the output $x^{N}$ of the encoder $f^{(N)}$ takes the form $x_{n} = f_{n}(m,s^{n})$ for all $1\leq n\leq N$ (i.e., $f_{n}$ is causal in $\{s_{n}\}$);
ii) the energy constraint \eqref{eq:EH-en-constr} is satisfied:
$\gamma(x_{n}) \leq s_{n}$ for all $n$.
\end{definition}

\begin{definition}
A block code $f^{(N)}$ of length $N$ for EH-SC2 is defined by a sequence of $N$ encoding functions
$\{f_{n}:\M\times\E_{B}\times\E_{H}^{n}\to\X\}_{n=1}^{N}$,
such that $\forall m\in\M$,  $\forall b_{1}\in\E_{B}$ and $\forall e^{N}\in\E_{H}^{N}$,
i) the output $x^{N}$ of the encoder $f^{(N)}$ takes the form
$x_{n} = f_{n}(m,b_{1},e^{n})$ for all $1\leq n\leq N$ (i.e., $f_{n}$ is causal in $\{e_{n}\}$);
ii) the energy constraint \eqref{eq:EH-en-constr} is satisfied.
\end{definition}

\begin{definition}
A block code for FSC-X takes the same form as in EH-SC1, except that the second requirement is changed to the input constraint $x_{n}\in\X(s_{n})$ for all $n$.
\end{definition}

As usual, the decoder for all the channels above is defined as $g^{(N)}: \Y^{N}\to\M$, which maps the output $y^{N}$ to an estimated message $\hm$. With the block codes properly defined, the definitions of probability of error, code rate, achievable rate, and channel capacity follow standard texts (see, e.g., \cite{Cover-Thomas-ElemIT}).

\begin{remark}\label{rmk:EH-SC1-2}
The capacity for EH-SC1 is smaller than or at most equal to EH-SC2, since the latter has more energy information at the transmitter, as mentioned above. Hence any capacity lower bound/achievable rate for EH-SC1 is also a capacity lower bound/achievable rate for EH-SC2, while any capacity upper bound of EH-SC2 is also a capacity upper bound of EH-SC1. That said, whether the first channel has a strictly smaller capacity is an open question and is not investigated in this paper. 
\end{remark}

\subsection{Equivalent Channels without Constraints}
\label{subsec:equiv-ch}

From the evolution of $\{S_{n},B_{n}\}$ in \eqref{eq:EH-en-constr}, the energy state $S_{n}$ depends on the full history of the harvested energy $E^{n}$, all the past transmitted symbols $X^{n-1}$, and the initial battery level $B_{1}$. This ever-growing memory of the energy constraint poses a major difficulty for the analysis of energy harvesting channels. If $\{E_{n}\}$ is i.i.d. and the battery capacity $\bc = 0$, then the system is actually memoryless, and it is easy to show that the channel for either scenario is simply equivalent to a DMC with an enlarged alphabet (similar to \cite{Shannon-CSIT}). However, if these conditions do not hold, then the system (under either scenario) has infinite memory and the analysis is much more involved---which is also the case for the channel FSC-X.

For such type of channels we use the approaches for channels with causal CSIT in \cite{Shannon-CSIT,Caire-Shamai-CSI} to convert them to equivalent channels without side information or constraints (which have enlarged input alphabets and still have memory). For each of our three models, the equivalent channel can be expressed as
\begin{equation}\label{eq:equivalent-channel}
\bW \triangleq \lt\{\,\U^{(N)},\ p(y^{N}\cn u^{N}),\ \Y^{N}\,\rt\}_{N=1}^{\infty},
\end{equation}
where we use $U_{n}$ and $Y_{n}$ to denote the new input and output symbols, respectively, and $p(y^{N}\cn u^{N})$ to denote the $N$-symbol channel transition probability. For each $N$ this new channel corresponds to $N$ operations of the original channel, starting from the beginning of transmission. The output alphabet $\Y^{N}$ is the same as before. The input alphabet $\U^{(N)}$, however, is different: a valid input symbol $U_{n}$ at each time $n$ is now a function of the causal side information, which respects the input constraints\footnote{Hence the input alphabet for block length $N$ is no longer the Cartesian product of $N$ single-channel-use alphabet, and thus it is denoted by $\U^{(N)}$ instead of $\U^{N}$.}. For each transmission cycle, such a function $U_{n}$ is sent to the channel input, which reads in the (causal) side information to produce a symbol $X_{n}\in\X$. This symbol is then sent to the original channel and an output symbol $Y_{n}\in\Y$ is received. The transition probabilities for $\bW$ are thus obtained by averaging those for the original channel over the randomness of the environment (or channel states). Below these definitions are made precise for each model, starting from the simplest case FSC-X.

\begin{definition}[Equivalent channel for FSC-X]\label{def:equiv-FSC-X}
The $n$-th input symbol is a function
$u_{n}:\St^{n}\to\X$,
which can also be viewed as a vector in $\X^{|\St|^{n}}$. The function needs to satisfy the input constraint
$u_{n}(s^{n}) \in\X(s_{n})$, $\forall s^{n}\in\St^{n}$, and so the input alphabets for time $n$ and for block length $N$ are, respectively,
\[ \U_{n} = \prod_{s\in\St}\X(s)^{|\St|^{n-1}},\qquad \U^{(N)} = \prod_{n=1}^{N}\U_{n}. \]
The $N$-symbol channel transition probability is determined as follows. First with the FSC probability model \eqref{eq:FSC-cond-prob} and the functional relation $x_{n} = u_{n}(s^{n})$,
\begin{align}
 p\lt(y^{N}s_{2}^{N+1}\,|\,u^{N}s_{1}\rt) &= \prod_{n=1}^{N}p\lt(y_{n}s_{n+1}\,|\,u^{N}s^{n}y^{n-1}\rt) \nonumber \\
 &= \prod_{n=1}^{N}p\lt(y_{n}s_{n+1}\,|\,x_{n} = u_{n}(s^{n}),s_{n}\rt).\label{eq:FSC-pYScondUs1}
\end{align}
Then as $S_{1}$ is independent of $U^{N}$: $p\lt(y^{N}\,|\,u^{N}\rt) = \sum_{s_{1}}p(s_1)\sum_{s_{2}^{N+1}}p\lt(y^{N}s_{2}^{N+1}\,|\,u^{N}s_{1}\rt)$.
\end{definition}

\begin{definition}[Equivalent channel for EH-SC1]
The input symbols/alphabets take the same forms as in the previous case, with $\X(s)$ defined by \eqref{eq:EH-SC1-alphabet}. For the channel transition probability, observe that i) $B_{1}$ and $E^{N}$ are independent, and are independent of $U^{N}$ (since they are unknown at the encoder); ii) $x^{N}$ is determined by $b_{1}$, $e^{N}$, and $u^{N}$ from the recursion
\[ \renewcommand{\arraystretch}{.9}
\lt\{\begin{array}{ccl}
s_{n} &=& S(b_{n},e_{n}) \\
x_{n} &=& u_{n}(s^{n}) \\
b_{n+1} &=& \min\left\{\, s_{n}-\gamma(x_{n}),\ \bc\, \right\}
\end{array}\rt. \]
and $s^{n}$ is a function of $b_{1}$, $e^{n}$ and $u^{n-1}$; iii) $y^{N}$ is produced by the DMC with input $x^{N}$. So
\begin{align}
p(y^{N}\cn u^{N}) 
&= \sum_{b_{1},e^{N}}p(b_{1})p(e^{N})p(y^{N}\cn b_{1}e^{N}u^{N}) \nonumber \\
&= \sum_{b_{1},e^{N}}p(b_{1})p(e^{N})\prod_{n=1}^{N}p\lt(\,y_{n}\cn x_{n} = u_{n}\lt(\,s^{n}\lt(b_{1},e^{n},u^{n-1}\rt)\,\rt)\,\rt).
\label{eq:EH-SC1-cond-prob}
\end{align}
\end{definition}

\begin{definition}[Equivalent channel for EH-SC2]\label{def:equiv-EH-SC2}
The $n$-th input symbol for the equivalent channel is a function $u_{n}: \E_{B}\times\E_{H}^{n}\to\X$, which is also a vector in $\X^{|\E_{B}|\cdot|\E_{H}|^{n}}$. The function $u_{n}$ needs to be compatible with the previous input symbols, $u^{n-1}$, in terms of the the energy constraint \eqref{eq:EH-en-constr}. In particular, for each block length $N$, a feasible input vector $u^{N}$ needs to satisfy $\gamma(u_{n}(b_{1},e^{n})) \leq s_{n}$ for all $1\leq n\leq N$, $\forall b_{1}\in\E_{B}$ and $\forall e^{N}\in \E_{H}^{N}$. Here $s_{n} = s_{n}(b_{1},e^{n},u^{n-1})$ is determined recursively by
\begin{equation}\label{eq:EH-SC2-recursion}
\renewcommand{\arraystretch}{.9}
\lt\{\begin{array}{ccl}
s_{n} &=& S(b_{n},e_{n}) \\
x_{n} &=& u_{n}(b_{1},e^{n}) \\
b_{n+1} &=& \min\left\{\, s_{n}-\gamma(x_{n}),\ \bc\, \right\}
\end{array}\rt..
\end{equation}
Note that the permitted function values of $u_{n}$ depends not only on the energy sequence $(b_{1},e^{n})$, but also on all previous input symbols $u^{n-1}$. So the input alphabet for time $n$ takes the form
\[ \U_{n}(u^{n-1}) =
\prod_{b_{1},e^{n}}\X\lt(s_{n}(b_{1},e^{n},u^{n-1})\rt), \]
where $\X(\cdot)$ is defined in \eqref{eq:EH-SC1-alphabet}; further, the input alphabet for block length $N$ is
\[ \U^{(N)} = \lt\{ (u_{1},\cdots,u_{N}): u_{n}\in\U_{n}(u^{n-1}),\  \forall 1\leq n\leq N \rt\}, \]
which is the collection of all vectors of $N$ causal functions on the energy sequence (and the initial battery) that are consistent with the energy constraint. For the channel transition probability, by the same arguments above (but with recursion \eqref{eq:EH-SC2-recursion} instead) we have
\begin{align}
p(y^{N}\cn u^{N}) 
&= \sum_{b_{1},e^{N}}p(b_{1})p(e^{N})p(y^{N}\cn b_{1}e^{N}u^{N}) \nonumber \\
&= \sum_{b_{1},e^{N}}p(b_{1})p(e^{N})\prod_{n=1}^{N}p\lt(\,y_{n}\cn x_{n} = u_{n}(b_{1},e^{n})\,\rt).
\label{eq:EH-SC2-cond-prob}
\end{align}
\end{definition}

Since there is no CSI or constraints for these new channels, the encoding maps now take the usual form $f^{(N)}:\M\to\U^{(N)}$, whereas the form of the decoders are not changed. For each model the new channel is equivalent to the original one, in the sense that they have the same capacity: in fact, as stated in \cite{Shannon-CSIT}, block codes for the original and equivalent channels can be easily translated into each other, which induce the same output distribution---hence using the same decoder, the same probability of error can be achieved.

\begin{remark}
The use of equivalent channels avoids the difficulty of dealing with either the CSIT or the input constraints, at the cost of more complicated input alphabets, whose sizes grow with $n$. Roughly speaking, the cardinality for the input alphabet at time $n$ grows double-exponentially (cf. \cite[Example~4.3.1]{Mao-thesis}).
\end{remark}

\section{Capacity Formulas}
\label{sec:channel-capacity}

To compute the capacity for a channel as general as \eqref{eq:equivalent-channel}, we need to invoke Verd\'u and Han's general capacity formula for arbitrary channels without feedback \cite{Verdu-Han-Capacity}. Define an \emph{input distribution process} $\bU$ to be a sequence of probability distributions defined on $\U^{(N)}$ for each $N$ (which need not have any relation among them). Equivalently, $\bU$ can be represented by a collection of random vectors $\lt\{U^{(N)}\rt\}_{N=1}^{\infty}$, where each $U^{(N)}$ is a random vector\footnote{We use $U^{(N)}$ instead of the usual $U^{N}$ here, since the first $N-1$ entries of $U^{(N)}$ need not agree with $U^{(N-1)}$.} in $\U^{(N)}$ that exactly has the $N$-th distribution of $\bU$. The corresponding \emph{output distribution process} $\bY = \lt\{Y^{(N)}\rt\}_{N=1}^{\infty}$ is the collection of random vectors $Y^{(N)}$ in $\Y^{N}$, where each $Y^{(N)}$ is induced by the input random vector $U^{(N)}$ and the $N$-symbol channel transition probability $p(y^{N}\cn u^{N})$. Furthermore, define the \emph{information density} between $U^{(N)}$ and $Y^{(N)}$ as
\[ i_{N}\lt(u^{N};y^{N}\rt) = i_{U^{(N)};Y^{(N)}}\lt(u^{N};y^{N}\rt) = \log\frac{p\left(y^{N}\cn u^{N}\right)}{p\left(y^{N}\right)} \]
for all $u^{N}\in\U^{(N)}$, $y^{N}\in\Y^{N}$. The \emph{inf-information rate} between $\bU$ and $\bY$ is then defined as 
\[ \infr(\bU;\bY) \triangleq \dpliminf{N}{\infty}\frac{1}{N}i_{N}\lt(U^{(N)};Y^{(N)}\rt), \]
where for any sequence of random variables $\lt\{A_{N}\rt\}_{N=1}^{\infty}$, define its \emph{liminf in probability} 
as the supremum of all the real numbers $\alpha$ for which $\Pr(A_{N}\leq\alpha)$ vanishes as $N\to\infty$:
\[ \dpliminf{N}{\infty} A_{N} \triangleq \sup\lt\{\alpha\in\mathbb{R}\ \big|\ \lim_{N\to\infty}\Pr(A_{N}\leq\alpha)= 0\rt\}. \]

\begin{theorem}[Verd\'u-Han formula\cite{Verdu-Han-Capacity}]
\label{thm:C-equivalent-channel}
The capacity of the channel $\bW$ in \eqref{eq:equivalent-channel} is given by
\begin{equation}\label{eq:C-equivalent-channel}
C = \sup_{\bU}\infr(\bU;\bY),
\end{equation}
where the supremum is taken over all input distribution processes $\bU$.
\end{theorem}

The channel capacities of the three models in this paper can all be obtained from their respective equivalent channels in Section~\ref{subsec:equiv-ch} and Theorem~\ref{thm:C-equivalent-channel}.
Despite its generality, however, the capacity formula \eqref{eq:C-equivalent-channel} has the following issues:
\begin{enumerate}
\item The supremum is taken over all possible input distribution processes, which is hard to enumerate/parameterize.
\item Given an arbitrary input distribution processes $\bU$, the inf-information rate is not always readily computable, as the asymptotic behavior for the corresponding random sequence might be unknown.
\item As the input alphabet size $\lt|\U^{(N)}\rt|$ grows double-exponentially (roughly), the computational complexity is also double-exponential when calculating either the information density distribution or the mutual information for a single block length $N$.
\end{enumerate}
Hence this formula is too complicated to evaluate in general. Nonetheless, it provides us a useful tool to analyze the channel capacities. In the next section we will try to resolve these difficulties under some simplifying conditions and assumptions to make the computation tractable. Such simplifications give us achievable rates for our channels, which are lower bounds of their respective capacities.

Note that in the following special cases, simpler (but still not computable) capacity formulas might be possible. For the binary noiseless EH-SC1, \cite{Tutuncuoglu-EnHarvTimingCh} proposes a single-letter capacity formula involving an auxiliary variable. More recently, for EH-SC2, \cite{Shaviv-Minh-Ozgur} shows that when the energy harvesting process $\{E_{n}\}$ is i.i.d., the channel capacity has a multi-letter mutual information expression: it can be written as the limit of maximum mutual information per channel use for the equivalent channel, as the block size tends to infinity. The achievability is proved using a complex transmission scheme, while the converse is given by Fano's inequality as in \cite{Verdu-Han-Capacity} (see also Section~\ref{subsec:UB-Gallager-type}). Furthermore, in \cite[Proposition~1]{Shaviv-Minh-Ozgur} the authors also prove that in this case the capacity does not depend on the initial battery value\footnote{Different from our setting, \cite{Shaviv-Minh-Ozgur} assumes that the initial battery is known at the transmitter. However, using this very Proposition~1 one can show that the channel capacity is identical to our model. The result in this proposition can also be extend to the general case when $\{E_{n}\}$ is stationary and ergodic, and for both energy harvesting scenarios EH-SC1 and EH-SC2.}. In Section~\ref{subsec:EH-UB} we show that with this proposition our Theorem~\ref{thm:LB-EH} can provide a much simpler achievability proof.

\section{Achievable Rates}
\label{sec:ach-rates}

%
%
%

To address the issues in computing the channel capacity \eqref{eq:C-equivalent-channel}, we restrict the input symbols of the channel model \eqref{eq:equivalent-channel} to a constant-sized subset of its alphabet and obtain a surrogate channel $\bW'$, whose capacity $C'$ provides a lower bound for the capacity $C$ of the channel $\bW$. In particular, instead of the full CSI history, the input functions now can only depend on a limited amount of the causal side information, with which  the transmitter can still compute the instantaneous input constraint.

Let $V_{n}$ denote the new input function at time $n$ and $\V$ denote its (constant-sized) alphabet. Similar to \eqref{eq:equivalent-channel}, the surrogate channel can be expressed as
\begin{equation}\label{eq:surrogate-channel}
\bW' \triangleq \lt\{\,\V^{N},\ p(y^{N}\cn v^{N}),\ \Y^{N}\,\rt\}_{N=1}^{\infty}. 
\end{equation}
It turns out that in many cases we are interested in, $\bW'$ becomes a finite state channel (FSC). The capacity of a general FSC is studied in \cite{Gallager-IT_Reliable,Chen-Permuter-Weissman-FSC-Bds}, both of which give two series as the capacity upper and lower bounds, in terms of the mutual information between the input and output for each block size $N$. When the FSC is \emph{indecomposable}\footnote{See Definition~\ref{def:Gallager-indecomposable} in Appendix~\ref{sec:ergodicity-Markov-channels} or \cite{Gallager-IT_Reliable}.}, the upper and lower bounds both converge to the capacity. However, these bounds are not very useful for us: i) since $C'$ is less than or equal to $C$, the lower bounds are genuine, but the upper bounds are not meaningful; ii) the computational complexity of such bounds is exponential in $N$; iii) the bounds in \cite{Gallager-IT_Reliable} are too loose for small $N$ and their convergence is slow (see \cite{Chen-Permuter-Weissman-FSC-Bds}); iv) the bounds in \cite{Chen-Permuter-Weissman-FSC-Bds} are supposed to be tighter, but the computation is not easy for a general $N$.

Another way of describing the capacity $C'$ is (again) through the Verd\'u-Han formula (cf. Theorem~\ref{thm:C-equivalent-channel}):
\begin{equation}\label{eq:C-surrogate-channel}
C' = \sup_{\bV}\infr(\bV;\bY),
\end{equation}
for which we define the same concepts and similar notations, as in Section~\ref{sec:channel-capacity}, with respect to the surrogate channel $\bW'$. The supremum in \eqref{eq:C-surrogate-channel} is taken over all input distribution processes $\bV$. Although in general this formula is still not computable, for any given input distribution process that yields a computable inf-information rate we can obtain an achievable rate for $\bW'$ (and hence also for $\bW$), which lower bounds the capacity $C'$ (and $C$). In particular, assume the input distribution process $\bV$ is induced by a source random process $\{V_{n}\}$, so that the $N$-th distribution of $\bV$ corresponds exactly to the random vector $V^{N}$ for each $N$. Assume further that the induced joint input-output process $\{V_{n},Y_{n}\}$ satisfies the Shannon-McMillan-Breiman (SMB) theorem (see Appendix~\ref{sec:SMB}), then the sample entropies for $\{V_{n},Y_{n}\}$ converge almost surely to their respective entropy rates. Accordingly, the normalized information density, which can be written as
\begin{equation}\label{eq:info-density-expansion}
\frac{1}{N}i_{N}\lt(V^{N};Y^{N}\rt) = \frac{1}{N}\log p\left(V^{N},Y^{N}\right) - \frac{1}{N}\log p\left(V^{N}\right) - \frac{1}{N}\log p\left(Y^{N}\right),
\end{equation}
converges almost surely to the mutual information rate (a.k.a. \emph{information rate})
\[ I(\V,\Y) \triangleq \lim_{N\to\infty}\frac{1}{N}I(V^{N};Y^{N}) = H(\V) + H(\Y) - H(\V,\Y), \]
where $H(\V)$, $H(\Y)$, and $H(\V,\Y)$ denote the (joint) entropy rates of $\{V_{n}\}$, $\{Y_{n}\}$, and $\{V_{n},Y_{n}\}$, respectively. As a result, the liminf in probability of $\lt\{\frac{1}{N}i_{N}\lt(V^{N};Y^{N}\rt)\rt\}_{N=1}^{\infty}$ evaluates to the same value $I(\V,\Y)$, and so the inf-information rate $\infr(\bV;\bY)$ becomes the mutual information rate, which yields, at least theoretically, a computable achievable rate. Alternatively, since AEP holds in this case (see Appendix~\ref{sec:SMB}), we can use the idea of typical set decoding as in \cite{Cover-Thomas-ElemIT} to directly prove the achievability of the rate $I(\V,\Y)$.

The Shannon-McMillan-Breiman theorem demands certain stationarity and ergodicity properties of the joint input-output process, which in turn require the source and channel to satisfy some conditions in that aspect. Specifically, the version of SMB theorem (Theorem~\ref{thm:SMB}) suitable for our models requires the joint process $\{V_{n},Y_{n}\}$ to be \emph{asymptotically mean stationary}\footnote{See the Appendices for this and other concepts in theories of stationarity and ergodicity.} (AMS) and ergodic. When the surrogate channel $\bW'$ is an FSC, it belongs to the category of \emph{Markov channels} and always produces an AMS joint input-state-output process for any AMS or stationary source. For such a channel $\bW'$, if (i) the source $\{V_{n}\}$ is stationary and ergodic while $\bW'$ satisfies some further ergodicity conditions with respect to the source, or (ii) the source $\{V_{n}\}$ is finite-order Markov and induces a joint source-channel Markov chain with some irreducibility condition, then the joint input-state-output process is AMS and ergodic, and so is the process $\{V_{n},Y_{n}\}$\footnote{See Appendix~\ref{sec:joint-marginal}.}. The descriptions of the specific conditions for each model are given in the next three subsections. Due to the technical nature, however, the exposition of the underlying stationarity and ergodicity theory is deferred to the Appendices. Such a theory is largely based on the theory of Markov channels developed in \cite{Kieffer-Rahe-AMS-Mrkv-Ch,Gray-Dunham-Gobbi} and the ergodic theory of stationary Markov chains in \cite{Walters-Ergodic-Theory}.

In practice, the computation of the mutual information rate $I(\V,\Y)$ for general source processes $\{V_{n}\}$ is a challenging problem. One can use the sequence of finite block length mutual information to approximate $I(\V,\Y)$, but since the alphabet sizes grow exponentially with the block length, so does the computational complexity. Moreover, the convergence of such a sequence is often rather slow. With the above stationarity and ergodicity conditions for the source and channel, however, we have the SMB theorem and so can estimate the information rate using the sample entropies (through \eqref{eq:info-density-expansion}) of a very long sample sequence, which can be computed using the transition probabilities in \eqref{eq:surrogate-channel} and the input distribution. In addition to that, when the source is a finite-order Markov process and the channel is an FSC, the computation of the sample entropies in \eqref{eq:info-density-expansion} has a complexity linear in $N$; in fact one can use the well-known BCJR algorithm\cite{BCJR-algorithm} (a.k.a. the sum-product algorithm\cite{Kschischang-Frey-Loeliger-SumProductAlg}) to compute them. This stochastic method for information rate computation was proposed independently in \cite{Arnold-Loeliger-info-rate,Sharma-Singh-info-rate,Pfister-et-al-info-rate}, and is summarized in \cite{Arnold-Simulation-SMB}.

So far by restricting the input alphabet and imposing extra stationarity and ergodicity conditions on the source and channel, we are able to resolve the issues 2) and 3) in Section~\ref{sec:channel-capacity} and efficiently compute some achievable rates for the channel $\bW$. If we further fix the order of a Markov input process, under some conditions (described below) we can maximize the achievable rate over a given set of transition probabilities for the Markov chain, thus also resolving the issue 1) in Section~\ref{sec:channel-capacity} to some extent. Specifically, we use the generalized Blahut-Arimoto algorithm (GBAA) for the achievable rate optimization\footnote{Apart from the GBAA, Han \cite{Han-RandomizedCapacity} also gives a stochastic method for the information rate optimization of a finite state channel. However, the assumptions on the channel are more stringent in \cite{Han-RandomizedCapacity} and so the algorithm is not used in our work.}, which is proposed by Vontobel~et al. in \cite{Vontobel-GBAA}. In their work, the traditional Blahut-Arimoto algorithm\cite{Blahut-BAA}, originally used for computing the capacity of a DMC, is generalized in the setting of an indecomposable FSC with a finite-order Markov input process, whose underlying chain is stationary, ergodic, and aperiodic, to optimize the information rate over a given set\footnote{This set may come from certain practical/physical constraints. It should also meet the above requirements for the chain.} of transition probabilities of the input Markov chain.\footnote{In fact, we found that the algorithm as it is in \cite{Vontobel-GBAA} is not applicable to all indecomposable FSC's, as the calculation of the critical $T$-values is erroneous for some channel models. However, surprisingly, this issue does not affect the correct calculation of the information rate at each iteration, but it only affects the selection of new optimization parameters for the next iteration. Furthermore, after we communicated with them, the authors corrected the $T$-values and fixed this issue.} The core part of the GBAA is to estimate the so-called ``$T$-values'' defined in \cite[Definition 41]{Vontobel-GBAA} through the algorithms in \cite[Lemma 70]{Vontobel-GBAA} in each iteration, which are then used both to calculate the information rate and to update the optimization parameters (i.e., the transition probabilities). As we examine the derivations and proofs in \cite{Vontobel-GBAA}, we find that, to the best of our knowledge\footnote{The details are not included in their related proofs.}, the sole purpose of both the indecomposable assumption of the FSC and the ergodicity and aperiodicity of the input Markov chain is to guarantee the almost sure convergence of the estimated $T$-values in \cite[Lemma~70]{Vontobel-GBAA}. Hence we speculate that the required convergence still holds as long as the joint input-state-output process satisfies the SMB theorem; in particular, when the joint process is AMS and ergodic. Such a requirement is fulfilled when the source is a stationary finite-order Markov process whose underlying chain is irreducible\footnote{A finite alphabet stationary Markov process is ergodic iff the chain is irreducible; see Theorem~\ref{thm:Markov-process-ergodic-irreducible}.}, while the channel is an FSC with the ergodicity conditions mentioned earlier (which are weaker than indecomposability). Consequently, we conjecture that the GBAA still works under these relaxed conditions. Besides, at the very least, we can use the GBAA primarily as a means to find a good set of input process parameters (i.e., the Markov transition probabilities); the resulting information rates can always be cross-checked using the stochastic methods described above, since the SMB theorem applies. Therefore, when these conditions hold, we can apply the GBAA to our surrogate channel $\bW'$ for each fixed Markov order of the input process\footnote{Recall that when the Markov order is $k$, the states of the underlying chain are the tuples of $k$ successive input symbols. Again, the optimization space is a subset of the transition probabilities (which satisfy the ergodicity conditions).} to find an optimized achievable rate.

In what follows we apply the above general methodology to each of our three channel models. First we describe the restriction on the input and show that the surrogate channel $\bW'$ is an FSC (under certain conditions), then give the stationarity and ergodicity conditions, with which the computation and optimization of achievable rates are possible. Some numerical examples are given in Section~\ref{sec:numerical} to illustrate the computation.

\subsection{FSC-X}
\label{subsec:ach-rate-FSC-X}

We restrict the input function $u_{n}$ to depend only on the $m$ most recent states, where $m>0$ is a fixed integer. To be specific, let $\V$ be the collection of all functions $v: \St^{m}\to \X$ such that
\[ v(s^{m}) \in\X(s_{m}),\qquad\forall s^{m}\in\St^{m}. \]
Therefore $\V = \prod_{s\in\St}\X(s)^{|\St|^{m-1}}$ has a constant alphabet size. We restrict $u_{n}$ in such a way that each $u_{n}$ is associated with a symbol $v_{n}\in\V$ and satisfies\footnote{For \eqref{eq:un-vn-FSC-X} to be meaningful when $m>1$ and $n<m$, we define the dummy variables $s_{-m+2},\cdots,s_{0}\in\St$ as the \emph{pre-historical states}, which are deterministic. These artificial states are only used in the arguments of $v_{n}$ for $n<m$, but do not affect the distribution of $S_{1}$ (which is determined by the environment/nature). See \cite{Mao-thesis} for a more detailed discussion.}
\begin{equation}\label{eq:un-vn-FSC-X}
u_{n}(s^{n}) = v_{n}(s_{n-m+1}^{n}),\qquad\forall s^{n}\in\St^{n}.
\end{equation}
With such a configuration we define a surrogate channel $\bW'$ with the input alphabet $\V$, whose transition probability is defined through the corresponding $u^{N}$ for each $N$. In other words, according to \eqref{eq:FSC-pYScondUs1},
\[ p\lt(y^{N}s_{2}^{N+1}\,|\,v^{N}s_{1}\rt) = \prod_{n=1}^{N}p\lt(y_{n}s_{n+1}\,|\,x_{n} = v_{n}(s_{n-m+1}^{n}),s_{n}\rt), \]
\begin{equation}\label{eq:FSC-X-surrogate-cond-prob}
p\lt(y^{N}\,|\,v^{N}\rt) = 
\sum_{s_{1}}p(s_1)\sum_{s_{2}^{N+1}} \prod_{n=1}^{N}p\lt(y_{n}s_{n+1}\,|\,x_{n} = v_{n}(s_{n-m+1}^{n}),s_{n}\rt).
\end{equation}

We claim that the channel $\bW'$ is an FSC for $n\geq m$,\footnote{This restriction does not affect the information rate computation by\cite[Lemma~3.4.1]{Gray-EIT}. See also \cite{Mao-thesis}.} whose state is defined as
\[ Z_{n} = S_{n-m+1}^{n}, \]
with alphabet $\Z = \St^{m}$. In fact, for $n\geq m$, the transition probability satisfies the following: if $z_{n}$ is compatible with $z_{n+1}$, i.e., for some $s^{n+1}\in\St^{n+1}$, $z_{n} = s^{n}_{n-m+1}$ while $z_{n+1} = s^{n+1}_{n-m+2}$, then by the FSC transition probability \eqref{eq:FSC-cond-prob},
\begin{align}
p(y_{n}z_{n+1}\cn v^{n}z^{n}y^{n-1}) &= p(y_{n}s_{n-m+2}^{n+1}\cn v^{n}s^{n}y^{n-1}) \nonumber \\
&= p(y_{n}s_{n+1}\cn v^{n}s^{n}y^{n-1},x_{n}= v_{n}(s_{n-m+1}^{n})) \nonumber\\
&= p(y_{n}s_{n+1}\cn x_{n}= v_{n}(s_{n-m+1}^{n}),s_{n}) \label{eq:FSC-X-FSC-cond-prob}\\
&= p(y_{n}z_{n+1}\cn v_{n}z_{n}).\nonumber
\end{align}
If $z_{n}$ is not compatible with $z_{n+1}$, then both the first and the last term are 0 and \eqref{eq:FSC-X-FSC-cond-prob} still holds.

For the required stationarity and ergodicity properties for the SMB theorem, we provide the following two sets of simple conditions. We also have some stronger but more complicated conditions, see Corollary~\ref{cor:Gray-Dunham-Gobbi-thm-2-extension-cor-Markov} and Lemma~\ref{lem:FSC-Markov-input} in the Appendices.

\begin{lemma}
Assume the input process $\{V_{n}\}$ of the surrogate channel $\bW'$ for FSC-X is stationary and ergodic. Then the joint process $\{V_{n},Y_{n}\}$ is AMS and ergodic, if any of the following holds.
\begin{enumerate}
\item[i)] $\bW'$ is indecomposable.  
\item[ii)] There is a finite vector $v_{m}^{N}$ with $\Pr(V_{m}^{N}=v_{m}^{N})>0$ satisfying the following property: given $V_{m}^{N}=v_{m}^{N}$, for any $z_{m},z_{m}'\in\Z$, there exists $y_{N}\in\Y$ and $z_{N+1}\in\Z$ such that when $Z_{m}=z_{m}$ or $z_{m}'$, we both have $Y_{N}Z_{N+1} = y_{N}z_{N+1}$ with positive probability.
\end{enumerate}
\end{lemma}

\begin{IEEEproof}
The first condition follows from Lemma~\ref{lem:Gray-Dunham-Gobbi-cor-2} and Theorem~\ref{thm:Gray-Dunham-Gobbi-thm-2-extension} (or Corollary~\ref{cor:Gray-Dunham-Gobbi-thm-2-extension-cor-Markov}), and the second follows from Corollary~\ref{cor:Gray-Dunham-Gobbi-thm-2-extension-cor-FSC}.
\end{IEEEproof}

\begin{lemma}
Assume the input process $\{V_{n}\}$ of the surrogate channel $\bW'$ for FSC-X is finite-order Markov, then so is the joint process $\{(V_{n},Y_{n},Z_{n+1})\}$. If the underlying Markov chain for the latter is irreducible, then $\{V_{n},Y_{n}\}$ is AMS and ergodic.
\end{lemma}

\begin{IEEEproof}
This lemma is a simple case of Lemma~\ref{lem:FSC-Markov-input}.
\end{IEEEproof}

\subsection{EH-SC1}
\label{subsec:ach-rate-EH-SC1}

Again we restrict the input function $u_{n}$ to depend only on the $m>0$ most recent (energy) states, and supply the dummy pre-historical states $s_{-m+2},\cdots,s_{0}\in\St$ when $m>1$. Then the surrogate channel $\bW'$ has the same input alphabet as in the previous section, with $\X(s)$ defined in \eqref{eq:EH-SC1-alphabet}. According to \eqref{eq:EH-SC1-cond-prob} the transition probabilities are\footnote{Note that $s_{-m+2}^{0}(\cdot)$ are given by the dummy variables.}
\[  p\lt(y^{N}\,|\,v^{N}\rt) = 
\sum_{b_{1},e^{N}}p(b_{1})p(e^{N})\prod_{n=1}^{N}p\lt(\,y_{n}\cn x_{n} = v_{n}\lt(\,s_{n-m+1}^{n}\lt(b_{1},e^{n},u^{n-1}\rt)\,\rt)\,\rt). \]

If the energy harvesting process $\{E_{n}\}$ is i.i.d., then, as shown in Section~\ref{sec:sys-model}, the channel EH-SC1 is an instance of FSC-X, and by the previous section $\bW'$ is an FSC with state variable $Z_{n}= S_{n-m+1}^{n}$. Note in passing that since the argument $s_{n-m+1}^{n}$ for $v_{n}$ is contained in $z_{n}$, by \eqref{eq:EH-SC1-FSC-cond-prob} and \eqref{eq:FSC-X-FSC-cond-prob} we have
\begin{equation}\label{eq:EH-FSC-separable}
p(y_{n}z_{n+1}\cn v_{n}z_{n}) = p(y_{n}\cn v_{n}z_{n})p(z_{n+1}\cn v_{n}z_{n}).
\end{equation}
More generally, if $\{E_{n}\}$ is Markov of order $r>0$, the surrogate channel is still an FSC for $n\geq \max\{m,r\}$, with the states
\[ Z_{n} = E_{n-r+1}^{n}S_{n-m+1}^{n}, \]
whose alphabet is $\Z=\E_{H}^{r}\times\St^{m}$. In fact, for $n\geq \max\{m,r\}$, the transition probability satisfies the following: if $z_{n}$ is compatible with $z_{n+1}$, i.e., for some $e^{n+1}\in\E_{H}^{n+1}$ and $s^{n+1}\in\St^{n+1}$, $z_{n} = e_{n-r+1}^{n}s^{n}_{n-m+1}$ while $z_{n+1} = e_{n-r+2}^{n+1}s^{n+1}_{n-m+2}$, then
\begin{align*}
p(y_{n}z_{n+1}\cn v^{n}z^{n}y^{n-1}) &= p(y_{n}e_{n-r+2}^{n+1}s_{n-m+2}^{n+1}\cn v^{n}e^{n}s^{n}y^{n-1}) \\
&= p(y_{n}e_{n+1}s_{n+1}\cn v^{n}e^{n}s^{n}y^{n-1},x_{n} = v_{n}(s_{n-m+1}^{n})) \\
&= p(y_{n}\cn x_{n})\cdot p(e_{n+1}\cn e_{n-r+1}^{n})\cdot \1{S(x_{n},s_{n},e_{n+1}) = s_{n+1}} \Big|_{x_{n} = v_{n}(s_{n-m+1}^{n})} \\
&= p(y_{n}z_{n+1}\cn v_{n}z_{n}),
\end{align*}
by the structure of the channel. If $z_{n}$ is not compatible with $z_{n+1}$, then both the first and the last term are 0. Again note that the argument $s_{n-m+1}^{n}$ for $v_{n}$ is contained in $z_{n}$, and so \eqref{eq:EH-FSC-separable} holds.

\begin{remark}
Observe that at time $n$, the energy state $S_{n}$ contains all the information about the energy constraint on the current immediate input symbol $X_{n}$, which is the only influence the full history of energy information has on the transmission. We conjecture that for the equivalent channel $\bW$, it is enough to only consider input functions $u_{n}$ that depends only on the current energy state $s_{n}$, as stated below formally. This form of optimal input function is conjectured for both channels EH-SC1 and EH-SC2, but we are not able to prove it yet.
\end{remark}
\begin{conjecture}
Setting $m=1$ in the surrogate channel $\bW'$ yields a capacity $C'=C$. 
\end{conjecture}

Next we give the stationarity and ergodicity conditions. Since \eqref{eq:EH-FSC-separable} is true, by Appendix~\ref{sec:joint-marginal} we can just consider a smaller FSC whose transition probability is $p(z_{n+1}\cn v_{n}z_{n})$. As before, we also have the following two set of simple conditions, as well as some stronger but more complicated ones---see Corollaries~\ref{cor:Gray-Dunham-Gobbi-thm-2-extension-cor-Markov} and~\ref{cor:FSC-separable-Markov-input} in the Appendices.

\begin{lemma}\label{lem:EH-SC1-erg-conds1}
Assume the input process $\{V_{n}\}$ of the surrogate channel $\bW'$ for EH-SC1 is stationary and ergodic. Then the joint process $\{V_{n},Y_{n}\}$ is AMS and ergodic, if any of the following holds.
\begin{enumerate}
\item[i)] $\bW'$ is indecomposable.  
\item[ii)] There is a finite vector $v_{m}^{N}$ with $\Pr(V_{m}^{N}=v_{m}^{N})>0$ satisfying the following property: given $V_{m}^{N}=v_{m}^{N}$, for any $z_{m},z_{m}'\in\Z$, there exists $z_{N+1}\in\Z$ such that when $Z_{m}=z_{m}$ or $z_{m}'$, we both have $Z_{N+1} = z_{N+1}$ with positive probability.
\end{enumerate}
\end{lemma}

\begin{lemma}
Assume the input process $\{V_{n}\}$ of the surrogate channel $\bW'$ for EH-SC1 is finite-order Markov, then so is the joint process $\{(V_{n},Z_{n+1})\}$. If the underlying Markov chain for the latter is irreducible, then $\{V_{n},Y_{n}\}$ is AMS and ergodic.
\end{lemma}

\begin{IEEEproof}
This lemma is a simple case of Corollary~\ref{cor:FSC-separable-Markov-input}.
\end{IEEEproof}

When we know more properties of the energy harvesting channel, we have more concrete conditions. Two such example theorems are:

\begin{theorem}\label{thm:erg-conds-Markov}
For the FSC $\bW'$ above, assume there exists $\alpha\in\E_{H}$ such that the order-$r$ Markov chain $\{E_{n}\}$ satisfies $\Pr(E_{r+1} =\alpha\cn E^{r} = e^{r}) > 0$ for all $e^{r}\in\E_{H}^{r}$. If, in addition, one of the following is satisfied, then $\bW'$ is indecomposable.
\begin{enumerate}
\item[i)] The energy model is \eqref{eq:EH-S-comb-2} and $\alpha \geq \bc$ .
\item[ii)] The energy model is \eqref{eq:EH-S-comb-1} or \eqref{eq:EH-S-comb-2}, and $\alpha > \max\{\gamma(x):x\in\X\}$.
\end{enumerate}
\end{theorem}

\begin{IEEEproof}
We prove $\bW'$ is indecomposable by showing the strong positive column property\footnote{See comments below Definition~\ref{def:Gallager-indecomposable} in the Appendix.} holds, i.e., there exist $N$ such that for any input sequence, there exist a state $z_{N}$ that can be reached from any initial state $z_{1}$. Now for i) let $N = \max\{m,r\} + 1$. We can see that $E_{2},\ldots,E_{N} = \alpha$ with a positive probability conditioned on any $z_{1}$, with the corresonding $S_{2},\ldots,S_{N} = \bc$. Hence the state $z_{N} = (\alpha,\ldots,\alpha;\bc,\ldots,\bc)$ can always reached for any initial state and any input sequence. For ii), let $N = \max\{m,r\} + \bc+1$. If $E_{2},\ldots,E_{N} = \alpha$, then starting from $n=2$, after at most $\bc$ transmission and energy replenishment cycles the battery is full, i.e., we have $B_{n} = \bc$ for all $n > \bc+1$. Now we can use an argument similar to i) to prove the result, where for energy model \eqref{eq:EH-S-comb-1} set $z_{N} = (\alpha,\ldots,\alpha;\bc+\alpha,\ldots,\bc+\alpha)$, and for energy model \eqref{eq:EH-S-comb-2} set $z_{N} = (\alpha,\ldots,\alpha;\bc,\ldots,\bc)$.
\end{IEEEproof}

\begin{theorem}\label{thm:erg-conds-iid}
For the FSC $\bW'$ with $m=1$, assume $\{E_{n}\}$ is i.i.d., the energy model is \eqref{eq:EH-S-comb-1} or \eqref{eq:EH-S-comb-2}, and the distribution of $E_{n}$ is supported on the full set $\E_{H}$. 
\begin{enumerate}
\item[i)] If there exists $N$ such that for each input sequence $\{v_{n}\}$ and any $S_{1}=s_{1}$, $B_{N} = \bc$ with a positive probability, then $\bW'$ is indecomposable.  
\item[ii)] If $\E_{H}$ is a continuous interval of non-negative integers and $\max\E_{H} - \min\E_{H} \geq \bc$ for the energy model \eqref{eq:EH-S-comb-1}, or $\max\E_{H} \geq \bc$ for the model \eqref{eq:EH-S-comb-2}, then $\bW'$ is indecomposable.
\item[iii)] If $\{V_{n}\}$ is stationary and ergodic, and there is $v^{N}$ with $\Pr(V^{N}=v^{N})>0$ such that for any $S_{1}=s_{1}$, either $B_{N} = 0$ or $B_{N} = \bc$ with positive probability, then $\{V_{n},Y_{n}\}$ is AMS and ergodic.
\item[iv)] Both i) and iii) hold if $\max\E_{H} > \max\{\gamma(x):x\in\X\}$.
\end{enumerate}
\end{theorem}

\begin{IEEEproof}
Note that in this case $Z_{n} = S_{n}$.

i): Whenever such $N$ exists, the strong positive column condition holds and so $\bW'$ is indecomposable.

ii): With a positive probability $S_{2}$ can always be boosted up to $s_{2} = \bc+\min\E_{H}$ for the model \eqref{eq:EH-S-comb-1}, or $\bc$ for the model \eqref{eq:EH-S-comb-2}, hence the strong positive column condition holds.

iii): This is a straightforward application of Lemma~\ref{lem:EH-SC1-erg-conds1}, condition ii).

iv): If $\max\E_{H} > \max\{\gamma(x):x\in\X\}$, then for any $\{v_{n}\}$ and $s_{1}$, at most after $n=\bc$ transmissions, $S_{n}-\gamma(X_{n})\geq \bc$ with a probability no smaller than ${[\Pr(E_{n}=\max\E_{H})]^{n}>0}$, in which case $B_{n+1} = \bc$.
\end{IEEEproof}

\begin{remark}
Note that there is some overlap between these two theorems. The conditions in Theorem~\ref{thm:erg-conds-Markov} and conditions~ii) and iv) of Theorem~\ref{thm:erg-conds-iid} are satisfied if $E_{n}$ can always reach a relatively high energy level (compared to $\X$ or $\bc$) with even a very small positive probability, which is not a harsh requirement for many natural energy sources. Alternatively, if the input process $\{V_{n}\}$ is stationary ergodic, and put a positive probability on a moderately long sequence of ``all zero'' functions (that is, $v_{n}(s_{n}) = 0$ for all $s_{n}$), or ``all-consuming'' functions (that is, $\gamma(v_{n}(s_{n})) = s_{n}$ for all $s_{n}$), then condition iii) of Theorem~\ref{thm:erg-conds-iid} is satisfied.
\end{remark}

\subsection{EH-SC2}
\label{subsec:ach-rate-EH-SC2}

As commented in Section~\ref{subsec:three-models}, EH-SC1 is a scenario with strictly less side information than EH-SC2.  Hence any further restriction on the input alphabet of EH-SC1 also works for EH-SC2, and hence all results from the previous subsection apply to the second scenario. In addition, more generally, since now we also have causal knowledge of $\{E_{n}\}$, we can restrict the input function to (essentially) depend only on the $m>0$ most recent energy states and an energy harvesting history of memory length $l\geq0$, to obtain a constant alphabet size. Consider such a special input symbol $u^{N} = (u_{1},\cdots,u_{N})$, whose $n$-th coordinate function $u_{n}$ is only a function of $S_{n-m+1}^{n}$ and $E_{n-l+1}^{n}$. To be precise, each $u_{n}$ is associated with an auxiliary function $v_{n}\in\V$, where $\V$ is the collection of all functions $v: \St^{m}\times\E_{H}^{l}\to \X$ such that
\[ v(s^{m},e^{l}) \in\X(s_{m}),\qquad\forall (s^{m},e^{l})\in\St^{m}\times\E_{H}^{l}, \]
with $\X(s)$ defined in \eqref{eq:EH-SC1-alphabet}. The input function $u_{n}$ is defined through $v_{n}$ in the following way: for each $(b_{1},e^{n})$, it first computes $s_{n} = s_{n}(b_{1},e^{n},u^{n-1})$ through the recursion \eqref{eq:EH-SC2-recursion}, then together with $e_{n-l+1}^{n}$ and the previously computed $s_{n-m+1}^{n-1}$ (which may also includes the dummy pre-historical states when necessary), $u_{n}$ assigns the function value
\[ u_{n}(b_{1},e^{n}) = v_{n}(s_{n-m+1}^{n},e_{n-l+1}^{n}). \]
Hence the vector $v^{N} = (v_{1},\cdots,v_{N})$ uniquely determines the input symbol $u^{N}$, and for each $N$ there is a one-to-one correspondence between $\V^{N}$ and the collection of all such special input symbols $u^{N}$.

Such a restriction again gives us a surrogate channel $\bW'$, whose channel transition probability is defined through \eqref{eq:EH-SC2-cond-prob}. Similar to the subsection above, we can show that if $\{E_{n}\}$ is Markov of order $r\geq0$ (including i.i.d.), then $\bW'$ is an FSC for $n\geq \max\{m,l,r\}$, with the states
\[ Z_{n} = E_{n-\max\{l,r\}+1}^{n}S_{n-m+1}^{n}. \]
Note that for this FSC \eqref{eq:EH-FSC-separable} still holds.

We can derive similar stationarity and ergodicity conditions as in the previous subsection, which is omitted. Also, in this case we have the same optimal input conjecture.

\begin{conjecture}
Setting $m=1$ and $l=0$ in the surrogate channel $\bW'$ yields a capacity $C'=C$. 
\end{conjecture}

\section{Capacity Bounds}
\label{sec:capacity-bounds}

Compared to the lower bounds/achievable rates, nontrivial capacity upper bounds are much more difficult to obtain in the study of energy harvesting channels. For the special case of binary noiseless EH-SC1, \cite{Tutuncuoglu-EnHarvTimingCh} derives an upper bound, assuming full CSI at the receiver (CSIR). \cite{Tutuncuoglu-EnHarvTimingCh2} tries to tighten this bound, though there appear to be gaps in the mathematical proofs. In this section we derive capacity bounds for our more general energy harvesting models, as well as the channel FSC-X. In particular, we obtain upper bounds for the energy harvesting channels when $\{E_{n}\}_{n=1}^{\infty}$ is finite-order Markov\footnote{Note that when $\{E_{n}\}$ is i.i.d., a proof idea for the same bounds also appears in \cite{Ozel-EH-CSIR-ESI}.}. These results are motivated by Gallager's study of finite state channels\cite{Gallager-IT_Reliable}, where as mentioned in the previous section, two convergent sequences in the form of maximized finite block length mutual information are shown to give series of upper and lower bounds of the channel capacity, respectively.

We begin our study by describing a general upper-bounding approach, which is based on techniques of Verd\'u and Han, and Gallager. Then we use this approach to derive the upper bounds for FSC-X, which also includes EH-SC1 as a special case, when the energy harvesting process is i.i.d. (as shown in Section~\ref{subsec:three-models}). After that we study the lower and upper bounds for EH-SC2. Note that all the bounds in this section are in the form of maximized block mutual information on the equivalent channels. Theoretically, these bounds are computable for each block length $N$; however, as $N$ grows the computation memory (and also time) increases rapidly, since the input alphabet size has a double-exponential growth rate. This complexity issue is addressed for the upper bounds in the next section.

\subsection{A General Gallager-type Upper Bound}\label{subsec:UB-Gallager-type}

Let $\left\{\X^{(N)}, p(y^{N}\,|\,x^{N}), \Y^{(N)}\right\}_{N=1}^{\infty}$ be a general channel without feedback, with input/output alphabets $\X^{(N)}$, $\Y^{(N)}$ and transition probabilities $p(y^{N}\,|\,x^{N})$ for each block length $N$. Using Fano's inequality Verd\'u and Han \cite{Verdu-Han-Capacity} showed that its capacity is upper bounded by
\begin{equation}\label{eq:UB-General}
\liminf_{N\to\infty}C_{N},\qquad C_{N} \triangleq \sup_{P_{X^{N}}}\frac{1}{N}I(X^{N};Y^{N}).
\end{equation}
In general, the upper bound is not easy to compute, since the limiting behavior of $C_{N}$ is unknown. On the other hand, Gallager \cite{Gallager-IT_Reliable} uses the following lemma to derive a series of computable upper bounds for finite state channels:

\begin{lemma}[Fekete's lemma]\label{lem:Fekete}
If the sequence $\{ a_n\}_{n=1}^\infty$ is subadditive, i.e., $a_{m+n}\leq a_{m}+a_{n}$ for all $m$ and $n$, then the limit $\lim_{n \to \infty} \frac{a_n}{n}$ exists and is equal to $\inf_{n} \frac{a_n}{n}$. Similarly, if the sequence is superadditive, then $\lim_{n \to \infty} \frac{a_n}{n} = \sup_{n}\frac{a_n}{n}$.
\end{lemma}

If we can show that for each $N$, there is a $\C_{N}$ such that
\begin{enumerate}[\IEEEsetlabelwidth{(R2)}]
\item[(R1)] $C_{N} \leq \C_{N}$,
\item[(R2)] $\{N\C_{N}\}_{N=1}^{\infty}$ is subadditive,
\end{enumerate}
then by Fekete's lemma, $\lim_{N \to \infty} \C_{N}$ exists and is equal to $\inf_{N} \C_{N}$. Hence \eqref{eq:UB-General} is upper bounded by $\liminf_{N\to\infty} \C_{N} = \inf_{N} \C_{N}$,
and so $\C_{N}$ is an upper bound for the general channel capacity for each finite $N$. In other words, the limiting process in \eqref{eq:UB-General} is not needed anymore, which greatly simplifies the computation of upper bounds, especially when such computable $\C_{N}$'s can be easily found.

\subsection{Upper Bounds for FSC-X / EH-SC1}
\label{subsec:FSC-X-UB}

We apply the technique above to the equivalent channel (see Definition~\ref{def:equiv-FSC-X}) to derive a series of Gallager-type upper bounds for FSC-X, which by Proposition~\ref{proposition:EH-SC1-FSC-X} also includes the channel EH-SC1 when the energy harvesting process is i.i.d.. To begin with, the capacity can be upper bounded by a system with full CSIR, and so in \eqref{eq:UB-General} we consider the mutual information $I(U^{N};Y^{N}S^{N+1})$. As $S_{1}$ is independent of $U^{N}$,
\begin{align*}
I\lt(U^{N};Y^{N}S^{N+1}\rt) &= I\lt(U^{N};Y^{N}S_{2}^{N+1}\,|\,S_{1}\rt)
\leq \max_{s_{1}}I\lt(U^{N};Y^{N}S_{2}^{N+1}\,|\,s_{1}\rt),
\end{align*}
where $I(\cdot\ ;\,\cdot\,|\,s_{1}) := I(\cdot\ ;\,\cdot\,|\,S_{1} = s_{1})$. Define
\begin{equation}\label{eq:UB-FSC-X}
\C_{N} = \max_{P_{U^{N}}}\max_{s_{1}}\frac{1}{N} I\lt(U^{N};Y^{N}S_{2}^{N+1}\,|\,s_{1}\rt),
\end{equation}
then $\C_{N}$ satisfies (R1) in the previous subsection. Furthermore, we have the following theorem:

\begin{theorem}\label{thm:UB-FSC-X}
For each $N$, $\C_{N}$ defined in \eqref{eq:UB-FSC-X} is an upper bound for the capacity of the channel FSC-X.
\end{theorem}

\begin{IEEEproof}
As described above, we can use \eqref{eq:UB-General} for the full CSIR case as an upper bound. Since $\C_{N}$ satisfies (R1) for this upper bound, if we can show it satisfies (R2) as well, then $\C_{N}$ is an upper bound for each $N$ by the argument above.

Let $N$ be arbitrary and let $m,n$ be positive integers that sum to $N$. In the following we will show that
\begin{equation}\label{eq:subadditivity}
N\C_{N} \leq  n\C_{n} + m\C_{m},
\end{equation}
i.e., $\{N\C_{N}\}_{N=1}^{\infty}$ is subadditive.
For any $P_{U^{N}}$ and $s_{1}$ consider the decomposition
\begin{IEEEeqnarray}{r/C/l}
I\lt(U^{N};Y^{N}S_{2}^{N+1}\mid s_{1}\rt) &=&  I\lt(U^{N};Y^{n}S_{2}^{n+1}\mid s_{1}\rt)
 + I\lt(U^{N};Y_{n+1}^{N}S_{n+2}^{N+1}\mid Y^{n}S_{2}^{n+1}s_{1}\rt) \nonumber\\
&=& I\lt(U^{n};Y^{n}S_{2}^{n+1}\mid s_{1}\rt) + I\lt(U_{n+1}^{N};Y^{n}S_{2}^{n+1}\mid U^{n}s_{1}\rt) \nonumber\\
&& +\: I\lt(U_{n+1}^{N};Y_{n+1}^{N}S_{n+2}^{N+1}\mid Y^{n}S_{2}^{n+1}s_{1}\rt) \nonumber\\
&& +\: I\lt(U^{n};Y_{n+1}^{N}S_{n+2}^{N+1}\mid U_{n+1}^{N}Y^{n}S_{2}^{n+1}s_{1}\rt) \nonumber\\
&=& I_{1} + I_{2} + I_{3} + I_{4}, \label{eq:I-decomp-FSC}
\end{IEEEeqnarray}
where $ I_{1}$--$I_{4}$ are respectively defined as the first to fourth terms in the line above them.
By the definition \eqref{eq:UB-FSC-X}, $I_{1}\leq n\C_{n}$. Next, using the property of FSC conditional probabilities as in \eqref{eq:FSC-pYScondUs1}, for $I_{2}$ and $I_{4}$ we respectively have
\begin{align*}
p\lt(y^{n}s_{2}^{n+1}\,|\,u_{n+1}^{N}u^{n}s_{1}\rt) 
= p\lt(y^{n}s_{2}^{n+1}\,|\,u^{n}s_{1}\rt),
\end{align*}
\begin{align*}
p\big(y_{n+1}^{N}s_{n+2}^{N+1}\,|\,u^{n}u_{n+1}^{N}y^{n}s_{2}^{n+1}s_{1}\big)
= p\lt(y_{n+1}^{N}s_{n+2}^{N+1}\,|\,u_{n+1}^{N}y^{n}s_{2}^{n+1}s_{1}\rt).
\end{align*}
Therefore $I_{2} = I_{4} = 0$. Furthermore,
\begin{equation}\label{eq:I3-FSC}
I_{3} = \sum_{y^{n}s_{2}^{n+1}}p\lt(y^{n}s_{2}^{n+1}\,|\,s_{1}\rt)I\lt(U_{n+1}^{N};Y_{n+1}^{N}S_{n+2}^{N+1}\,|\,y^{n}s^{n+1}\rt).
\end{equation}

Fix $y^{n}s^{n+1}$. For each $u_{n+1}^{N}$ and $k = 1,\ldots,m$, define
$ {\tu_{k}:\St^{k}\to\X} $
to be the projection $\tu_{k}\lt(\cdot\rt) = u_{n+k}\lt(s^{n},\cdot\rt)$, i.e.,
\[ \tu_{k}\lt(t^{k}\rt) = u_{n+k}\lt(s^{n},t^{k}\rt),\qquad\forall t^{k}\in\St^{k}. \]
Then $\forall t^{k}\in\St^{k}$, $\tu_{k}\lt(t^{k}\rt)\in\X(t_{k})$ and so $\tu_{k}\in\U_{k}$. By \eqref{eq:FSC-pYScondUs1} again
\begin{align*}
p\big(y_{n+1}^{N}s_{n+2}^{N+1}|\,u_{n+1}^{N}y^{n}s^{n+1}\big) 
&= p\big(y_{n+1}^{N}s_{n+2}^{N+1}|\,u_{n+1}^{N}s^{n+1}\big) \\
&= p\big(y_{n+1}^{N}s_{n+2}^{N+1}|\,u_{n+1}^{N}(s^{n},\cdot),s_{n+1}\big) \\
&= Q\lt(y_{n+1}^{N}s_{n+2}^{N+1}|\,\tu^{m}s_{n+1}\rt),
\end{align*}
where $Q \triangleq P_{Y^{m}S_{2}^{m+1}\,|\,U^{m}S_{1}}$ is the $m$-block channel transition probability given $S_{1}$. Denote the projection map
$ T:u_{n+1}^{N}\mapsto\tu^{m}$,
which depends on $s^{n}$. Then $T$ and $P_{U_{n+1}^{N}\,|\,y^{n}s^{n+1}}$ induce a probability distribution $\tP$ on $\U^{(m)}$: for all $\tu^{m}\in\U^{(m)}$,
\begin{align*}
\tP\lt(\tu^{m}\rt) &= \Pr\lt(T\lt(U_{n+1}^{N}\rt) = \tu^{m}\,\big|\,Y^{n}S^{n+1} = y^{n}s^{n+1}\rt)
= \sum_{u_{n+1}^{N}:T\lt(u_{n+1}^{N}\rt) = \tu^{m}}p\lt(u_{n+1}^{N}\,|\,y^{n}s^{n+1}\rt).
\end{align*}
Now it is easy to verify that
\begin{align*}
p\big(y_{n+1}^{N}s_{n+2}^{N+1}\,|\,y^{n}s^{n+1}\big)
&= \sum_{u_{n+1}^{N}}p\lt(u_{n+1}^{N}\,|\,y^{n}s^{n+1}\rt)p\lt(y_{n+1}^{N}s_{n+2}^{N+1}\,|\,u_{n+1}^{N}y^{n}s^{n+1}\rt) \\
&= \sum_{\tu^{m}}\tP\lt(\tu^{m}\rt)Q\lt(y_{n+1}^{N}s_{n+2}^{N+1}\,|\,\tu^{m}s_{n+1}\rt) \\
&= \tR\left(y_{n+1}^{N}s_{n+2}^{N+1}\,|\,s_{n+1}\right),
\end{align*}
where $\tR(\cdot\,|\,s_{n+1})$ is the $m$-block channel output distribution given $S_{1} = s_{n+1}$, induced by $\tP$ and the channel $Q$. Thus if we denote the relative entropy
\[ D_{u_{n+1}^{N}\,|\,y^{n}s^{n+1}}\triangleq
D\left(P_{Y_{n+1}^{N}S_{n+2}^{N+1}\,|\,u_{n+1}^{N}y^{n}s^{n+1}}\, \big\|\, P_{Y_{n+1}^{N}S_{n+2}^{N+1}\,|\,y^{n}s^{n+1}}\right), \]
then
$ D_{u_{n+1}^{N}\,|\,y^{n}s^{n+1}} = D\big(Q\lt(\,\cdot\,|\,\tu^{m}s_{n+1}\rt)\, \|\, \tR(\cdot\,|\,s_{n+1})\big)$. 
Therefore we can write
\begin{align*}
I\big(U_{n+1}^{N};Y_{n+1}^{N}S_{n+2}^{N+1}\,|\,y^{n}s^{n+1}\big)
&= \sum_{u_{n+1}^{N}}p\lt(u_{n+1}^{N}\,|\,y^{n}s^{n+1}\rt)D_{u_{n+1}^{N}\,|\,y^{n}s^{n+1}} \\
&= \sum_{\tu^{m}}\tP(\tu^{m})D\big(Q(\,\cdot\,|\,\tu^{m}s_{n+1})\, \|\, \tR(\cdot\,|\,s_{n+1})\big) \\
&= I_{\tP}\lt(U^{m};Y^{m}S_{2}^{m+1}\,|\,S_{1}=s_{n+1}\rt) \\
&\leq \max_{s_{1}} I_{\tP}\lt(U^{m};Y^{m}S_{2}^{m+1}\,|\,s_{1}\rt)\\
&\leq m\C_{m},
\end{align*}
where $I_{\tP}$ denotes the mutual information induced by the input distribution $\tP$. Since this inequality holds for all $y^{n}s^{n+1}$, by \eqref{eq:I3-FSC} we have
$ I_{3} \leq m\C_{m}$. 

Combining the results for $I_{1}$--$I_{4}$ with \eqref{eq:I-decomp-FSC}, we have
\[ I\lt(U^{N};Y^{N}S_{2}^{N+1}\,|\,s_{1}\rt) \leq n\C_{n} + m\C_{m}. \]
This inequality is true for all $P_{U^{N}}$ and $s_{1}$, so it must be true for the maximization over them, and thus \eqref{eq:subadditivity} holds.
\end{IEEEproof}

\begin{remark}\label{rem:UB-FSC-X}
Note that since the order of maximization in \eqref{eq:UB-FSC-X} can be exchanged, $\C_{N}$ can be calculated by finding the capacities of $|\St|$ discrete memoryless channels (DMC), which can be efficiently computed using the Blahut-Arimoto algorithms (see, e.g., \cite{Blahut-BAA}).
\end{remark}

\subsection{Bounds for EH-SC2}\label{subsec:EH-UB}

For this channel we also use the equivalent channel model (Definition~\ref{def:equiv-EH-SC2}) to derive series of capacity bounds, including a new type of lower bounds and the Gallager-type upper bounds. Note that the energy harvesting process and the initial battery level are independent, and are both independent of the input in the equivalent channel. Let us define a notation $e^{-r} := e_{-r+1}^{0}$ for $r\geq0$. We have
\begin{equation}\label{eq:EH-pYEcondUb1e}
 p\lt(y^{N}e^{N}|u^{N}b_{1}e^{-r}\rt) = p(e^{N}|e^{-r})\prod_{n=1}^{N}p\lt(y_{n}|x_{n} = u_{n}(b_{1},e^{n})\rt).
\end{equation}

We first develop some preliminary results on the input alphabet and block conditional mutual information. Recall that $\U^{(N)}$ is the collection of all causal mappings ${\E_{B}\times\E_{H}^{N}\to\X^{N}}$ that are consistent with the energy constraint \eqref{eq:EH-en-constr}. Let $\V^{(N)}$ denote all causal mappings $\E_{H}^{N}\to\X^{N}$, and define $\U^{(N)}_{b_{1}}$ as the ``$b_{1}$-th section of $\U^{(N)}$'', which consists of all mappings in $\V^{(N)}$ that together with $B_{1}=b_{1}$ satisfy the energy constraint, i.e.,
\[ \U_{b_{1}}^{(N)} \triangleq \lt\{v^{N} = u^{N}(b_{1},\cdot)\mid u^{N}\in\U^{(N)}\rt\}. \]
Let $b_{1}\leq b'_{1}$. For each $e^{N}$, $x^{N}$ satisfies \eqref{eq:EH-en-constr} with $b'_{1}$ whenever it does with $b_{1}$, so $\U_{b_{1}}^{(N)}\subseteq\U_{b'_{1}}^{(N)}$. In particular, 
\begin{equation}\label{eq:U-b1-inclusion}
\U_{b_{1}}^{(N)}\subseteq\U_{\bc}^{(N)},\qquad \forall b_{1}\in\E_{B}.
\end{equation}

Now fix $b_{1}$. Define the projection map $T:\U^{(N)}\to\V^{(N)}$ with $T(u^{N}) = u^{N}(b_{1},\cdot)$, and denote $\hu^{N}=T\lt(u^{N}\rt)$. Then the image of $T$ is in $\U_{b_{1}}^{(N)}$.
Furthermore, for $v^{N}\in\V^{(N)}$ and $r\geq0$ we define
\begin{equation}\label{eq:EH-pYEcondVe}
p\lt(y^{N}e^{N}\,|\,v^{N}e^{-r}\rt) = p\lt(e^{N}\,|\,e^{-r}\rt)\prod_{n=1}^{N}p\lt(y_{n}\,|\,x_{n} = v_{n}(e^{n})\rt)
\end{equation}
and $p\lt(y^{N}\,|\,v^{N}\rt) = \sum_{e^{N}}p\lt(y^{N}e^{N}\,|\,v^{N}\rt)$
through $p(e^{N})$ and $p(y|x)$. Then by \eqref{eq:EH-pYEcondUb1e} we have 
\begin{align*}
p\lt(y^{N}e^{N}\,|\,u^{N}b_{1}e^{-r}\rt) &= p\lt(y^{N}e^{N}\,|\,\hu^{N}e^{-r}\rt), \\
p\lt(y^{N}\,|\,u^{N}b_{1}\rt) &= p\lt(y^{N}\,|\,\hu^{N}\rt).
\end{align*}
By the same argument as in the proof of Theorem~\ref{thm:UB-FSC-X},
\begin{align}
I(U^{N};Y^{N}E^{N}\,|\,b_{1}e^{-r}) &= I(\hat{U}^{N};Y^{N}E^{N}\,|\,e^{-r}), \\
I(U^{N};Y^{N}\,|\,b_{1}) &= I(\hat{U}^{N};Y^{N}), \label{eq:EH-I-U-V}
\end{align}
where the distribution of $\hU^{N} = T(U^{N})$ is supported on $\U_{b_{1}}^{(N)}$.

\begin{lemma}\label{lem:I-U-V-max}
Let $\Ps_{b_{1}}^{(N)}$ denote the family of all probability distributions on $\U_{b_{1}}^{(N)}$. We have
\begin{align*}
\max_{P_{U^{N}}}I(U^{N};Y^{N}E^{N}|b_{1}e^{-r}) &= \max_{P_{V^{N}}\in\Ps_{b_{1}}^{(N)}} I(V^{N};Y^{N}E^{N}|e^{-r}), \\
\max_{P_{U^{N}}}I(U^{N};Y^{N}\,|\,b_{1}) &= \max_{P_{V^{N}}\in\Ps_{b_{1}}^{(N)}} I(V^{N};Y^{N}).
\end{align*}
\end{lemma}

\begin{IEEEproof}
We only prove the second equation since the proof of the first is essentially the same. Denote the LHS and RHS by $C_{U}$ and $C_{V}$, respectively. For any $P_{U^{N}}$, we have $P_{\hU^{N}}\in\Ps_{b_{1}}^{(N)}$ and so
$I(U^{N};Y^{N}\,|\,b_{1})\leq C_{V}$
by \eqref{eq:EH-I-U-V}. Hence $C_{U}\leq C_{V}$. On the other hand, for every $v^{N}\in\U_{b_{1}}^{(N)}$ we have $T^{-1}\lt(v^{N}\rt) \neq \emptyset$. Thus for any $P_{V^{N}}\in\Ps_{b_{1}}^{(N)}$, define a $P_{U^{N}}$ such that
\[ P_{U^{N}}\lt(T^{-1}\lt(v^{N}\rt)\rt) = P_{V^{N}}\lt(v^{N}\rt) \]
for all $v^{N}\in\U_{b_{1}}^{(N)}$, then $P_{V^{N}}$ is induced by $P_{U^{N}}$ and $T$. Then by \eqref{eq:EH-I-U-V} again,
$C_{U} \geq I(V^{N};Y^{N})$
and so $C_{U}\geq C_{V}$.
\end{IEEEproof}

We are now ready to present the capacity bounds.

\begin{theorem}\label{thm:LB-EH}
For the channel EH-SC2, if $\{E_{n}\}_{n=1}^{\infty}$ is i.i.d., then for each $N$
\[ \uC_{N} := \max_{P_{U^{N}}}\frac{1}{N}I(U^{N};Y^{N}\,|\,B_{1} = 0) \]
is a capacity lower bound. Moreover, $\lim_{N\to\infty}\uC_{N} = \sup_{N}\uC_{N}$.
\end{theorem}

\begin{IEEEproof}
Consider using the channel in blocks of length $N$ and restrict the input functions to those that i) ignore the initially stored energy in the battery, and ii) essentially comprise concatenations of functions in $\U_{0}^{(N)}$. That is, for $k>0$ the input $u^{kN}$ is only a function of $e^{kN}$ and can be identified with the collection
$\big\{\bv_{i}\in\U_{0}^{(N)}, 1\leq i\leq k\big\}$,
where for any $b_{1}$ and $e^{kN}$,
\[ u^{kN}\lt(b_{1},e^{kN}\rt) = \lt(\bv_{1}\lt(e^{N}\rt),\ldots,\bv_{k}\lt(e_{(k-1)N+1}^{kN}\rt)\rt). \]
It is a legitimate input symbol since between the transition of blocks the function ignores the remaining battery energy, thus is always compatible with the energy constraint \eqref{eq:EH-en-constr}.

Let $\bx_{i}$, $\by_{i}$ and $\be_{i}$ denote $x_{(i-1)N+1}^{iN}$, $y_{(i-1)N+1}^{iN}$ and $e_{(i-1)N+1}^{iN}$, respectively. By \eqref{eq:EH-pYEcondUb1e} and the i.i.d. assumption for $\{E_{n}\}$,
\[ p\lt(y^{kN}e^{kN}\,|\,u^{kN}b_{1}\rt) = p(e^{kN})\prod_{n=1}^{kN}p\lt(y_{n}|x_{n} = u_{n}(b_{1},e^{n})\rt)
= \prod_{i=1}^{k}p(\be_{i})p(\by_{i}\,|\,\bx_{i} = \bv_{i}(\be_{i})). \]
Hence if we define the transition probability $Q_{i}\lt(\by_{i}\,|\,\bv_{i}\rt) = \sum_{\be_{i}}p(\be_{i})p(\by_{i}\,|\,\bx_{i} = \bv_{i}(\be_{i}))$, then
\[ p\lt(y^{kN}\,|\,u^{kN}\rt) = \prod_{i=1}^{k}Q_{i}\lt(\by_{i}\,|\,\bv_{i}\rt). \]
Since $\{E_{n}\}$ is i.i.d. and $p(\by_{i}\,|\,\bx_{i})$ is obtained from the DMC $p(y|x)$, $Q_{i}$ is the same for each $i$ and hence is denoted by $Q$. Thus $kN$ times of using the original channel in the specified manner is equivalent to $k$ times of using a DMC $Q\lt(\by\,|\,\bv\rt)$ with input alphabet $\U_{0}^{(N)}$, whose capacity is
\begin{equation}\label{eq:max-LB-EH}
\max_{P_{V^{N}}\in\Ps_{0}^{(N)}} I(V^{N};Y^{N}).
\end{equation}
By Lemma~\ref{lem:I-U-V-max} and considering the block length $N$, $\uC_{N}$ is achievable.

Finally we use Lemma~\ref{lem:Fekete} to prove $\lim_{N\to\infty}\uC_{N} = \sup_{N}\uC_{N}$. It suffices to show that $\{N\uC_{N}\}_{N=1}^{\infty}$ is superadditive. Fix $N$ and let $m,n>0$ with $m+n=N$. As $N\uC_{N}$ can be written as \eqref{eq:max-LB-EH}, let $P^{*}_{V^{m}}$ and $P^{*}_{V^{n}}$ be the distributions that achieve the maximum of \eqref{eq:max-LB-EH} for block lengths $m$ and $n$, respectively. For block length $N$ consider the subset $\V'$ of $\U_{0}^{(N)}$ that comprises all concatenations of functions in $\U_{0}^{(m)}$ and $\U_{0}^{(n)}$. Specifically, each input function $v^{N}$ in $\V'$ can be represented by a pair $(\bv_{1},\bv_{2})\in\U_{0}^{(m)}\times\U_{0}^{(n)}$ such that
$v^{N}\lt(e^{N}\rt) = \lt(\bv_{1}\lt(e^{m}\rt),\bv_{2}\lt(e_{m+1}^{N}\rt)\rt)$
for any $e^{N}$, and for each such pair there is a corresponding function $v^{N}\in\V'$. Then similar to the argument above, with \eqref{eq:EH-pYEcondVe} we can show that
\[ p\lt(y^{N}\,|\,v^{N}\rt) = p\lt(y_{1}^{m}\,|\,\bv_{1}\rt)p\lt(y_{m+1}^{N}\,|\,\bv_{2}\rt),\qquad \forall v^{N}\in\V'. \]
Define a distribution $P_{V^{N}}\in\Ps_{0}^{(N)}$ that satisfies
$P_{V^{N}}(v^{N}) = P^{*}_{V^{m}}(\bv_{1})\cdot P^{*}_{V^{n}}(\bv_{2})$ for all $v^{N}\in\V'$, and $P_{V^{N}}(v^{N}) = 0$ for $v^{N}\notin\V'$, then
\[ I(V^{N};Y^{N})\Big|_{P_{V^{N}}} = I(V^{m};Y^{m})\Big|_{P^{*}_{V^{m}}} +\, I(V^{n};Y^{n})\Big|_{P^{*}_{V^{n}}}. \]
Since in the equation above, $LHS\leq N\uC_{N}$ while $RHS = m\uC_{m}+n\uC_{n}$, superadditivity holds.
\end{IEEEproof}

\begin{remark}
As mentioned at the end of Section~\ref{sec:channel-capacity}, Theorem~\ref{thm:LB-EH} can serve as a simple achievability proof for the multi-letter mutual information capacity formula (18) in \cite{Shaviv-Minh-Ozgur}. In fact, from above we know $\lim_{N\to\infty}\uC_{N}$ is achievable for any initial battery distribution. Using Verd\'u and Han's method (see the beginning part of Section~\ref{subsec:UB-Gallager-type}, or \cite{Shaviv-Minh-Ozgur}), one can show that $\liminf_{N\to\infty}\uC_{N}$ is also a capacity upper bound when the initial battery $B_{1} = 0$. Thus the channel capacity for the case $B_{1}=0$ is
\[ \lim_{N\to\infty}\uC_{N} = \lim_{N\to\infty}\frac{1}{N}\max_{P_{V^{N}}\in\Ps_{0}^{(N)}} I(V^{N};Y^{N}). \]
The result for arbitrary $B_{1}$ follows from \cite[Proposition~1]{Shaviv-Minh-Ozgur}.
\end{remark}

\begin{theorem}\label{thm:UB-EH}
If $\{E_{n}\}_{n=1}^{\infty}$ is a homogeneous Markov chain of order $r\geq0$, then for each $N$
\[ \C_{N} := \max_{P_{U^{N}}}\max_{e^{-r}}\frac{1}{N}I(U^{N};Y^{N}\,|\,E^{N},B_{1} = \bc,e^{-r}) \]
is an upper bound of the channel capacity for EH-SC2.
\end{theorem}

\begin{IEEEproof}
We use the same upper bounding technique as in the FSC-X / EH-SC1 case and the proof parallels that of Theorem~\ref{thm:UB-FSC-X}. By providing full CSIR to the receiver, in \eqref{eq:UB-General} we consider
\begin{align*}
I\lt(U^{N};Y^{N}E_{-r+1}^{N}B_{1}\rt) &= I\lt(U^{N};Y^{N}E^{N}\,|\,B_{1}E^{-r}\rt) 
\leq \max_{b_{1},e^{-r}}I\lt(U^{N};Y^{N}E^{N}\,|\,b_{1}e^{-r}\rt),
\end{align*}
due to the independence between $B_{1}E^{-r}$ and $U^{N}$. Now define
\begin{equation}\label{eq:UB-EH}
\C_{N} = \max_{P_{U^{N}}}\max_{b_{1},e^{-r}}\frac{1}{N}I\lt(U^{N};Y^{N}E^{N}\,|\,b_{1}e^{-r}\rt).
\end{equation}
We will show that it is equivalent to the definition in the theorem. For each $b_{1}e^{-r}$, by Lemma~\ref{lem:I-U-V-max}, \eqref{eq:U-b1-inclusion} and the independence between $\{E_{n}\}$ and the input symbols,
\begin{align*}
\max_{P_{U^{N}}}I(U^{N};Y^{N}E^{N}\,|\,b_{1}e^{-r}) &\leq \max_{P_{V^{N}}\in\Ps_{\bc}^{(N)}} I(V^{N};Y^{N}E^{N}\,|\,e^{-r}) \\
& = \max_{P_{U^{N}}}I(U^{N};Y^{N}E^{N}\,|\,B_{1} = \bc,e^{-r}) \\
& = \max_{P_{U^{N}}}I(U^{N};Y^{N}\,|\,E^{N},B_{1} = \bc,e^{-r})
\end{align*}
with the equality attained when $b_{1} = \bc$. Now, taking the maximum of both sides over $e^{-r}$ and exchanging the order of maximization, we see the equivalence of both definitions.

From the analysis above $\C_{N}$ satisfies (R1) in Section~\ref{subsec:UB-Gallager-type}. Next we will show the subadditivity \eqref{eq:subadditivity} and then the theorem is proved. Let $N$ be arbitrary and let $m,n$ be positive integers that sum to $N$. We have the decomposition
\begin{IEEEeqnarray}{r/C/l}
I\lt(U^{N};Y^{N}E^{N}\,|\,b_{1}e^{-r}\rt)
&=& I\lt(U^{n};Y^{n}E^{n}\,|\,b_{1}e^{-r}\rt) + I\lt(U_{n+1}^{N};Y^{n}E^{n}\,|\,U^{n}b_{1}e^{-r}\rt) \nonumber\\
&& +\: I\lt(U_{n+1}^{N};Y_{n+1}^{N}E_{n+1}^{N}\,|\,Y^{n}E^{n}b_{1}e^{-r}\rt) \nonumber\\
&& +\: I\lt(U^{n};Y_{n+1}^{N}E_{n+1}^{N}\,|\,U_{n+1}^{N}Y^{n}E^{n}b_{1}e^{-r}\rt) \nonumber\\
&=& I_{1} + I_{2} + I_{3} + I_{4}, \label{eq:I-decomp-EH}
\end{IEEEeqnarray}
where $ I_{1}$--$I_{4}$ are respectively defined as the first to fourth terms above.
By the definition \eqref{eq:UB-EH}, $I_{1}\leq n\C_{n}$. Next using \eqref{eq:EH-pYEcondUb1e} we can show that $I_{2} = I_{4} = 0$. Furthermore,
\begin{equation}\label{eq:I3-EH}
I_{3} = \sum_{y^{n}e^{n}}p\lt(y^{n}e^{n}\,|\,b_{1}e^{-r}\rt)I\lt(U_{n+1}^{N};Y_{n+1}^{N}E_{n+1}^{N}\,|\,y^{n}b_{1}e_{-r+1}^{n}\rt).
\end{equation}

Fix $y^{n}b_{1}e_{-r+1}^{n}$. For each $u_{n+1}^{N}$ define the projection map
\[ {u_{n+1}^{N}\mapsto\tu^{m}} := u_{n+1}^{N}(b_{1}e^{n},\cdot). \]
Since $u_{n+1}^{N}$ is extracted from a legal input function $u^{N}\in\U^{(N)}$, for any $e_{n+1}^{N}$ the output $\tu^{m}(e_{n+1}^{N}) = u_{n+1}^{N}(b_{1}e^{n},e_{n+1}^{N})$ needs to satisfy \eqref{eq:EH-en-constr} with the intermediate battery level $b_{n+1}$, which is determined by $u^{n}$ and $b_{1}e^{n}$. Hence
$\tu^{m}\in\U_{b_{n+1}}^{(m)}\subseteq\U_{\bc}^{(m)}$
by \eqref{eq:U-b1-inclusion}. Now by \eqref{eq:EH-pYEcondUb1e}
\begin{align*}
p\big(y_{n+1}^{N}e_{n+1}^{N}\,|\,u_{n+1}^{N}y^{n}b_{1}e_{-r+1}^{n}\big)
&= p(e_{n+1}^{N}\,|\,e_{n-r+1}^{n})\cdot p\big(y_{n+1}^{N}\,|\,u_{n+1}^{N}(b_{1}e^{n},e_{n+1}^{N})\big) \\
&= p(e_{n+1}^{N}\,|\,e_{n-r+1}^{n})\cdot p\big(y_{n+1}^{N}\,|\,\tu^{m}(e_{n+1}^{N})\big) \\
&= P_{Y^{m}E^{m}\,|\,V^{m}E^{-r}}\lt(y_{n+1}^{N}e_{n+1}^{N}\,|\,\tu^{m}e_{n-r+1}^{n}\rt),
\end{align*}
where we used the Markov property of $E_{n}$ and $P_{Y^{m}E^{m}\,|\,V^{m}E^{-r}}$ is defined by \eqref{eq:EH-pYEcondVe}. Again similar to Theorem~\ref{thm:UB-FSC-X}, for an induced distribution $\tP$ on $\U_{\bc}^{(m)}$
\begin{align*}
I\big(U_{n+1}^{N};Y_{n+1}^{N}E_{n+1}^{N}\,|\,y^{n}b_{1}e_{-r+1}^{n}\big) &= I_{\tP}\big(V^{m};Y^{m}E^{m}\,|\,e_{n-r+1}^{n}\big) \\
&\leq \max_{P_{V^{m}}\in\Ps_{\bc}^{(m)}} I(V^{m};Y^{m}E^{m}\,|\,e_{n-r+1}^{n}) \\
&= \max_{P_{U^{m}}}I(U^{m};Y^{m}E^{m}|B_{1} = \bc, e_{n-r+1}^{n}) \\
&\leq m\C_{m},
\end{align*}
where we used Lemma~\ref{lem:I-U-V-max}. Since this inequality holds for all $y^{n}b_{1}e_{-r+1}^{n}$, by \eqref{eq:I3-EH} we have
$I_{3} \leq m\C_{m}$.

Combining the results for $I_{1}$--$I_{4}$ with \eqref{eq:I-decomp-EH}, we have
\[ I\lt(U^{N};Y^{N}E^{N}\,|\,b_{1}e^{-r}\rt) \leq n\C_{n} + m\C_{m} \]
for arbitrary $P_{U^{N}}$ and $b_{1}e^{-r}$, thus \eqref{eq:subadditivity} holds.
\end{IEEEproof}

\begin{remark}
As stated in Remark~\ref{rem:UB-FSC-X}, $\C_{N}$ can be computed by finding the capacities of a finite number of DMC's. 
\end{remark}

\section{Simplification and Relaxation of Upper Bounds}
\label{sec:simplified-upper-bounds}

In the equivalent channel models, the input alphabet size for each channel use grows double exponentially, and so does the (spatial) computational complexity of the bounds in the previous section. To address this problem, in this section we rewrite the upper bounds in the form of maximized directed information on the original channels, which has a constant input alphabet size and the complexity becomes exponential. It turns out that for the case of FSC-X / EH-SC1, this new formulation also allows for a nice dynamic programming recursion, which only has a linear complexity. For EH-SC2, we need to loosen the upper bounds a bit to obtain a similar recursion. If we relax these bounds even further, the recursions can be solved analytically. Although such relaxed upper bounds are looser than the original ones for each block length $N$, since we can compute them for very large $N$, the results are sometimes tighter (as verified by the numerical results in next section).

\subsection{Upper Bounds for FSC-X / EH-SC1}\label{subsec:FSC-X-linUB}

First we introduce some notations. The directed information between $X^{N}$ and $Y^{N}$ is defined as
\[ I(X^{N}\to Y^{N}) \triangleq \sum_{n=1}^{N}I(X^{n};Y_{n}\cn Y^{n-1}). \]
The directed information $I(\cdot\to\cdot\cn s_{1})$ conditioned on a initial state $S_{1}=s_{1}$ is defined similarly. For the channel FSC-X, a collection of conditional input distributions $\{p(x_{n}\cn x^{n-1}s^{n})\}_{n=1}^{N}$ is called \emph{legal} if it puts zero probability on $x_{n}\notin \X(s_{n})$, $1\leq n\leq N$. Fix $s_{1}$ and consider the mutual information $I\big(U^{N};Y^{N}S_{2}^{N+1}\cn s_{1}\big)$ in the context of feedback channel \cite{Tatikonda-Mitter-Feedback}, where $U^{N}$ is the code function. At each time $n$  the output is $Y_{n}S_{n+1}$ and the feedback is $S_{n}$. Then similar to Lemmas~5.1, 5.2, and 5.4 of \cite{Tatikonda-Mitter-Feedback}, we can show the following for FSC-X.
\begin{enumerate}
\item Any input function distribution $P_{U^{N}}$ on $\U^{(N)}$ induces a collection of legal conditional input distributions $\{p(x_{n}\cn x^{n-1}s^{n})\}_{n=1}^{N}$ \footnote{Defined on a set of measure 1.}. (We say $P_{U^{N}}$ and $\{p(x_{n}\cn x^{n-1}s^{n})\}_{n=1}^{N}$ are compatible in this case.) Conversely, if the collection $\{p(x_{n}\cn x^{n-1}s^{n})\}_{n=1}^{N}$ is legal, one can construct a distribution on $\U^{(N)}$ that is compatible with it.
\item We have the relation
\[ I\big(U^{N};Y^{N}S_{2}^{N+1}\cn s_{1}\big) = I\big(X^{N}\to Y^{N}S_{2}^{N+1}\cn s_{1}\big), \]
where the directed information is determined by the induced collection $\{p(x_{n}\cn x^{n-1}s^{n})\}_{n=1}^{N}$.
\end{enumerate}
As a result we can rewrite the upper bound in Theorem~\ref{thm:UB-FSC-X} as
\begin{equation}\label{eq:UB-FSC-X-directed-info}
\C_{N} = \frac{1}{N} \max_{s_{1}}\max_{\{p(x_{n}\cn x^{n-1}s^{n})\}_{n=1}^{N}\text{ l.g.}}I\big(X^{N}\to Y^{N}S_{2}^{N+1}\cn s_{1}\big),
\end{equation}
where ``l.g.'' stands for ``legal''.

Next we show that this expression can be simplified further to allow for a dynamical programming recursion similar to \cite{Chen-Berger-FSC-FB}. Let us start with a few more notations. For each $s\in\St$ define $\Pst_{s}$ to be the set of all probability distributions on $\X(s)$ and $\Pst = \prod_{s\in\St}\Pst_{s}$. We say the conditional distribution $P_{X|S}\in\Pst$ iff $P_{X|S}(\cdot\,|\,s) \in\Pst_{s}$ for all $s\in\St$. Let $p_{n} \triangleq P_{X_{n}|S_{n}}$, we write $\{p_{n}\}_{n=1}^{N}\subset\Pst$ if $p_{n}\in\Pst$ for all $1\leq n\leq N$. For a fixed $S_{1} = s_{1}$, we can write
\[ I\big(X^{N}\to Y^{N}S_{2}^{N+1}\cn s_{1}\big) = \sum_{n=1}^{N}I\big(X^{n};Y_{n}S_{n+1}\,|\,Y^{n-1}S^{n}\big). \]
Observe that every legal collection $\{p(x_{n}\cn x^{n-1}s^{n})\}_{n=1}^{N}$ together with $P_{S_{1}}$ determines a random tuple $(X^{N},S^{N+1})$, which further induces a set of conditional probabilities $\{p_{n}\}_{n=1}^{N}\subset\Pst$. Following the argument for eq. (117) in \cite{Permuter-FSC-FB} we have
\[ \max_{\{p(x_{n}\cn x^{n-1}s^{n})\}_{n=1}^{N}\text{ l.g.}}\sum_{n=1}^{N}I\big(X^{n};Y_{n}S_{n+1}\,|\,Y^{n-1}S^{n}\big) \leq \max_{\{p_{n}\}_{n=1}^{N}\subset\Pst}\sum_{n=1}^{N}I(Y_{n}S_{n+1};X_{n}\,|\,S_{n}). \]
On the other hand, by setting $p(x_{n}\cn x^{n-1}s^{n}) = p_{n}(x_{n}\cn s_{n})$ we see that $\{p_{n}\}_{n=1}^{N}\subset\Pst$ indeed belongs to the family of legal conditional input distributions. Thus by the argument for eq. (120) in \cite{Permuter-FSC-FB}, in fact the above inequality holds with equality.

Summarizing the discussion above we have the following theorem.\footnote{In \cite{Ozel-EH-CSIR-Discrete} it is claimed that the results in \cite{Permuter-FSC-FB} can be directly applied to the channel EH-SC1 with CSIR and so $\lim_{N\to\infty}\C_{N}$ is actually the capacity in this case. However, \cite{Permuter-FSC-FB} only deals with FSC's without input constraints. Given that the results therein are built up gradually through a series of sophisticated theorems and lemmas, they should be re-proved (if this is indeed possible) for the case with input constraints before being applied to EH-SC1.}

\begin{theorem}\label{thm:FSC-linUB}
For each $N$, the FSC-X capacity upper bound in Theorem~\ref{thm:UB-FSC-X} can be written as
\[ \C_{N} = \frac{1}{N}\max_{s\in\St}\tc_{N,s},\qquad
\tc_{N,s} \triangleq \max_{\{p_{n}\}_{n=1}^{N}\subset\Pst}\sum_{n=1}^{N}I(Y_{n}S_{n+1};X_{n}\,|\,S_{n})\bigg|_{S_{1}=s}. \]
\end{theorem}

The terms $\tc_{N,s}$ can be calculated using a dynamic programming recursion similar to \cite{Chen-Berger-FSC-FB}. To see that denote $Q = P_{S_{n+1}\mid X_{n}S_{n}}$ and thus
$Q(s_{n+1}\,|\,x_{n}s_{n}) = \sum_{y_{n}}p(y_{n}s_{n+1}\,|\,x_{n}s_{n})$.
Moreover, for any conditional distribution $p = P_{X|S}$ define
\[ I(p,s) = I(Y_{n}S_{n+1};X_{n}\,|\,S_{n}=s)\big|_{p_{n} = p}. \]

\begin{theorem}\label{thm:FSC-linUB-Recursion}
Let $S_{1}$ have an arbitrary distribution $\pi$, define
\[ \tc_{N}(\pi) \triangleq \max_{\{p_{n}\}_{n=1}^{N}\subset\Pst}\sum_{n=1}^{N}I(Y_{n}S_{n+1};X_{n}\,|\,S_{n})\bigg|_{S_{1}\sim\pi}. \]
Then $\tc_{N}(\pi) = \sum_{s\in\St}\pi(s)\cdot\tc_{N,s}$ where for each $s$
\[ \tc_{N,s} = \max_{p(\cdot|s)\in\Pst_{s}}\lt[ I(p,s) + \sum_{x}p(x|s)\sum_{t}Q(t|xs)\cdot\tc_{N-1,t} \rt], \]
with the initial condition $\tc_{0,s} = 0$, $\forall s\in\St$.
\end{theorem}

\begin{IEEEproof}
From the definitions, $\tc_{N,s} = \tc_{N}(\delta_{s})$ where $\delta_{s}$ puts probability 1 on $s$. For $N=1$, the theorem is true (note that the optimization for $\tc_{1}(\pi)$ is over $\{p_{1}(\cdot|s):s\in\St\}$, which can be separated). Assume it is true for $N=k$, then for $N = k+1$,
\[ \tc_{k+1}(\pi) = \max_{p_{1}\in\Pst}\lt[ \sum_{s}\pi(s)I(p_{1},s)
+ \max_{\{p_{n}\}_{n=2}^{k+1}\subset\Pst}\sum_{n=2}^{k+1}I(Y_{n}S_{n+1};X_{n}\,|\,S_{n}) \rt]. \]
Given $\pi$ and $p_{1}$,
\[ P_{S_{2}}(t) = \sum_{x,s}\pi(s)p_{1}(x|s)Q(t|xs). \]
Now define
$(\hX^{k},\hS^{k+1},\hY^{k}) = (X_{2}^{k+1},S_{2}^{k+2},Y_{2}^{k+1})$
and so
\[ \hp_{n} \triangleq P_{\hX_{n}|\hS_{n}} = p_{n+1},\quad 1\leq n\leq k. \]
Using the theorem for $N=k$, we have
\begin{align*}
\max_{\{p_{n}\}_{n=2}^{k+1}\subset\Pst}\sum_{n=2}^{k+1}I(Y_{n}S_{n+1};X_{n}\,|\,S_{n})
&=\max_{\{\hp_{n}\}_{n=1}^{k}\subset\Pst}\sum_{n=1}^{k}I(\hY_{n}\hS_{n+1};\hX_{n}\,|\,\hS_{n}) \\
= \tc_{k}(P_{\hS_{1}})
&= \tc_{k}(P_{S_{2}}) 
= \sum_{t\in\St}P_{S_{2}}(t)\cdot\tc_{k,t},
\end{align*}
as the value of $\sum_{n=1}^{k}I(\hY_{n}\hS_{n+1};\hX_{n}\,|\,\hS_{n})$ is uniquely determined by $P_{\hS_{1}}$, $\{\hp_{n}\}_{n=1}^{k}$ and the time-invariant transition probabilities $p(y_{n}s_{n+1}\,|\, x_{n}s_{n})$. Thus
\begin{align*}
\tc_{k+1}(\pi) &= \max_{p_{1}\in\Pst}\lt[ \sum_{s}\pi(s)I(p_{1},s) + \sum_{t}P_{S_{2}}(t)\cdot\tc_{k,t} \rt] \\
&= \max_{p_{1}\in\Pst}\lt[ \sum_{s}\pi(s)I(p_{1},s) + \sum_{s}\pi(s)\sum_{x}p_{1}(x|s)\sum_{t}Q(t|xs)\cdot\tc_{k,t} \rt] \\
&= \sum_{s}\pi(s)\!\!\max_{p_{1}(\cdot|s)\in\Pst_{s}}\lt[ I(p_{1},s) + \sum_{x}p_{1}(x|s)\sum_{t}Q(t|xs)\cdot\tc_{k,t} \rt].
\end{align*}
Letting $\pi = \delta_{s}$ for each $s$ we obtain the statement for $\tc_{k+1,s}$, which can be plugged back into the expression above to obtain the result for $\tc_{k+1}(\pi)$. So the theorem is true for $N = k+1$ and hence true for all $N$.
\end{IEEEproof}

\begin{remark}
Note that for every recursion we only need to maximize the sum of a concave function $I(\cdot,s)$ and a linear term over the same space $\Pst_{s}$, which is simple to compute using convex optimization. Also note that the alphabet size is constant and the computational complexity is linear.
\end{remark}

The recursion can even be solved analytically if we relax $\C_{N}$ further. Using the inequality
\begin{equation}\label{eq:FSC-I-H}
I(Y_{n}S_{n+1};X_{n}\,|\,S_{n}) \leq H(X_{n}\,|\,S_{n}),
\end{equation}
we can replace the mutual information in the expressions of $\tc_{N,s},\C_{N}$ and $\tc_{N}(\pi)$ in Theorems~\ref{thm:FSC-linUB} and \ref{thm:FSC-linUB-Recursion} by the corresponding conditional entropies to define $\tc'_{N,s},\tC'_{N}$ and $\tc'_{N}(\pi)$ and obtain a corresponding new theorem:

\begin{theorem}\label{thm:FSC-linUB-Analytical}
Assume the base of $\log$ is $e$. We have
$\C_{N} \leq \tC'_{N}$, 
$\quad\tc'_{N}(\pi) = \sum_{s\in\St}\pi(s)\cdot\tc'_{N,s}\ $ and
\[ \tc'_{N,s} = \log\sum_{x\in\X(s)}\exp\lt[\sum_{t}Q(t|xs)\cdot\tc'_{N-1,t}\rt], \]
with the initial condition $\tc'_{0,s} = 0$, $\forall s\in\St$.
\end{theorem}

\begin{IEEEproof}
By \eqref{eq:FSC-I-H}, $\C_{N} \leq \tC'_{N}$. Using arguments similar to Theorem~\ref{thm:FSC-linUB-Recursion} and defining
\[ H(p,s) = H(X_{n}\,|\,S_{n}=s)\big|_{p_{n} = p}, \]
we have 
$\tc'_{N}(\pi) = \sum_{s\in\St}\pi(s)\cdot\tc'_{N,s}$ with
\[ \tc'_{N,s} = \max_{p(\cdot|s)\in\Pst_{s}}\lt[ H(p,s) + \sum_{x}p(x|s)\sum_{t}Q(t|xs)\cdot\tc'_{N-1,t} \rt]. \]
Denote
$\alpha_{x} = \sum_{t}Q(t|xs)\cdot\tc'_{N-1,t}$ and $r_{x} = p(x|s)$.
The optimization problem above can be written as
\[ \begin{array}{rc}
\text{maximize} & - \sum_{x}r_{x}\log r_{x} + \sum_{x}r_{x}\alpha_{x} \\
\text{s.t.} & \sum_{x}r_{x} = 1 \\
& r_{x} \geq 0,\ \forall x\in\X(s)
\end{array}, \]
whose solution $r^{*}$ can be easily found using KKT conditions:
\[ r^{*}_{x} = \frac{e^{\alpha_{x}}}{\sum_{x'\in\X(s)}e^{\alpha_{x'}}}. \]
Plugging into the objective function, we obtain the desired formula for $\tc'_{N,s}$.
\end{IEEEproof}

\begin{remark}\label{rmk:FSC-I-H-exact}
When $X_{n}$ is uniquely determined by $Y_{n}$ (e.g., $Y_{n}=X_{n}$), \eqref{eq:FSC-I-H} holds with equality and so $\tC'_{N}=\C_{N}$.
\end{remark}

\subsection{Upper Bounds for EH-SC2}
\label{subsec:EH-FB-linUB}

First let us rewrite the upper bounds in Theorem~\ref{thm:UB-EH} in the form of maximized directed information. From the proof of Theorem~\ref{thm:UB-EH}, the upper bound can be expressed as
\begin{equation}\label{eq:UB-EH-alternative}
\C_{N} = \frac{1}{N} \max_{e^{-r}} \max_{P_{V^{N}}\in\Ps_{\bc}^{(N)}} I\big(V^{N};Y^{N}E^{N}\,|\,e^{-r}\big).
\end{equation}
Since $E_{1}$ is independent of $V^{N}$ given $E^{-r} = e^{-r}$, also $E_{N+1}$ is independent of $V^{N}$ and $Y^{N}$ given $E^{N}$ and $E^{-r} = e^{-r}$, we know that $I\big(V^{N};E_{1}\,|\,e^{-r}\big)$ and $I\big(V^{N};E_{N+1}\,|\,Y^{N}E^{N},e^{-r}\big)$ are both 0. Thus
\[ I\big(V^{N};Y^{N}E^{N}\,|\,e^{-r}\big) = I\big(V^{N};Y^{N}E_{2}^{N}\,|\,E_{1},e^{-r}\big) = I\big(V^{N};Y^{N}E_{2}^{N+1}\,|\,E_{1},e^{-r}\big). \]
Now in the context of feedback channel in \cite{Tatikonda-Mitter-Feedback}, consider $V^{N}$ as the code function, $Y_{n}E_{n+1}$ as the channel output and $E_{n}$ as the feedback at each time $n$. Define a collection of conditional input distributions $\{p(x_{n}\cn x^{n-1}e^{n})\}_{n=1}^{N}$ to be \emph{legal} if for $1\leq n\leq N$, the conditional probability is zero whenever $x^{n}$ does not satisfy the energy constraint \eqref{eq:EH-en-constr} with $e^{n}$ and $B_{1} =\bc$. Then similar to Lemmas~5.1, 5.2, and 5.4 of \cite{Tatikonda-Mitter-Feedback} and the previous subsection, we can show:
\begin{enumerate}
\item Any input function distribution $P_{V^{N}}\in\Ps_{\bc}^{(N)}$ induces a collection of legal conditional input distributions $\{p(x_{n}\cn x^{n-1}e^{n})\}_{n=1}^{N}$ (almost everywhere), in which case we say they are compatible. Conversely, if the collection $\{p(x_{n}\cn x^{n-1}e^{n})\}_{n=1}^{N}$ is legal, one can construct a distribution on $\U_{\bc}^{(N)}$ that is compatible with it.
\item We have the relation
\[ I\big(V^{N};Y^{N}E_{2}^{N+1}\,|\,E_{1},e^{-r}\big) = I\big(X^{N}\to Y^{N}E_{2}^{N+1}\cn E_{1},e^{-r}\big), \]
where the directed information is determined by the induced collection $\{p(x_{n}\cn x^{n-1}e^{n})\}_{n=1}^{N}$.
\end{enumerate}

\begin{theorem}\label{thm:EH-FB-UB}
For each $N$, the capacity upper bound in Thereom~\ref{thm:UB-EH} can be rewritten as
\begin{equation}\label{eq:UB-EH-directed-info}
\C_{N} = \frac{1}{N} \max_{e^{-r}}\max_{\{p(x_{n}\cn x^{n-1}e^{n})\}_{n=1}^{N}\text{ l.g.}}I\big(X^{N}\to Y^{N}E_{2}^{N+1}\cn E_{1},e^{-r}\big).
\end{equation}
\end{theorem}

\begin{remark}
An extended Blahut-Arimoto algorithm is proposed in \cite{Naiss-Permuter-EBAA} to maximize the directed information for feedback channels. This algorithm can be adapted to compute the inner maximization of the directed information in \eqref{eq:UB-EH-directed-info}
. In fact, by \cite[Lemma~1]{Permuter-POST-FB}, the causal conditioning distributions form a polyhedron in $\R^{|\X|^{N}|\E_{H}|^{N}}$. Adding energy constraints (i.e., requiring the conditional distributions to be legal) forces some coordinates to be zero, which imposes some extra linear equalities on this set. Therefore, the resulting collection of distributions still form a polyhedron, which is convex. Thus \cite[Lemma~1]{Naiss-Permuter-EBAA} guarantees that the corresponding alternating maximization procedure converges to the global maximum. Furthermore, examining the algorithm in \cite{Naiss-Permuter-EBAA} we see that if we start with a conditional distribution that satisfies the energy constraint, then every iteration returns a legal collection of conditional distributions. Hence indeed this algorithm can be used to compute $\C_{N}$, and the (spatial) computational complexity is exponential in $N$.
\end{remark}

Next we want to relax the upper bounds $\C_{N}$ to obtain a dynamic programming recursion similar to the case FSC-X / EH-SC1. Recall that the energy harvesting process $\{E_{n}\}_{n=1}^{\infty}$ is a homogeneous Markov chain of order $r\geq0$. Define the ``overall'' state
\[ Z_{n} \triangleq E_{n-r+1}^{n}S_{n}, \]
with alphabet $\Z \triangleq \E_{H}^{r}\times\St$. When $r=0$, $E_{n}$ is i.i.d. and $Z_{n} = S_{n}$, whose transition probability $p(z_{n+1}\,|\, x_{n}z_{n})$ can be obtained through \eqref{eq:EH-SC1-FSC-cond-prob}. Now assume $r > 0$. For each $z_{n} = e_{n-r+1}^{n}s_{n}\in\Z$, if $z_{n+1} = e_{n-r+2}^{n+1}s_{n+1}$ for some $e_{n+1}s_{n+1}$, then
\[ p(z_{n+1}\,|\, x_{n}z_{n}) = p(e_{n+1}\,|\,e_{n-r+1}^{n})\cdot \1{S(x_{n},s_{n},e_{n+1})=s_{n+1}}, \]
otherwise
$p(z_{n+1}\,|\, x_{n}z_{n}) = 0$.
Note that these transition probabilities are independent of $n$. Furthermore, let $\pi_{b_{1},e^{-r}}$ denote the distribution of $Z_{1}$ when $B_{1} = b_{1}$ and $E^{-r} = e^{-r}$, which is determined by
$p(e_{1}s_{1} | b_{1}e^{-r}) = p(e_{1}\,|\,e^{-r})\cdot\1{S(b_{1},e_{1}) = s_{1}}$.

We relax $\C_{N}$ by providing more energy information to the receiver in \eqref{eq:UB-EH-alternative}. For fixed $B_{1} = \bc$ and $E^{-r} = e^{-r}$, we know $E_{1}S_{1}$ is conditionally independent of $V^{N}$ and so
\[ I\big(V^{N};Y^{N}E^{N}\,|\,e^{-r}\big) \leq I\big(V^{N};Y^{N}E^{N+1}S^{N+1}\,|\,e^{-r}\big) = I\big(V^{N};Y^{N}E_{2}^{N+1}S_{2}^{N+1}\,|\,E_{-r+1}^{1}S_{1}\big). \]
Now in the setting of feedback channel view $Y_{n}E_{n+1}S_{n+1}$ as the channel output and $E_{n}$ as the feedback. Then similar to 2) above we can show
\begin{align*}
I\big(V^{N};Y^{N}E_{2}^{N+1}S_{2}^{N+1}\,|\,E_{-r+1}^{1}S_{1}\big) &= I\big(X^{N}\to Y^{N}E_{2}^{N+1}S_{2}^{N+1}\cn E_{-r+1}^{1}S_{1}\big) \\
&= \sum_{n=1}^{N}I\big(X^{n}; Y_{n}E_{n+1}S_{n+1}\cn Y^{n-1}E_{-r+1}^{n}S^{n}\big),
\end{align*}
where $X^{N}=V^{N}(E^{N})$. Since given $X_{n}Z_{n}$, $\ Y_{n}E_{n+1}S_{n+1}$ is independent of all other previous random variables, we have
\begin{align*}
H\big(Y_{n}E_{n+1}S_{n+1}\mid Y^{n-1}E_{-r+1}^{n}S^{n}X^{n}\big)
&= H\big(Y_{n}E_{n+1}S_{n+1}\mid X_{n}Z_{n}\big) 
= H\big(Y_{n}E_{n+1}Z_{n+1}\mid X_{n}Z_{n}\big),
\end{align*}
\[ H\big(Y_{n}E_{n+1}S_{n+1}\mid Y^{n-1}E_{-r+1}^{n}S^{n}\big) \leq H(Y_{n}E_{n+1}Z_{n+1}\mid Z_{n}), \]
and as a result of these (in)equalies
\begin{equation}\label{eq:EH-linUB-auxiliary}
I\big(V^{N};Y^{N}E^{N}\,|\,e^{-r}\big) \leq \sum_{n=1}^{N}I(Y_{n}E_{n+1}Z_{n+1}; X_{n}\mid Z_{n})
\bigg|_{Z_{1}\sim\,\pi_{\bc,e^{-r}}}.
\end{equation}

For $z\in\Z$ define $s(z)$ to be the $\St$-component of $z$, and let
\[ \X(z) \triangleq \{x\in\X: \gamma(x)\leq s(z)\},\quad \forall z\in\Z. \]
Similar to the previous subsection, we define $\Pst_{z},\Pst,p_{n}$, and $Q$ w.r.t. the state $Z_{n}$, and write
\[ I(p,z) = I(Y_{n}E_{n+1}Z_{n+1};X_{n}\mid Z_{n}=z)\big|_{p_{n} = p}. \]
Moreover, for an arbitrary distribution $\pi$ on $\Z$, define
\[ \tc_{N}(\pi) = \max_{\{p_{n}\}_{n=1}^{N}\subset\Pst}\sum_{n=1}^{N} I(Y_{n}E_{n+1}Z_{n+1}; X_{n}\mid Z_{n})\bigg|_{Z_{1}\sim\pi}. \]
Observe that when $B_{1} = \bc$ and $E^{-r} = e^{-r}$ are fixed, for any $P_{V^{N}}$, $\big(V^{N},E^{N+1},S^{N+1},Y^{N}\big)$ induces a random tuple $(X^{N},E_{2}^{N+1},Z^{N+1},Y^{N})$. It further induces a set of conditional probabilities $\{p_{n}\}_{n=1}^{N}\subset\Pst$, which together with $P_{Z_{1}} = \pi_{\bc,e^{-r}}$ and $p(y_{n}e_{n+1}z_{n+1}\cn x_{n}z_{n})$ uniquely determines the RHS of \eqref{eq:EH-linUB-auxiliary} (cf. \cite[App. VIII]{Permuter-FSC-FB}). Thus for any $P_{V^{N}}\in\Ps_{\bc}^{(N)}$ we have
\[ I\big(V^{N};Y^{N}E^{N}\,|\,e^{-r}\big) \leq \tc_{N}\big(\pi_{\bc,e^{-r}}\big), \]
and so with \eqref{eq:UB-EH-alternative} we can establish the following theorems.

\begin{theorem}\label{thm:EH-linUB}
Assume $\{E_{n}\}_{n=1}^{\infty}$ is Markov of order $r\geq0$. Then for each $N$,
\[ \tC_{N}\triangleq \frac{1}{N}\max_{e^{-r}}\tc_{N}\big(\pi_{\bc,e^{-r}}\big) \geq \C_{N} \]
is an upper bound for the channel capacity of EH-SC2.
\end{theorem}

\begin{theorem}\label{thm:EH-linUB-Recursion}
Define $\tc_{N,z} = \tc_{N}(\delta_{z})$ for all $z\in\Z$. Then
$\tc_{N}(\pi) = \sum_{z\in\Z}\pi(z)\cdot\tc_{N,z}$ and
\[ \tc_{N,z} = \max_{p(\cdot|z)\in\Pst_{z}}\lt[ I(p,z) + \sum_{x}p(x|z)\sum_{w}Q(w|xz)\cdot\tc_{N-1,w} \rt], \]
with the initial condition $\tc_{0,z} = 0$, $\forall z \in \Z$.
\end{theorem}

Since $E_{n+1}$ is independent of $X_{n}$ given $Z_{n}$, we can rewrite the expression
\[ I(p,z) = I(Y_{n}S_{n+1}; X_{n}\,|\,E_{n+1},Z_{n}=z) \]
with $P_{X_{n}|E_{n+1}Z_{n}} = p_{n}$ to further simplify the computation. Also, since
\begin{equation}\label{eq:EH-I-H}
I(Y_{n}E_{n+1}Z_{n+1}; X_{n}\mid Z_{n}) \leq H(X_{n}\,|\,Z_{n}),
\end{equation}
we can replace the mutual information in the definitions of $\tc_{N,z},\tC_{N}$ and $\tc_{N}(\pi)$ by the corresponding conditional entropies to define $\tc'_{N,z},\tC'_{N}$ and $\tc'_{N}(\pi)$ and obtain:

\begin{theorem}\label{thm:EH-linUB-Analytical}
Assume the base of $\log$ is $e$. We have
$\tC_{N} \leq \tC'_{N}$,
$\quad \tc'_{N}(\pi) = \sum_{z\in\Z}\pi(z)\cdot\tc'_{N,z}\ $ and
\[ \tc'_{N,z} = \log\sum_{x\in\X(z)}\exp\lt[\sum_{w}Q(w|xz)\cdot\tc'_{N-1,w}\rt], \]
with the initial condition $\tc'_{0,z} = 0$, $\forall z \in \Z$.
\end{theorem}

The proofs for Theorems~\ref{thm:EH-linUB-Recursion} and~\ref{thm:EH-linUB-Analytical} are similar to Theorems~\ref{thm:FSC-linUB-Recursion} and \ref{thm:FSC-linUB-Analytical}, respectively, and hence are omitted.

\begin{remark}\label{rmk:EH-I-H-exact}
When $X_{n}$ is uniquely determined by $Y_{n}$ (e.g., $Y_{n}=X_{n}$), we have $\tC'_{N}=\tC_{N}$.
\end{remark}

\section{Numerical Results}
\label{sec:numerical}

\subsection{Achievable Rates}
\label{subsec:numerical-ach-rates}

\begin{figure}[t]
  \centering
  \includegraphics[width=\textwidth]{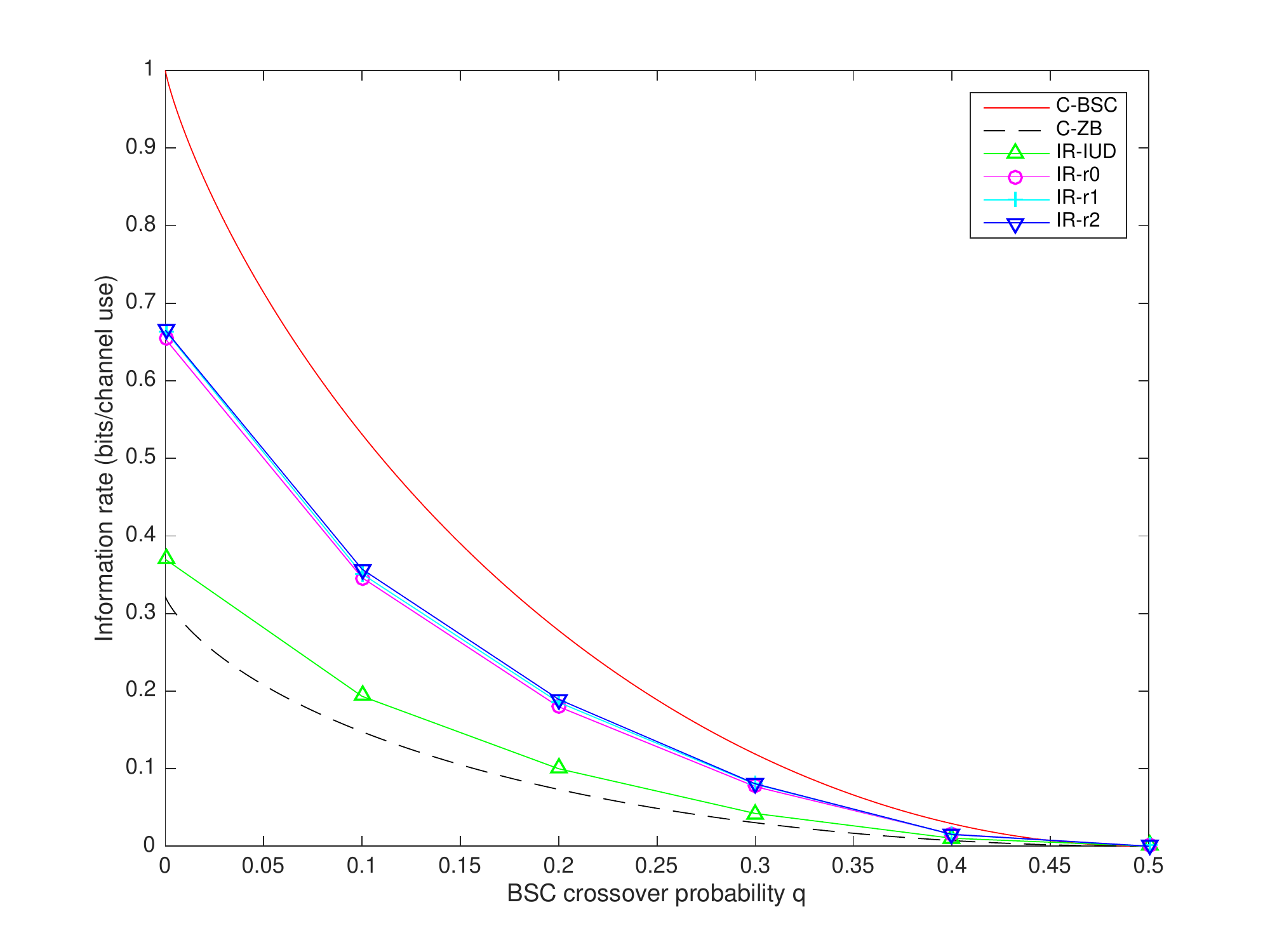}
  \caption{The information rates. C-BSC = BSC capacity, C-ZB = zero battery capacity, IR-IUD = i.u.d. information rate; IR-r0, -r1, -r2 = optimized information rates for Markov input processes of order 0,1,2, resp.}
  \label{fig:rates}
\end{figure}
 
We illustrate the computation and optimization of the achievable rates in Section~\ref{sec:ach-rates} with the following energy harvesting example. Let $\{E_{n}\}$ be an i.i.d. Bernoulli(0.5) process with $\E_{H} = \X = \{0,1\}$, and assume the battery limit $\bc = 1$. The energy model satisfies \eqref{eq:EH-S-comb-1} and \eqref{eq:EH-cost-quadratic}, i.e., the harvested energy is immediately available and the energy cost is quadratic. Since $\bc = 1$, the energy states take value in $\St = \{0,1,2\}$. The DMC in the model is BSC($q$), i.e., the binary symmetric channel with crossover probability $q$. The energy information scenario can be either EH-SC1 or EH-SC2, but we only use results for the former to compute achievable rates (which work in both cases). Let $m=1$ in Section~\ref{subsec:ach-rate-EH-SC1}, then case~ii) of Theorem~\ref{thm:erg-conds-iid} is satisfied and so $\bW'$ is indecomposable. Furthermore, the input alphabet of the surrogate channel is $\V = \{v_{a},v_{b},v_{c},v_{d}\}$ with
\[ v_{a} = (0,0,0),\ v_{b} = (0,0,1),\ v_{c} = (0,1,0),\ v_{d} = (0,1,1), \]
where we use vectors in $\X^{|\St|}$ to represent functions. The channel state transition probability is given by $p(s_{n+1}|v_{n}s_{n}) = P^{(i)}(s_{n}+1,s_{n+1}+1)$, where $i=1$ for $v_{n} \in \{v_{a},v_{b}\}$ and $i=2$ for $v_{n}\in \{v_{c},v_{d}\}$, and
\[ P^{(1)} = \begin{bmatrix}
0.5 & 0.5 & 0 \\
0 & 0.5 & 0.5 \\
0 & 0.5 & 0.5
\end{bmatrix},\qquad
P^{(2)} = \begin{bmatrix}
0.5 & 0.5 & 0 \\
0.5 & 0.5 & 0 \\
0 & 0.5 & 0.5
\end{bmatrix}. \]

For the surrogate channel $\bW'$ we compute the i.u.d. rate, which is the information rate for the i.i.d. uniform input process, and optimize the information rate over Markov input processes of order 0 (which is i.i.d.), 1 and 2. The results of these computations are shown in Fig.~\ref{fig:rates}. For comparison, in the same figure we also show the capacities for the same BSC without energy constraint, and with zero battery. The BSC capacity is $1-H(q)$, which is an upper bound for the case of infinite battery, and this bound is tight when $p\geq 0.5$ (using an argument similar to \cite{Ozel-Ulukus-AWGN}). The zero battery capacity, as commented in Section~\ref{subsec:equiv-ch}, can be obtained by constructing an equivalent DMC using Shannon's method \cite{Shannon-CSIT}. The new input alphabet $\U = \{u_{a}, u_{b}\}$, where $u_{a} = (0,0)$ and $u_{b} = (0,1)$, both of which are functions of $E_{n}$. The transition probability is
\[ p(y|u) = \sum_{e\in\E_{H}}p(e)p(y|u(e)),\quad \forall y\in\Y,\ u\in\U. \]
In particular, $p(y|u_{a}) = p(y|0)$ and $p(y|u_{b}) = 0.5$. The capacity of this DMC is
\[ H\left((1+\alpha)^{-1}\right) - 1 + r(1-H(q)), \]
where $\alpha = 2^{-\frac{1-H(q)}{0.5-q}}$ and $r = \frac{(1+\alpha)^{-1}-0.5}{0.5-q}$.

Observing from Fig.~\ref{fig:rates}, we have the following remarks for this energy harvesting channel:
 \begin{enumerate}
 \item Compared to the zero battery case, using the minimum non-zero battery (whose energy storage is just enough to transmit any single symbol) can obtain a remarkable channel capacity gain; it even achieves a significant fraction (around 70\%) of the capacity for the infinite battery case.
\item The optimized Markov input processes (including the iid case) achieve much higher information rates than the i.u.d. input. However, while the information rates are higher for higher order Markov processes, the increase is quite limited (even slight in many cases). This phenomenon is also observed in the numerical simulation results in \cite{Vontobel-GBAA}.
 \end{enumerate}

The numerical results can provide some guidance in the design of an energy harvesting communication system. For example, if battery storage is expensive, a small non-zero battery might be desirable; as it can provide a significant capacity gain over a system with no battery, while investing in a big battery can only yield a limited increase in channel capacity. Furthermore, when designing channel codes for transmission, observation 2) may suggest that i.i.d. random codes (with an optimized distribution) could achieve a good data rate.

\subsection{Capacity Bounds}

\begin{figure}[!t]
\centering
\includegraphics[width=\textwidth]{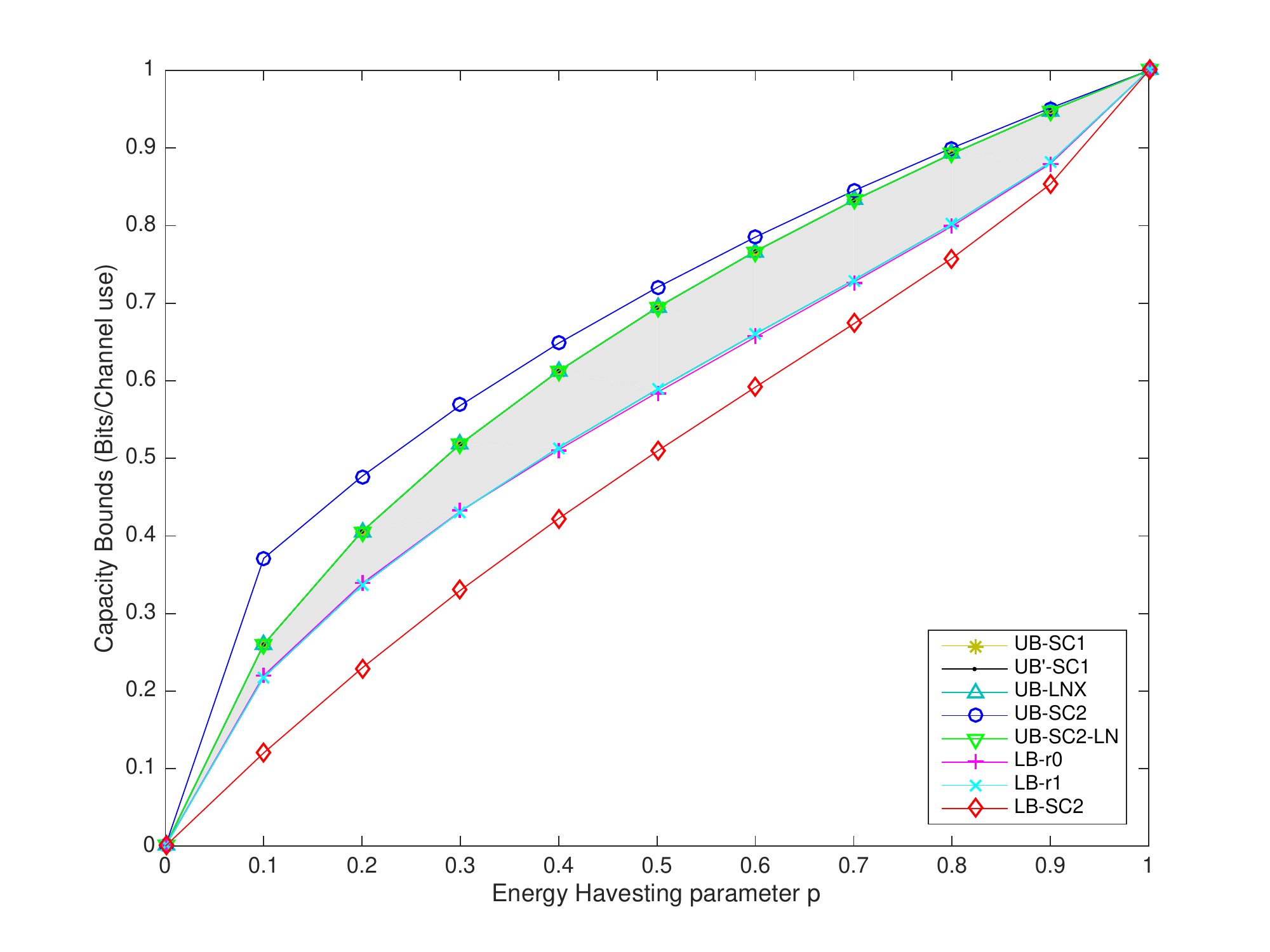}
\caption{Capacity bounds for $q=0$. UB-SC1, UB-LNX, and UB-SC2-LN = upper bounds from Theorems~\ref{thm:FSC-linUB}, \ref{thm:FSC-linUB-Analytical}/\ref{thm:EH-linUB-Analytical}, and \ref{thm:EH-linUB}, resp. ($N=10^{4}$); UB-SC2 = upper bound from Theorem~\ref{thm:EH-FB-UB} ($N=16$); UB$'$-SC1 = upper bound from \cite{Tutuncuoglu-EnHarvTimingCh}; LB-r0, -r1 = optimized information rates for Markov order 0 and 1, resp.; LB-SC2 = lower bound from Theorem~\ref{thm:LB-EH} ($N=4$).}
\label{fig:numerical-bounds-q0}
\end{figure}

\begin{figure}[!t]
\centering
\includegraphics[width=\textwidth]{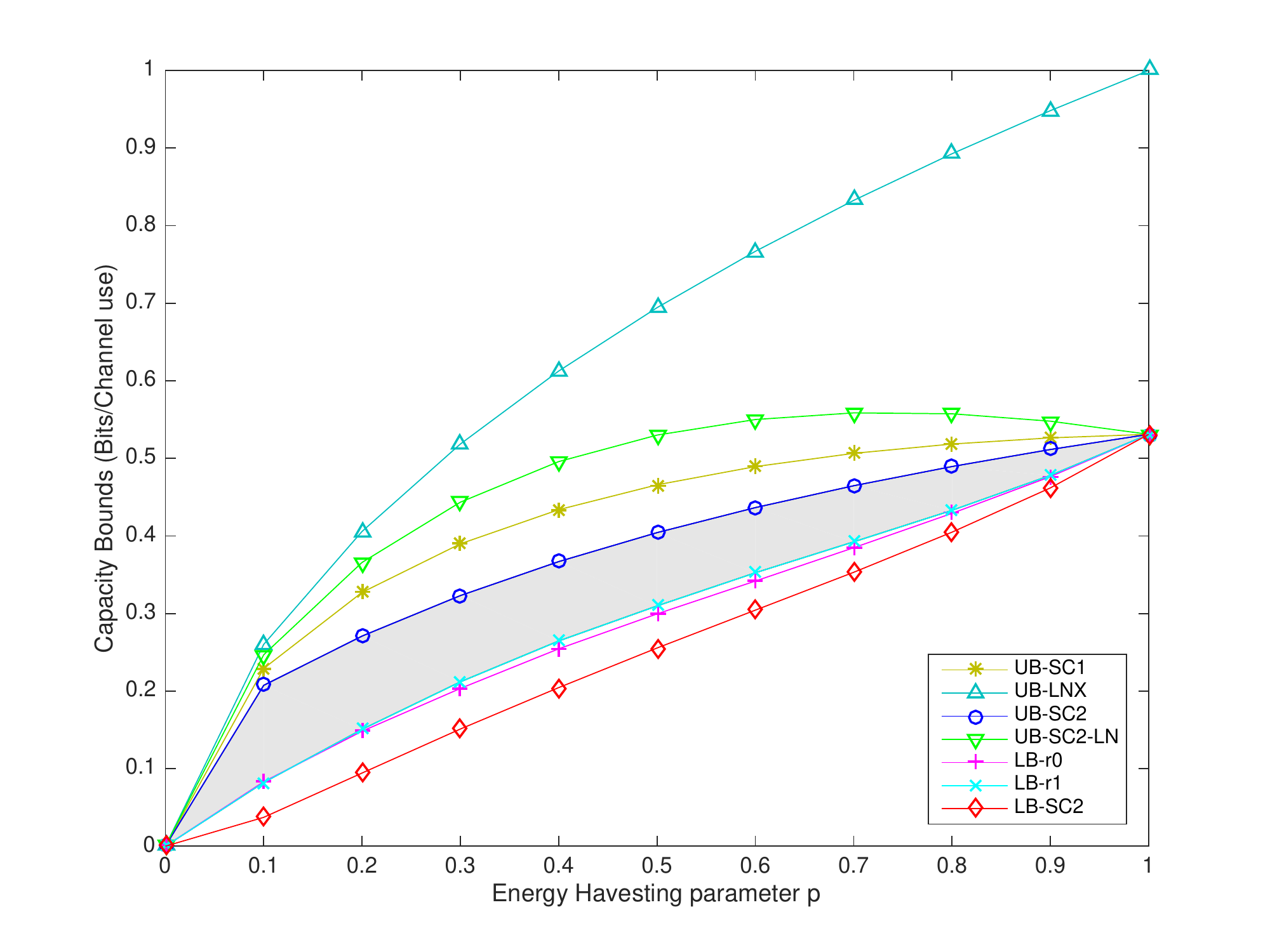}
\caption{Capacity bounds for $q=0.1$. The notations are the same as Fig.~\ref{fig:numerical-bounds-q0}.}
\label{fig:numerical-bounds-q0dot1}
\end{figure}

We use a slightly different energy harvesting example to demonstrate the computation of the bounds in Sections~\ref{sec:capacity-bounds} and~\ref{sec:simplified-upper-bounds}. In this model $E_{n}$ is i.i.d. Bernoulli($p$), $\bc=1$, $\X=\Y = \{0,1\}$. The DMC is BSC($q$) and we require $E_{n}$ to be stored in the battery first, i.e., the energy model satisfies \eqref{eq:EH-S-comb-2} and \eqref{eq:EH-cost-quadratic}.

Fig.~\ref{fig:numerical-bounds-q0} and Fig.~\ref{fig:numerical-bounds-q0dot1} show various capacity bounds and achievable rates for $q = 0$ and $q=0.1$, respectively. The notations are explained under Fig.~\ref{fig:numerical-bounds-q0}. Each bound is prefixed by UB or LB, to denote whether it is an upper or lower bound. In many notations we also explicitly indicate the scenario from which the corresponding bounds are derived; but also recall that the lower bounds/achievable rates for EH-SC1 also work for EH-SC2, and that the upper bounds for the latter are also upper bounds for the former. With that in mind, we can see that for both $q = 0$ and $q=0.1$, the true channel capacity for either scenario actually lies in the shaded region, which happens to be the area between the smallest upper bound and the largest lower bound in the respective figures. The optimized achievable rates LB-r0, -r1 are calculated in the setting of Section~\ref{subsec:ach-rate-EH-SC1} (i.e., EH-SC1) with $m=1$, as in the previous subsection (where they are denoted by IR-r0, -r1). The relaxed upper bound from Theorem~\ref{thm:EH-linUB} is denoted by UB-SC2-LN to emphasize its linear complexity. The notation UB-LNX for the upper bounds from Theorem~\ref{thm:FSC-linUB-Analytical}/\ref{thm:EH-linUB-Analytical} is similarly defined, where X denotes the relaxation from conditional mutual information to conditional entropy of $X_{n}$ in \eqref{eq:FSC-I-H} or \eqref{eq:EH-I-H}. Since $E_{n}$ is i.i.d., the overall state $Z_{n}$ in Section~\ref{subsec:EH-FB-linUB} become $S_{n}$, and in this example one can show that these two theorems indeed give the same bound. In addition, UB$'$-SC1 (from \cite{Tutuncuoglu-EnHarvTimingCh}) is an upper bound for EH-SC1 when $q=0$, whereas when $q>0$ this result does not apply.

\begin{figure}[!t]
\centering
\includegraphics[width=\textwidth]{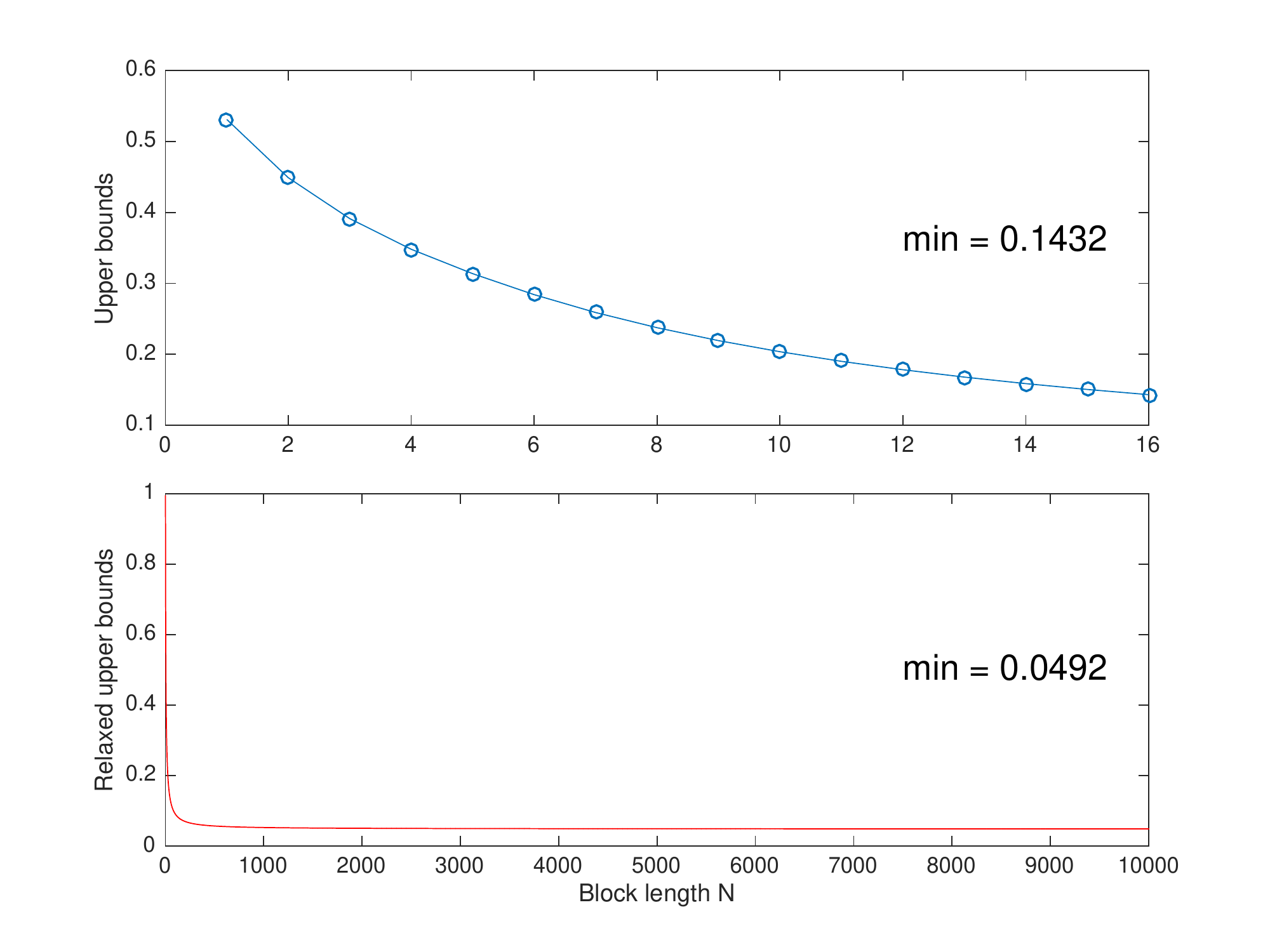}
\caption{Comparison of upper bounds for $q=0.1$, $p=0.01$, as a function of block length. Upper plot = UB-SC2, lower plot = UB-SC2-LN.}
\label{fig:numerical-ub-fb-lin-conv}
\end{figure}

For the linear complexity bounds UB-SC1, UB-SC2-LN, and UB-LNX, we can easily compute their values for block length $N=10^{4}$, when the bounds seem to have converged to their respective limits (cf. Fig.~\ref{fig:numerical-ub-fb-lin-conv}). For the exponential complexity bound UB-SC2, we are able to compute it up to $N=16$; for the double-exponential complexity bound LB-SC2, however, we are only able to compute it up to $N = 4$.

In the noiseless case Fig.~\ref{fig:numerical-bounds-q0}, the four upper bounds UB-SC1, UB-LNX, UB-SC2-LN and UB$'$-SC1 collapse to a single one. The coincidence of UB-SC1 (resp. UB-SC2-LN) and UB-LNX is guaranteed by Remark~\ref{rmk:FSC-I-H-exact} (resp. Remark~\ref{rmk:EH-I-H-exact}). The coincidence of UB-SC1 and UB$'$-SC1 hints that the best full-CSIR upper bound in EH-SC1 can be achieved when taking the bound in Theorem~\ref{thm:FSC-linUB} (or equivalently Theorem~\ref{thm:UB-FSC-X}) to its limit/infimum. Furthermore, the coincidence of UB-SC1 and UB-SC2-LN suggests that when supplying the information of $S_{n}$ to the receiver, the information of $E_{n}$ is not necessary for the noiseless case. Also from Fig.~\ref{fig:numerical-bounds-q0} we can observe that UB-SC2 (considered as an upper bound for EH-SC1) appears uniformly looser than UB-SC1, but it is unclear whether this is still the case when $N\to\infty$, since we are only able to compute the former up to $N = 16$. For the same reason, although UB-SC2-LN and UB-LNX are both relaxations of UB-SC2 and yield looser bounds for every fixed $N$, the end results in Fig.~\ref{fig:numerical-bounds-q0} are indeed tighter when we compute them for a much larger $N$.
 
In the noisy case Fig.~\ref{fig:numerical-bounds-q0dot1}, all curves are separated. The upper bound UB-LNX is the same as the case $q=0$, which can be seen from the relaxation \eqref{eq:FSC-I-H} or \eqref{eq:EH-I-H}. The bound UB-SC2-LN is looser than UB-SC1, which suggests that supplying the information of $E_{n}$ in addition to $S_{n}$ at the receiver helps identifying the source message in the noisy case. Moreover, the bound UB-SC2 is now tighter than UB-SC1, despite the fact that the latter is computed at a much larger block length. This phenomenon may hint that providing $S_{n}$ to the receiver gives away more information of the source than providing $E_{n}$. Furthermore, from Fig.~\ref{fig:numerical-bounds-q0dot1} the relaxations UB-LNX and UB-SC2-LN now both appear uniformly looser than the bound UB-SC2. However, when $p$ is small, the advantage of linear complexity can still yield a tighter bound, as illustrated for the case $p=0.01$ in Fig.~\ref{fig:numerical-ub-fb-lin-conv}: we can only compute the exponential-complexity upper bound UB-SC2 up to $N=16$ and the resulting best bound is 0.1432 bit/channel use, whereas UB-SC2-LN gives a much smaller bound 0.0492 at block length $N=10^{4}$.

\section{Conclusions}
\label{sec:conclusions}

We study the channel capacity problem of a discrete energy harvesting channel with a finite battery in its full generality. After introducing two energy harvesting channel models and a related finite state channel model, we convert them into their respective equivalent channels and express the capacities using Verd\'u-Han's formula. Then some simplifying restrictions are imposed on the inputs to give a surrogate channel for each equivalent model. Such types of channels allow us to use the Shannon-McMillan-Breiman theorem to compute some achievable rates, under the necessary stationarity and ergodicity conditions. These rates are then optimized using the GBAA algorithm. Following that we utilize Gallager's technique and Verd\'u-Han's bounds to arrive at series of capacity upper bounds by providing channel side information to the receiver. In addition, a lower bound in a similar form is also derived. The upper bounds are further simplified and relaxed to reduce the computational complexity.

Our results can be extended to several directions. The theory in Section~\ref{sec:ach-rates} can be generalized to continuous energy harvesting channels, especially when the input alphabet is finite, e.g., an AWGN channel with discrete input. Although the FSC and Markov channel results both can only work in the finite alphabet case, we can consider just the state process itself as the output of an Markov channel, and then connect a continuous memoryless channel to its output (cf. Appendix~\ref{sec:joint-marginal}). For such a case we can still derive ergodicity results and apply the SMB theorem. In addition, we can extend the results for both the achievable rates and capacity bounds to certain discrete energy harvesting channels with channel memory. For example, if the channel is not DMC, but an FSC, we can easily incorporate the channel state of the FSC in the derivations in both Section~\ref{sec:ach-rates} and Section~\ref{sec:capacity-bounds} and obtain corresponding results.

The results in this paper can also provide some guidelines in the design of practical energy harvesting communication systems. For example, one implication of our numerical results is that it might be a good strategy to invest in a small but nonzero battery if battery storage is costly. Furthermore, if the conjectures in Section~\ref{sec:ach-rates} are true, namely, the optimal input functions depend only on the current energy state, then designing energy harvesting channel codes can be greatly simplified---we would only need to consider codewords whose input symbols take this special form. The numerical results further suggest that i.i.d. random codes with an optimized distribution can have a good performance.


%

\appendices

\section*{Appendices: A Theory of Stationarity and Ergodicity}

To apply the Shannon-McMillan-Breiman theorem in our channel models, we need to derive the required stationarity and ergodicity conditions. For that purpose in the appendices of this paper we introduce the theory of stationarity and ergodicity for Markov channels, mostly established by \cite{Kieffer-Rahe-AMS-Mrkv-Ch,Gray-Dunham-Gobbi} (also see Gray's books \cite{Gray-PRE1,Gray-EIT}). It turns out, however, that some results in Gray et~al. \cite{Gray-Dunham-Gobbi} are inaccurate and/or not properly proved; thus before making use of them we must fix these issues first. In addition, besides the existing theory we also want to develop some extended results tailored for our own purposes, especially for the application in the finite state channels arising from energy harvesting systems. Hence in the following sections we first present the necessary background and preliminary results, then state the relevant theory of Markov channels from the literature, correct or supplement it if necessary, and in the meantime derive some extended or additional stationarity/ergodicity results of our own. In addition, we study the special case of a finite state channel with a finite-order Markov input, and obtain some other ergodicity conditions using results from \cite{Walters-Ergodic-Theory} for stationary Markov chains. Following that we present the Shannon-McMillan-Breiman theorem in the setting of an AMS ergodic process, and then develop some specific results for the models used in this work.

Throughout the appendices we follow the notations in \cite{Kieffer-Rahe-AMS-Mrkv-Ch,Gray-Dunham-Gobbi}, which uses a convention different from our main text. The reason is, we want to make the notations consistent with the related literature to facilitate a coherent understanding of the material. Due to space limitation, we omit some long proofs and detailed derivations and refer the interested readers to \cite{Mao-thesis}.

\section{Preliminaries for the Theory}

In this section we gather the most frequently used notations, concepts and preliminary results for the theory, and refer the interested readers to the original works on Markov channels \cite{Gray-Dunham-Gobbi,Kieffer-Rahe-AMS-Mrkv-Ch} for the rest. Most of the terminology and basic results can also be found in Gray's books \cite{Gray-EIT,Gray-PRE1}.

\subsection{General Properties}

Let $(\Omega,\bF)$ be a measurable space and $T:\Omega\to\Omega$ be a measurable mapping on it. Define a probability measure $\mu$ on $(\Omega,\bF)$ to be \emph{stationary}\footnote{This and many subsequent notions are defined \emph{with respect to $T$}. But for conciseness we usually omit this modifying phrase.} if
\[ \mu(T^{-1}F) = \mu(F),\qquad \forall F\in\bF. \]
For a probability measure $\mu$ on $(\Omega,\bF)$, if
\[ \lim_{n\to\infty} \frac{1}{n}\sum_{i=0}^{n-1}\mu(T^{-i}F) \]
exists for all $F\in\bF$, we say $\mu$ is \emph{asymptotically mean stationary} (AMS). The above equation also defines a stationary probability measure on $(\Omega,\bF)$, which is called the \emph{stationary mean} of $\mu$ and is usually denoted by $\omu$. Define an event $F$ to be \emph{invariant} if $T^{-1}F = F$. $\mu$ is \emph{ergodic} if $\mu(F)$ is either 0 or 1 for all invariant events $F$. Note that an AMS measure is ergodic iff its stationary mean is\cite[Lemma 6.7.1]{Gray-PRE1}.

We say a dynamical system $(\Omega,\bF,\mu,T)$ is stationary, AMS, or ergodic if the measure $\mu$ is. The following lemmas provide some useful results regarding the AMS property of a dynamical system (see \cite[Sec.~6.2--6.3]{Gray-PRE1}):

\begin{lemma}\label{lem:PRE-ams-iff}
$(\Omega,\bF,\mu,T)$ is AMS iff there exists a probability measure $\omu$ on $(\Omega,\bF)$ which is stationary and which agrees with $\mu$ on each invariant event.
\end{lemma}

\begin{lemma}\label{lem:PRE-ams-if}
$(\Omega,\bF,\mu,T)$ is AMS if there exists a stationary probability measure $\omu$ on $(\Omega,\bF)$ such that for any invariant $F\in\bF$, $\mu(F)=0$ whenever $\omu(F)=0$.
\end{lemma}

\subsection{Sources, Channels, and Hookups}
\label{subsec:source-channel-hookup}

The dynamical systems we are interested in are sources and source-channel hookups, both of which can be either two-sided or one-sided. Let $(A,\bA)$ be a measurable space, on which we want to define the one- and two-sided sequence spaces and sources. Let $(\osd{A},\osd{\bA})$ denote the measurable space of one-side sequences from alphabet $A$, whose sample space is composed of all sequences $(x_{1},x_{2},\cdots)$ from $A$ and whose $\sigma$-field $\osd{\bA}$ is the usual product $\sigma$-field of $\osd{A}$. Let $T$ be the left shift on $\osd{A}$, i.e.,
\[ T:(x_{1},x_{2},\cdots) \mapsto (x_{2},x_{3},\cdots), \]
which is a measurable map. A dynamical system $(\osd{A},\osd{\bA},\mu,T)$ of this form is called a one-sided \emph{source}, or \emph{process}, and is abbreviated to $[A,\mu]$. A two-sided source $(A^{\infty}, \bA^{\infty},\mu,T)$ is defined analogously: the sample space $A^{\infty}$ consists of all two-sided sequences $(\cdots,x_{-1},x_{0},x_{1},\cdots)$ from $A$ and the $\sigma$-field $\osd{\bA}$ is the corresponding product $\sigma$-field. Again, $T$ is the left shift, which maps a sequence $x=\{x_{i}\}_{i=-\infty}^{\infty}\in A^{\infty}$ to $Tx\in A^{\infty}$, where
\[ (Tx)_{i} = x_{i+1},\qquad \forall i\in\z. \]
Note that in this case $T$ has an inverse (the right shift), and both $T$ and $T^{-1}$ are measurable.

The same notation $T$ is used for the left shifts on both spaces, but context should make clear what the underlying space is. Furthermore, for unified treatment of both cases, let $(\Sigma_{A},\bSig_{A})$ denote the one- or two-sided sequence space of $(A,\bA)$, and let $\bI$ denote the time index set, which equals $\z^{+}\triangleq\{1,2,\cdots\}$ or $\z$ for the one- or two-sided cases, respectively. Recall that the basic events of the sequence spaces are the (finite dimensional) \emph{rectangles}, also called the \emph{cylinder sets}, which are subsets $F$ of the form
\[ F = \lt\{\, x\in\Sigma_{A}: x_{i}\in F_{i},\ \forall i\in\bJ\,\rt\}, \]
where $\bJ$ is a finite subset of $\bI$ and $F_{i}\in\bA$ for all $i\in\bJ$. The sets $F_{i}$, $i\in\bJ$ are called the \emph{coordinate events}. When $F_{i}$ is a singleton for each $i\in\bJ$, $F$ is called a \emph{thin cylinder}.

A \emph{channel} $[A,\nu,B]$ with input alphabet $A$ and output alphabet $B$ is defined by a family of probability measures $\{\nu_{x}:x\in\Sigma_{A}\}$ on $(\Sigma_{B},\bSig_{B})$ such that for each event $F\in\bSig_{B}$, the map
\[ x \mapsto \nu_{x}(F) \]
from $(\Sigma_{A},\bSig_{A})$ into $[0,1]$ with its Borel $\sigma$-field is measurable. A channel is called one- or two-sided if the underlying sequence space is. Given a source $[A,\mu]$ and a channel $[A,\nu,B]$, the \emph{source-channel hookup}, or the \emph{input-output process}, is the process $[A\times B,\mu\nu]$, where the measure $\mu\nu$ is defined by
\[ \mu\nu(F) = \int_{\Sigma_{A}}\nu_{x}(F_{x})\ud\mu(x),\qquad \forall F\in\bSig_{A\times B}, \]
with $F_{x}$ being the section of $F$ at $x$:
\[ F_{x} \triangleq \{y\in\Sigma_{B}\mid(x,y)\in F\}. \]
The corresponding left shift for this process is still denoted by $T$, with
\[ T(x,y) = (Tx,Ty),\qquad \forall (x,y)\in\Sigma_{A}\times\Sigma_{B}. \]

Sometimes when the alphabets are understood we simply denote the above source, channel, and their hookup by the corresponding measures $\mu$, $\nu$, and $\mu\nu$, respectively. As usual, the random processes corresponding to the source and hookup can be denoted by their respective sequences of coordinate random variables $\{X_{n}\}_{n\in\bI}$ and $\{(X_{n},Y_{n})\}_{n\in\bI}$, where for any $n$ we define
\begin{align*}
&X_{n}:\Sigma_{A}\to A,\quad x \mapsto x_{n} \\
&Y_{n}:\Sigma_{B}\to B,\quad y \mapsto y_{n}.
\end{align*}
Sometimes we also drop the subscript $n\in\bI$ when there is no confusion. We say these random processes are stationary, AMS, or ergodic if the underlying dynamic systems are. Furthermore, for convenience we define the projection map $\pi$ between the one- and two-sided spaces on $A$ as
\[ \begin{array}{c}
\pi:A^{\infty}\to\osd{A}\\
\phantom{\pi:}\ x\ \mapsto \osd{x}
\end{array}, \]
where $x=\{x_{i}\}_{i=-\infty}^{\infty}$ and
\[ x_{1}^{\infty}\triangleq (x_{1},x_{2},\cdots). \]
Similarly define the projection maps for the alphabets $B$ and $A\times B$, which are still denoted by $\pi$. It is easy to verify that $\pi$ is always measurable and \emph{stationary}, namely, $\pi T=T\pi$.

A channel $[A,\nu,B]$ is said to be \emph{stationary} if $\forall x\in\Sigma_{A}$, $\forall F\in\bSig_{B}$
\[ \nu_{Tx}(F) = \nu_{x}(T^{-1}F). \]
The term ``stationary'' is justified by \cite[Lemma~9.3.1]{Gray-EIT}, which shows that connecting a stationary source to a stationary channel yields a stationary input-output process. The channel is said to be \emph{AMS} if, for every AMS source, the source-channel hookup is AMS. An AMS channel $\nu$ is \emph{ergodic} if the hookup $\mu\nu$ is ergodic whenever $\mu$ is AMS and ergodic.

A simple example of stationary channels is the family of \emph{stationary memoryless channels}\footnote{In \cite{Gray-EIT} such channels are simply called memoryless channels.}. Every channel $[A,\nu,B]$ in this family is associated with a collection of probability measures $\{q_{a}:a\in A\}$ on $(B,\bB)$, such that for each output rectangle $F\in\bSig_{B}$,
\[ \nu_{x}(F) = \prod_{i\in\bJ}q_{x_{i}}(F_{i}), \]
where $\bJ$ is the index set and $F_{i}$, $i\in\bJ$ are the coordinate events of $F$. When $A$ and $B$ are finite sets, $\nu$ is called a \emph{discrete memoryless channel} (DMC).

\subsection{Markov Channels and Finite State Channels}
\label{subsec:Markov-Finite-State-channels}

Fix the input and output measurable spaces $(A,\bA)$ and $(B,\bB)$, where $(A,\bA)$ is arbitrary, but $B$ is a finite set with cardinality $K$ and $\bB$ consists of all subsets of $B$. Let $\bP$ denote the space of all $K\times K$ stochastic matrices $P$, whose $(i,j)$-th entry is denoted by $P(i,j)$ for $1\leq i,j\leq K$. Using the Euclidean metric on $\bP$ we can construct its Borel $\sigma$-field to form a measurable space, which in turn induces a one- or two-sided sequence space $(\Sigma_{P},\bSig_{P})$. Given a sequence $P\in\Sigma_{P}$, let $\bM(P)$ denote the set of all probability measures on $(\Sigma_{B},\bSig_{B})$ with respect to which $Y_{m},Y_{m+1},\cdots$ forms a (non-homogeneous) Markov chain with transition matrices $P_{m},P_{m+1},\cdots$ for any integer $m\in\bI$. That is, $\lambda\in\bM(P)$ iff $\forall m\in\bI$, $\forall n>m$, and $\forall y_{m},\cdots,y_{n}\in B$,
\[ \lambda(Y_{m}=y_{m},\cdots,Y_{n}=y_{n}) = \lambda(Y_{m}=y_{m})\prod_{i=m}^{n-1}P_{i}(y_{i},y_{i+1}). \]
In the one-sided case only $m=1$ need be verified.

As before we say a map $\phi:\Sigma_{A}\to\Sigma_{P}$ is \emph{stationary} if $\phi T=T\phi$. A channel $[A,\nu,B]$ is called \emph{Markov} if there exists a stationary measurable map $\phi:\Sigma_{A}\to\Sigma_{P}$ such that
\[ \nu_{x}\in\bM(\phi(x)),\qquad \forall x\in\Sigma_{A}. \]
The major results proved in \cite{Kieffer-Rahe-AMS-Mrkv-Ch} by Kieffer and Rahe for Markov channels is summarized in the following theorem:

\begin{theorem}\label{thm:Markov-channel-AMS}
Every one- and two-sided Markov channel is AMS.
\end{theorem}

Now let $A$ also be finite and let $\{P_{a}:a\in A\}\subset\bP$. If a one-sided Markov channel $[A,\nu,B]$ satisfies
\[ \phi(x)_{n} \triangleq [\phi(x)]_{n} = P_{x_{n}},\qquad \forall n>0, \]
then $\nu$ is called a \emph{finite state channel}. In this case, the matrix produced by $\phi$ at time $n$ depends only on the input at that time, $x_{n}$. This definition is equivalent to Gallager's finite state channel (FSC) defined in\cite{Gallager-IT_Reliable} (see Definition~\ref{def:Gallager-FSC}), in terms of channel transitions. In fact, for the latter definition we have finite input, output, and state alphabets with respective symbols $X_{n},Y_{n}$, and $S_{n}$ that fulfill the conditional probability requirement\footnote{As in the main text, the state index is increased by 1 compared to the original definition in \cite{Gallager-IT_Reliable}.}
\begin{equation}\label{eq:FSC-cond-prob-App}
\Pr\lt(\begin{array}{c}
Y_{n}=y_{n},\\
S_{n+1}=s_{n+1}
\end{array} \lt| \begin{array}{l}
Y_{i}=y_{i},\, 0<i<n; \\
S_{j}=s_{j},\, 0<j\leq n; \\
X_{k}=x_{k},\, k>0
\end{array}\rt.\rt) = p(y_{n}s_{n+1}\cn x_{n}s_{n}).
\end{equation}
In other words, conditioned on $(X_{n},S_{n})$, the pair $(Y_{n},S_{n+1})$ is independent of all prior inputs, outputs, and states\footnote{Actually from \eqref{eq:FSC-cond-prob-App}, $(Y_{n},S_{n+1})$ is also conditionally independent of the future inputs, i.e., the channel is causal. It is also implicitly assumed when computing the block conditional probability in \cite{Gallager-IT_Reliable} (equation (4.6.1)). This condition is indeed satisfied by the FSC models we study. (See \cite{Mao-thesis} for more discussion).}. If we define the new output $Y'_{n}$ of the channel as the output-state pair $(Y_{n-1},S_{n})$ with
\[ P_{x_{n}}\lt(y'_{n},y'_{n+1}\rt) = 
P_{x_{n}}\lt(\,(y_{n-1},s_{n}),(y_{n},s_{n+1})\,\rt) \triangleq p(y_{n}s_{n+1}\cn x_{n}s_{n}), \]
then Gallager's model fits in the definition here. The other direction is obvious if we define $S_{n}=Y_{n}$. In light of their equivalence, we do not explicitly distinguish the two definitions in this paper. Most of the time we will find out that it is more convenient to work with the first one when studying the general theory, while the second one provides more flexibility when dealing with specific channel models.

\subsection{Constructions by Kieffer and Rahe}
\label{subsec:Kieffer-Rahe-auxilliary-construction}

To prove Theorem~\ref{thm:Markov-channel-AMS}, Keiffer and Rahe establish some intermediate source and channel constructions in \cite{Kieffer-Rahe-AMS-Mrkv-Ch}, which we will need for the relevant ergodicity results and are summarized below.

Let $[A,\mu]$ be an AMS source and $[A,\nu,B]$ be a Markov channel, with $\phi$ being the corresponding stationary map. Since $\mu$ is AMS, by Lemma~\ref{lem:PRE-ams-iff} there is a stationary measure $\omu$ on $(\Sigma_{A},\bSig_{A})$ that agrees with $\mu$ on each invariant event in $\bSig_{A}$. ($\omu$ can be simply taken to be the stationary mean of $\mu$.) Define a two-sided stationary source $[A,\omus]$ as follows: if the original source is two-sided, then $\omus=\omu$; otherwise let $\omus$ be the two-sided stationary extension\footnote{Such an extension is always possible and unique by the Kolmogorov extension theorem if the measurable space $(A,\bA)$ is \emph{standard}, which is true for countable or Euclidean spaces. Interested readers may consult \cite[Ch.~2,3]{Gray-PRE1} for details.} of the one-sided measure $\omu$, which is specified by
\[ \omus(\,(X_{m},X_{m+1},\cdots)\in F\,) = \omu(F),\quad \forall m\in\z,\ \forall F\in\osd{\bA}. \]
In particular, considering $m=1$ we have
\[ \omus(\pi^{-1}F) = \omu(F). \]
Also, define a two-sided stationary map $\phi'$ by setting $\phi'=\phi$ if the original system is two-sided, and defining
\[ \phi'(x)_{i} = \phi(x_{i}^{\infty})_{1}\qquad \forall i\in\z,\ \forall x\in A^{\infty} \]
otherwise, where $x_{i}^{\infty} \triangleq (x_{i},x_{i+1},\cdots)$. In particular, for the latter case
\[ \osd{\phi'(x)} = \phi(\osd{x}) = \phi(\pi(x)). \]
Furthermore, \cite{Kieffer-Rahe-AMS-Mrkv-Ch} constructs a measurable subset $R\subset\bP^{\infty}$ and proves that the measurable set
\[ R' = (\phi')^{-1}(R) = \{ x\in A^{\infty}: \phi'(x) \in R \} \]
is invariant and has probability 1 under any stationary probability measure on $(A^{\infty},\bA^{\infty})$, in particular
\[ \omus(R')=1. \]

With these constructions Kieffer and Rahe define a two-sided channel $[A,\hnu,B]$ which has the following properties:
\begin{enumerate}
\item $\hnu$ is stationary and hence so is the input-output process $\omus\hnu$.
\item $\hnu_{x}\in\bM(\phi'(x))$ for $x\in R'$, so $\hnu$ has the same transition structure as $\nu$, $\omus$-a.e.
\end{enumerate}
Besides, if the original system is two-sided, then $\mu\nu$ is absolutely continuous w.r.t. $\omus\hnu$. In particular, for any invariant event $F\in \bA^{\infty}\times\bB^{\infty}$, $\mu\nu(F)=0$ whenever $\omus\hnu(F)=0$, whereas if $\nu$ is one-sided, \cite{Kieffer-Rahe-AMS-Mrkv-Ch} defines the ``one-sided restriction'' of the two-sided measure $\omus\hnu$ as
\[ (\omus\hnu)' \triangleq (\omus\hnu)\pi^{-1}, \]
which is also stationary since $\pi$ is. Moreover, if $F\in \osd{\bA}\times\osd{\bB}$ is invariant and $(\omus\hnu)'(F) = 0$, then also $\mu\nu(F)=0$. Therefore in both cases $\mu\nu$ is AMS by Lemma~\ref{lem:PRE-ams-if}, and so is $\nu$.

\begin{remark}\label{rmk:R'-issue}
In \cite{Gray-Dunham-Gobbi} property 2) of $\hnu$ is assumed to be true for all $x\in A^{\infty}$, which is not the case in the original construction of \cite{Kieffer-Rahe-AMS-Mrkv-Ch}. This misrepresentation is one source of inaccuracy for Lemma~2 and the proof of Theorem~2 in \cite{Gray-Dunham-Gobbi}, which we will fix in later sections.
\end{remark}

From these facts we can also obtain the following two results regarding the ergodicity of certain related processes, which are indispensable in current approaches for proving ergodicity of Markov channels. Although their proofs are not difficult and \cite{Kieffer-Rahe-AMS-Mrkv-Ch} uses these results without explicitly proving them, we provide the proofs below for the sake of clarity and completeness.

\begin{lemma}\label{lem:mu-auxiliary-ergodicity}
If $\mu$ is ergodic, then so is the auxiliary measure $\omus$ for both one- and two-sided systems.
\end{lemma}

\begin{IEEEproof}
By construction $\omu$ is ergodic iff $\mu$ is, so for the two-sided case we are done. For the one-sided case, by the generating field structure of $\bA^{\infty}$ and \cite[Lemma~6.7.4]{Gray-PRE1} it is enough to prove that
\begin{equation}\label{eq:mu-auxiliary-ergodicity}
\lim_{n\to\infty} \frac{1}{n}\sum_{i=0}^{n-1}\omus(T^{-i}F\cap G) = \omus(F)\omus(G)
\end{equation}
for all rectangles $F,G\in\bA^{\infty}$ when $\mu$ is ergodic. But by the stationarity of $\omus$, without loss of generality we can assume the relevant coordinates for the rectangles $F$ and $G$ are positive. Thus there exists rectangles $F',G'\in\osd{\bA}$ such that $F = \pi^{-1}F'$ and $G = \pi^{-1}G'$. Now by the relation of $\omu$ and $\omus$ and the stationarity of $\pi$, \eqref{eq:mu-auxiliary-ergodicity} becomes
\[ \lim_{n\to\infty} \frac{1}{n}\sum_{i=0}^{n-1}\omu(T^{-i}F'\cap G') = \omu(F')\omu(G'), \]
which is true by \cite[Lemma~6.7.3]{Gray-PRE1} when $\omu$ is ergodic.
\end{IEEEproof}

\begin{lemma}\label{lem:mu-nu-auxiliary-ergodicity}
If the auxiliary measure $\omus\hnu$ is ergodic, then so is $\mu\nu$ for both one- and two-sided systems.
\end{lemma}

\begin{IEEEproof}
Observe that the complement of an invariant event is also invariant. In the two-sided case, if $\omus\hnu(F) = 1$ for an invariant $F$, then $\omus\hnu(F^{c}) = 0$ and so $\mu\nu(F^{c}) = 0$, and thus $\mu\nu(F) = 1$. Hence ergodicity of $\omus\hnu$ implies ergodicity of $\mu\nu$. For the one-sided case, let $F\in\osd{\bA}\times\osd{\bB}$ be invariant, then $\pi^{-1}F\in \bA^{\infty}\times\bB^{\infty}$ is also invariant as $\pi$ is stationary. Assume $\omus\hnu$ is ergodic, then
\[ (\omus\hnu)'(F) = [(\omus\hnu)\pi^{-1}](F) = \omus\hnu(\pi^{-1}F), \]
which is either 1 or 0. Again by the same argument, $\mu\nu(F) = 1$ or 0 and hence $\mu\nu$ is also ergodic.
\end{IEEEproof}

\section{Ergodicity Results for Markov Channels}
\label{sec:ergodicity-Markov-channels}

We are now ready to present the relevant results in \cite{Gray-Dunham-Gobbi}, together with our comments, amendments, and corrections. In the meantime, we will develop some supplementary or extended results to apply in our own work.

\subsection{Weak Ergodicity of Markov Channels}

Assume the same setting as the previous section, where we have an AMS source $[A,\mu]$ and a Markov channel $[A,\nu,B]$ with the corresponding auxiliary constructions. For any $m\in\bI$, $\forall n>m$ and $\forall x\in\Sigma_{A}$, we denote the output transition probability matrix for $\nu$ from time $m$ to $n$ by $H_{mn}(x) = H_{m,n}(x)$. In other words, for $1\leq j,k\leq K$,
\[ [H_{mn}(x)]_{jk} \triangleq \nu_{x}(Y_{n}=b_{k}\cn Y_{m}=b_{j}), \]
where we fix an ordered enumeration $\{b_{1},b_{2},\cdots,b_{K}\}$ of $B$. Since $\nu_{x}\in\bM(\phi(x))$,
\begin{equation}\label{eq:H-phi-original}
H_{mn}(x) = \prod_{i=m}^{n-1}\phi(x)_{i},\qquad \forall x\in\Sigma_{A}.
\end{equation}
Similarly, for the auxiliary two-sided channel $\hnu$, for any $m<n\in\z$ and $\forall x\in A^{\infty}$ we define the matrix $H^{*}_{mn}(x) = H^{*}_{m,n}(x)$ by
\[ [H^{*}_{mn}(x)]_{jk} \triangleq \hnu_{x}(Y_{n}=b_{k}\cn Y_{m}=b_{j}) \]
for $1\leq j,k\leq K$. Since $\hnu_{x}\in\bM(\phi'(x))$ on $R'$, we have
\begin{equation}\label{eq:H-phi-auxiliary}
H^{*}_{mn}(x) = \prod_{i=m}^{n-1}\phi'(x)_{i},\qquad \forall x\in R'.
\end{equation}
Thus if $\nu$ is two-sided, then $\phi'=\phi$ and so
\begin{equation}\label{eq:H-twoside}
H^{*}_{mn}(x) = H_{mn}(x),\qquad \forall x\in R',\ \forall n\geq m\in\z,
\end{equation}
whereas if $\nu$ is one-sided, as $\osd{\phi'(x)} = \phi(\pi(x))$ for all $x\in A^{\infty}$,
\begin{equation}\label{eq:H-oneside}
H^{*}_{1n}(x) = H_{1n}(\pi(x)),\qquad \forall x\in R',\ \forall n\geq 1.
\end{equation}

\begin{definition}\label{def:weakly-ergodic}
Let $H_{mn}$ denote the transition matrix from time $m$ to $n$ for a non-homogeneous Markov chain with $K$ states, for $0<m<n$. The Markov chain is called \emph{weakly ergodic} if
\begin{equation}\label{eq:weakly-ergodic-Markov-chain}
\lim_{n\to\infty} \lt| (H_{mn})_{ij} - (H_{mn})_{kj} \rt| = 0,\qquad \forall m>0,\ \forall1\leq i,j,k\leq K.
\end{equation}
A Markov channel $\nu$ is \emph{weakly ergodic} if for all $x\in\Sigma_{A}$,
\begin{equation}\label{eq:weakly-ergodic-channel}
\lim_{n\to\infty} \lt| [H_{mn}(x)]_{ij} - [H_{mn}(x)]_{kj} \rt| = 0,\qquad \forall m\in\bI,\ \forall1\leq i,j,k\leq K.
\end{equation}
We also say it is \emph{weakly ergodic on a set $F$} if \eqref{eq:weakly-ergodic-channel} holds for all $x\in F$. Furthermore, $\nu$ is called \emph{weakly ergodic $\mu$-a.e.} for a probability measure $\mu$ if it is weakly ergodic on a set with $\mu$-measure 1. 
\end{definition}

Since $\phi$ is stationary, by \eqref{eq:H-phi-original}
\[ H_{mn}(x)=H_{1,(n-m+1)}(T^{m-1}x). \]
This relation is true for both one- and two-sided channels for all $m\in\bI$, noting that in the latter case $T$ is invertible and so $T^{m-1}x$ is always a single point. Hence we only need to verify \eqref{eq:weakly-ergodic-channel} for the special case $m=1$ to prove the weak ergodicity of a Markov channel. Similarly, for the almost everywhere definition we have

\begin{lemma}[Lemma~1 in \cite{Gray-Dunham-Gobbi}]\label{lem:Gray-Dunham-Gobbi-lem-1}
Suppose $\mu$ is a stationary source. Then a Markov channel $\nu$ is weakly ergodic $\mu$-a.e. iff for $m=1$, \eqref{eq:weakly-ergodic-channel} holds with $\mu$-probability 1.
\end{lemma}

Given a $K\times K$ stochastic matrix $P$, define
\[ \delta(P) = \max_{s,t}\sum_{1\leq k\leq K}(P_{tk}-P_{sk})^{+}, \]
where $(a)^{+} \triangleq \max\{0,a\}$. It is the maximum total variation distances between the rows of $P$, with $0\leq \delta(P)\leq 1$. $P$ is called \emph{scrambling} if $\delta(P)<1$, which holds iff for any two rows $i$ and $k$ there is at least one column $j$ for which both $P_{ij}>0$ and $P_{kj}>0$; or equivalently, no two rows of $P$ are orthogonal. Moreover, for any stochastic matrices $P$ and $Q$,
\begin{equation}\label{eq:delta-PQ}
\delta(PQ)\leq\delta(P)\delta(Q).
\end{equation}

Observe that for any fixed $m$, \eqref{eq:weakly-ergodic-Markov-chain} is true iff
\[ \lim_{n\to\infty}\delta(H_{mn}) = 0. \]
This gives an equivalent definition for the weak ergodicity of a non-homogeneous Markov chain. By the same token we have the following lemma. Its first part comes from \cite[Lemma~2]{Gray-Dunham-Gobbi} with the issue of $\hnu$ (metioned in Remark~\ref{rmk:R'-issue}) fixed, while the second part comprises two statements supplemented by ourselves.

\begin{lemma}[Lemma~2 in \cite{Gray-Dunham-Gobbi}, amended for $R'$ and extended]\label{lem:Gray-Dunham-Gobbi-lem-2}
A Markov channel $\nu$ is weakly ergodic iff
\[ \lim_{n\to\infty}\delta(H_{1n}(x)) = 0,\qquad \forall x\in\Sigma_{A}. \]
In this case, the induced channel $\hnu$ is weakly ergodic on $R'$. Given a source $[A,\mu]$, a Markov channel $\nu$ is weakly ergodic $\mu$-a.e. iff the event
\[ F \triangleq \lt\{x\in \Sigma_{A}: \lim_{n\to\infty}\delta(H_{mn}(x)) = 0,\ \forall m\in\bI \rt\} \]
has $\mu$-probability 1. If the $\mu$ is stationary, then only $m=1$ need be considered. Furthermore, if $\mu$ is AMS, then $\nu$ is weakly ergodic $\mu$-a.e. iff $\omu$-a.e., in which case $\hnu$ is also weakly ergodic on a subset of $R'$ with $\omus$-probability 1.
\end{lemma}

\begin{IEEEproof}
See \cite{Mao-thesis}.
\end{IEEEproof}

The first main result in \cite{Gray-Dunham-Gobbi} provides an alternative characterization of a.e. weakly Markov channels. Let $\Ex{\cdot}$ denote expectation, i.e., the integration w.r.t. the corresponding measure.

\begin{theorem}[Theorem~1 in \cite{Gray-Dunham-Gobbi}]\label{thm:Gray-Dunham-Gobbi-thm-1}
A necessary condition for a Markov channel $\nu$ to be weakly ergodic $\mu$-a.e. for a stationary measure $\mu$ is that there exists an $N$ such that
\begin{equation}\label{eq:Gray-et-al-thm-1}
\Ex{\ln\delta(H_{1N}(X))} < 0.
\end{equation}
A sufficient condition for $\nu$ to be weakly ergodic $\mu$-a.e. for a stationary and ergodic measure $\mu$ is that there exists an $N$ such that \eqref{eq:Gray-et-al-thm-1} holds.
\end{theorem}

Gray et al. further derive three corollaries of this theorem in \cite{Gray-Dunham-Gobbi}. However, all of them are inaccurate in that they all require an additional condition to hold: the source $\mu$ need be ergodic, apart from being stationary. That is because essentially the proofs all need to use the sufficient condition of the theorem. Below we state these corollaries as lemmas, together with the corrections and some extended results.

\begin{lemma}[Corollary~1 in \cite{Gray-Dunham-Gobbi}, corrected and amended]\label{lem:Gray-Dunham-Gobbi-cor-1}
Given a Markov channel $\nu$ and a stationary ergodic source $\mu$ the following conditions are equivalent.
\begin{enumerate}
\item[a)] The channel is weakly ergodic $\mu$-a.e..
\item[b)] For $\mu$-a.e. each $x$, $\exists n$ such that no two rows of $H_{1n}(x)$ are orthogonal; or equivalently, $H_{1n}(x)$ is scrambling, i.e., $\delta(H_{1n}(x))<1$.
\item[c)] The channel has the ``positive column property'' $\mu$-a.e.; that is, for $\mu$-a.e. each $x$ there is an $n$ for which $H_{1n}(x)$ has a positive column.
\end{enumerate}
\end{lemma}

\begin{IEEEproof}
The proof provided in \cite{Gray-Dunham-Gobbi} is mostly correct, except that the result that b) implies a) does require the sufficient condition of Theorem~\ref{thm:Gray-Dunham-Gobbi-thm-1}. To prove that result, assume b) is true but a) is false. Then $\Ex{\ln\delta(H_{1n}(X))} = 0$ for all $n$, otherwise by the sufficient condition $\nu$ is indeed weakly ergodic $\mu$-a.e.. As $\ln\delta(\cdot)\leq0$, for each $n$ we must have $\ln\delta(H_{1n}(x))=0$ on a set $F_{n}$ with $\mu$-probability 1. Thus the intersection $\bigcap_{n>1}F_{n}$ also has $\mu$-probability 1, on which $\delta(H_{1n}(x))=1$ for all $n$. As a result, the set
\[ E \triangleq \Big(\bigcap_{n>1}F_{n}\Big)^{c} = \lt\{\,x\in\Sigma_{A}: \exists n>1\text{ s.t. } \delta(H_{1n}(x))<1\,\rt\} \]
is null, i.e., $\mu(E)=0$. This is a contradiction, since $\mu(E)=1$ by b).
\end{IEEEproof}

From the proof above, the contradiction still exists as long as $E$---the set on which the requirement for b) holds---has a positive $\mu$-probability. Also, for each point $x$ the requirement for b) is implied by that of c). Hence we can relax the conditions b) and c), to only requiring them to hold on a set with positive $\mu$-probability, and the lemma is still correct. However, actually this is not a true relaxation, in view of our next lemma.

\begin{lemma}\label{lem:Gray-Dunham-Gobbi-cor-1-relaxation}
Let $\mu$ be stationary and ergodic. The corresponding requirement for each condition of Lemma~\ref{lem:Gray-Dunham-Gobbi-cor-1} holds $\mu$-a.e. iff it holds on a set of positive $\mu$-probability.
\end{lemma}

\begin{IEEEproof}
See \cite{Mao-thesis}.
\end{IEEEproof}

Furthermore, note that for both conditions b) and c), the corresponding properties only need to hold on a finite segment of a sequence. Combining this observation with the definition of finite state channels, we have the following corollary.

\begin{corollary}\label{cor:Gray-Dunham-Gobbi-cor-1-relaxation-cor}
Let $\mu$ be a stationary ergodic source and $\nu$ be a Markov channel. For either condition b) or c) of Lemma~\ref{lem:Gray-Dunham-Gobbi-cor-1}, if there exists a finite-dimensional rectangle $F$ possessing positive $\mu$-probability such that the corresponding requirement holds for all $x\in F$, then $\nu$ is weakly ergodic $\mu$-a.e. In particular, when $\nu$ is a finite state channel and $F$ is a thin cylinder, we have a specific result: let $(a_{1},\cdots,a_{n})\in A^{n}$, if
\begin{enumerate}
\item $\mu(X_{1}=a_{1},\cdots,X_{n}=a_{n}) > 0$,
\item $\prod_{i=1}^{n}P_{a_{i}}$ is scrambling, or has a positive column,
\end{enumerate}
then $\nu$ is weakly ergodic $\mu$-a.e.
\end{corollary}

\begin{IEEEproof}
The first statement follows from the two lemmas above. For a finite state channel $\nu$, let $F$ be the thin cylinder with coordinate events $F_{i} = \{a_{i}\}$ for $1\leq i\leq n$. Then by \eqref{eq:H-phi-original}
\[ H_{1,(n+1)}(x) = \prod_{i=1}^{n}\phi(x)_{i} = \prod_{i=1}^{n}P_{a_{i}},\qquad \forall x\in F. \]
Hence the second statement holds as a special case of the first one.
\end{IEEEproof}

The second corollary of Theorem~\ref{thm:Gray-Dunham-Gobbi-thm-1} deals with Gallager's concept of indecomposable finite state channels\cite{Gallager-IT_Reliable}, which is generalized to all Markov channels in \cite{Gray-Dunham-Gobbi} as follows.

\begin{definition}\label{def:Gallager-indecomposable}
A Markov channel $\nu$ is \emph{indecomposable in the Gallager sense}\footnote{In the main text we only use the term \emph{indecomposability} in the context of an FSC and it refers exclusively to this definition.} if for every $\epsilon>0$ there is an $N$ such that for all $n\geq N$
\[ \lt| [H_{1n}(x)]_{ij} - [H_{1n}(x)]_{kj} \rt| <\epsilon,\qquad \forall x\in\Sigma_{A},\ \forall1\leq i,j,k\leq K. \]
\end{definition}

\begin{remark}
For a Markov channel both the indecomposability in the Gallager sense and the weak ergodicity require that asymptotically the rows of the transition matrix become more and more alike. However, the former requires uniform convergence for all input sequences $x$ while the latter does not.
\end{remark}

If a Markov channel $\nu$ is indecomposable in the Gallager sense, then $\nu$ has the \emph{strong positive column property}, that is, there is an $n$ such that $H_{1n}(x)$ has a positive column for every $x$. If $\nu$ is a finite state channel, then \cite{Gallager-IT_Reliable} shows that the relation is indeed \emph{if and only if}. Since obviously strong positive column property implies positive column property, by Lemma~\ref{lem:Gray-Dunham-Gobbi-cor-1} we have the following lemma.

\begin{lemma}[Corollary~2 in \cite{Gray-Dunham-Gobbi}, corrected]\label{lem:Gray-Dunham-Gobbi-cor-2}
A sufficient condition for a Markov channel to be weakly ergodic $\mu$-a.e. for a stationary and ergodic source $\mu$ is that it is indecomposable in the Gallager sense $\mu$-a.e.
\end{lemma}

The third corollary of Theorem~\ref{thm:Gray-Dunham-Gobbi-thm-1} is not used in our work and requires some extra definitions, hence we only correct it below and refer the interested readers to the original paper of Gray et al. for the concept of indecomposability for a Markov channel (which is different from Definition~\ref{def:Gallager-indecomposable}). 

\begin{lemma}[Corollary~3 in \cite{Gray-Dunham-Gobbi}, corrected]\label{lem:Gray-Dunham-Gobbi-cor-3}
A sufficient condition for a Markov channel to be weakly ergodic $\mu$-a.e. for a stationary and ergodic source $\mu$ is that it is indecomposable $\mu$-a.e.
\end{lemma}

\begin{remark}\label{rmk:Gray-Dunham-Gobbi-cor2-3-relaxation}
Since Lemma~\ref{lem:Gray-Dunham-Gobbi-cor-2} and \ref{lem:Gray-Dunham-Gobbi-cor-3} essentially use Lemma~\ref{lem:Gray-Dunham-Gobbi-cor-1}, by Lemma~\ref{lem:Gray-Dunham-Gobbi-cor-1-relaxation} we only need their corresponding conditions to hold on a set of positive probability.
\end{remark}

\subsection{Mixing and Ergodic Markov Channels}

Before presenting the main ergodicity results for Markov channels, we require yet another definition of a class of channels, which was first introduced by Adler in \cite{Adler-Ergodic-Mixing}.

\begin{definition}\label{def:strongly-mixing}
A channel $\nu$ is called \emph{strongly mixing}, or \emph{output mixing}\cite{Gray-EIT}, or \emph{asymptotically
independent of the remote past}\cite{Adler-Ergodic-Mixing} if for all output rectangles $F$ and $G$ and all input sequences $x$
\begin{equation}\label{eq:strongly-mixing}
\lim_{n\to\infty}\lt| \nu_{x}(F\cap T^{-n}G) - \nu_{x}(F)\nu_{x}(T^{-n}G) \rt| = 0.
\end{equation}
It is called \emph{strongly mixing $\mu$-a.e.} for a probability measure $\mu$ if the above condition holds for all $x$ in a set of $\mu$-measure 1.
\end{definition}

\begin{remark}
Immediately from the definition we can see that stationary memoryless channels are strongly mixing. In fact, the strongly mixing channels are proposed in \cite{Adler-Ergodic-Mixing} to generalize the idea of channels with finite memory (which obviously include the memoryless channels).
\end{remark}

The importance of strongly mixing channels lies in the following theorem, which is adapted from \cite{Adler-Ergodic-Mixing} and \cite[Lemma~9.4.3]{Gray-EIT}.

\begin{theorem}[Adler's Theorem]\label{thm:Adler}
Let $\nu$ be a stationary channel. If $\mu$ is a stationary ergodic source and $\nu$ is strongly mixing $\mu$-a.e., then $\mu\nu$ is also stationary and ergodic. Similarly, if $\mu$ is AMS ergodic and $\nu$ is strongly mixing $\mu$-a.e., then $\mu\nu$ is also AMS and ergodic.
\end{theorem}

\begin{IEEEproof}
For the statement with stationary $\mu$, see \cite{Adler-Ergodic-Mixing} or \cite[Lemma~9.4.3]{Gray-EIT} for a proof. For the AMS case the proof can be easily adapted from the stationary case with \cite[Lemma~9.3.2]{Gray-EIT}.
\end{IEEEproof}

The following lemma connects the a.e. weak ergodicity and a.e. strongly mixing property of Markov channels.

\begin{lemma}[Lemma~3 in \cite{Gray-Dunham-Gobbi}, corrected]\label{lem:Gray-Dunham-Gobbi-lem-3}
Given a stationary source $\mu$, if a Markov channel is weakly ergodic $\mu$-a.e., then it is also strongly mixing $\mu$-a.e.
\end{lemma}

\begin{remark}
The original statement of Lemma~3 in \cite{Gray-Dunham-Gobbi} claims that the reverse direction is also true. However, the proof for this direction has a missing link: equation (12) in \cite{Gray-Dunham-Gobbi} is not necessarily true when $\nu_{x}(F) = 0$, thus one cannot deduce weak ergodicity from strongly mixing property by (12). Nevertheless, since the reverse direction is not used in our work, we will not discuss possible fixes of that proof.
\end{remark}

The proof of the above lemma in \cite{Gray-Dunham-Gobbi} indeed gives the following specific pointwise result, which we will use later.

\begin{lemma}\label{lem:Gray-Dunham-Gobbi-lem-3-enhanced}
Let $[A,\nu,B]$ be a channel (not necessarily Markov) and $x\in\Sigma_{A}$. If $\nu_{x}$ corresponds to a weakly ergodic Markov chain, namely, \eqref{eq:weakly-ergodic-channel} is true for $x$, then \eqref{eq:strongly-mixing} holds for $x$ for all output rectangles $F$ and $G$.
\end{lemma}

Next we state the second main result in \cite{Gray-Dunham-Gobbi}.

\begin{theorem}[Theorem~2 in \cite{Gray-Dunham-Gobbi}]\label{thm:Gray-Dunham-Gobbi-thm-2}
If a stationary Markov channel $\nu$ is weakly ergodic $\mu$-a.e. for a stationary and ergodic source $\mu$, then $\mu\nu$ is stationary and ergodic. A Markov channel is ergodic if it is weakly ergodic $\mu$-a.e. with respect to all stationary measures $\mu$ (e.g., if it is weakly ergodic everywhere).
\end{theorem}

\begin{remark}
In fact the condition for the second statement can be weakened to just requiring $\nu$ to be weakly ergodic $\mu$-a.e. with respect to all stationary and ergodic measures $\mu$.
\end{remark}

The proof of this theorem in \cite{Gray-Dunham-Gobbi} is mostly correct, except that the proof for the second statement has the issue of $\hnu$ mentioned in Remark~\ref{rmk:R'-issue}. Also it is too sketchy. In the following we use the same proof idea to extend this theorem to a more specific one tailored for our own purposes. Its proof not only rigorously assembles various results built up in the Appendices, but also demonstrates the proper treatment of the corresponding measurable sets on which the desired properties hold. In particular, the above issue of $\hnu$ is fixed in this proof.

\begin{theorem}\label{thm:Gray-Dunham-Gobbi-thm-2-extension}
Let $\nu$ be a Markov channel and $\mu$ be an AMS ergodic source. If $\nu$ is weakly ergodic $\mu$-a.e., then the input-output process $\mu\nu$ is also AMS and ergodic.
\end{theorem}

\begin{IEEEproof}
Construct the auxiliary measures/processes $\omu$ and $\omus$ and the auxiliary two-sided channel $\hnu$ as in Section~\ref{subsec:Kieffer-Rahe-auxilliary-construction}. First from Theorem~\ref{thm:Markov-channel-AMS} we know $\mu\nu$ is AMS and by Lemma~\ref{lem:mu-auxiliary-ergodicity} the stationary measure $\omus$ is also ergodic. Next, as $\nu$ is weakly ergodic $\mu$-a.e. and $\mu$ is AMS, $\hnu$ is weakly ergodic on a subset $R^{*}\subseteq R'$ with $\omus$-probability 1 by Lemma~\ref{lem:Gray-Dunham-Gobbi-lem-2}. Hence by Lemma~\ref{lem:Gray-Dunham-Gobbi-lem-3-enhanced} the condition in Definition~\ref{def:strongly-mixing} for the channel $\hnu$ holds for all $x\in R^{*}$, so $\hnu$ is strongly mixing $\omus$-a.e. Now as $\hnu$ is also stationary while $\omus$ is stationary and ergodic, $\omus\hnu$ is also stationary and ergodic by Theorem~\ref{thm:Adler}. Finally, $\mu\nu$ is also ergodic by Lemma~\ref{lem:mu-nu-auxiliary-ergodicity}.
\end{IEEEproof}

\begin{corollary}\label{cor:Gray-Dunham-Gobbi-thm-2-extension-cor-Markov}
Let $\nu$ be a Markov channel and $\mu$ be a stationary ergodic source. If any one of the conditions in Lemmas~\ref{lem:Gray-Dunham-Gobbi-cor-1}, \ref{lem:Gray-Dunham-Gobbi-cor-2}, and \ref{lem:Gray-Dunham-Gobbi-cor-3} holds on a set of positive $\mu$-probability, then $\mu\nu$ is AMS and ergodic.
\end{corollary}

\begin{IEEEproof}
The result is obtained by combining Lemmas~\ref{lem:Gray-Dunham-Gobbi-cor-1}--\ref{lem:Gray-Dunham-Gobbi-cor-3}, and Remark~\ref{rmk:Gray-Dunham-Gobbi-cor2-3-relaxation} together with Theorem~\ref{thm:Gray-Dunham-Gobbi-thm-2-extension}.
\end{IEEEproof}

\begin{corollary}\label{cor:Gray-Dunham-Gobbi-thm-2-extension-cor-FSC}
Let $\nu$ be a finite state channel and $\mu$ be a stationary ergodic source. Let $(a_{1},\cdots,a_{n})\in A^{n}$, if
\begin{enumerate}
\item $\mu(X_{1}=a_{1},\cdots,X_{n}=a_{n}) > 0$,
\item $\prod_{i=1}^{n}P_{a_{i}}$ is scrambling, or has a positive column,
\end{enumerate}
then $\mu\nu$ is AMS and ergodic.
\end{corollary}

\begin{IEEEproof}
The result is obtained by combining Corollary~\ref{cor:Gray-Dunham-Gobbi-cor-1-relaxation-cor} and Theorem~\ref{thm:Gray-Dunham-Gobbi-thm-2-extension}.
\end{IEEEproof}

\section{Results for Finite State Channels with Markov sources}
\label{sec:ergodicity-FSC-Markov-source}

In this section we specialize to the case of connecting a finite-order Markov input process to a finite state channel, and obtain some stationarity and ergodicity results. These results provide an alternative set of sufficient conditions for the Shannon-McMillan-Breiman theorem. We start our development with the ergodicity of finite-order Markov processes, and then extend to finite state channels with finite-order Markov sources. The main theoretical tool is the following theorem for the ergodicity of stationary Markov chains from \cite{Walters-Ergodic-Theory}.

\begin{theorem}[Theorem~1.19 in \cite{Walters-Ergodic-Theory}]
\label{thm:Markov-process-ergodic-irreducible}
Consider a Markov chain on a finite state space $\{1,2,\cdots,K\}$ with transition matrix $P$. Assume the initial distribution $\pi$ is a positive stationary distribution for this chain, namely, $\pi P = P$ and $\pi_{i}>0$ for all  $1\leq i\leq K$. Then the corresponding stationary random process is ergodic iff $P$ is irreducible, in which case $\pi$ is the unique stationary distribution for $P$.
\end{theorem}

Assume $\{X_{n}\}_{n>0}$ is a Markov process of order $k$, with a finite alphabet $A$. Let $W_{n}$ denote the state $(X_{n-k+1},\cdots,X_{n})$ of the underlying Markov chain for $n\geq k$. The state process $\{W_{n}\}_{n\geq k}$ and the original process $\{X_{n}\}_{n>0}$ uniquely determine each other, and the stationarity, AMS property, or ergodicity of one process implies the same property for the other.\footnote{See \cite{Mao-thesis} for a more detailed discussion.} Let $P$ denote the transition matrix of the Markov chain. The process measure $\eta$ of $\{W_{n}\}$ is determined by $P$ and the initial distribution, and is AMS by \cite[Theorem~9]{Kieffer-Rahe-AMS-Mrkv-Ch}. Let $\oeta$ be the stationary mean of $\eta$ and $\pi$ be the initial distribution for $\oeta$, then $\pi$ is a stationary distribution of $P$.\footnote{A stationary distribution always exists for any finite-state Markov chain\cite{Ephraim-Merhav-HMP}.} Denote the support of $\pi$ by $\Gamma$, which is called the \emph{contingent stationary support} of the Markov process $\{W_{n}\}$ (since it depends on the initial distribution). It is easy to see that $\Gamma$ is a \emph{closed} subset of $A^{k}$, that is, $P_{ij} = 0$ for all $i\in\Gamma, j\notin\Gamma$.

Now assume that the Markov chain $P$ is irreducible on $\Gamma$. As the conditions for Theorem~\ref{thm:Markov-process-ergodic-irreducible} are satisfied on $\Gamma$ with the initial distribution $\pi$, the stationary measure $\oeta$ is ergodic, and so is $\eta$ (see \cite[Lemma 6.7.1]{Gray-PRE1}). Hence $\{W_{n}\}$ and $\{X_{n}\}$ are AMS ergodic processes. Conversely, if $\{X_{n}\}$ or $\{W_{n}\}$ is ergodic, then $\eta$, and so $\oeta$ are ergodic, and by Theorem~\ref{thm:Markov-process-ergodic-irreducible}, $P$ is irreducible on $\Gamma$.

Moreover, when either of the above conditions holds, Theorem~\ref{thm:Markov-process-ergodic-irreducible} states that $\pi$ is the unique stationary distribution for the chain on $\Gamma$. Thus if another initial distribution on the Markov chain induces a process measure $\teta$, whose stationary mean has a (stationary) initial distribution $\tpi$ that is also supported on $\Gamma$, then necessarily $\tpi = \pi$ and the stationary mean is $\oeta$. In particular, if $\Gamma = A^{k}$, or equivalently, (the full matrix) $P$ is irreducible, then the stationary process measures for $\{W_{n}\}$ and $\{X_{n}\}$ are unique.


Summarizing the discussions above we have the following lemma.

\begin{lemma}\label{lem:finite-order-Markov-ergodicity}
Let $\{X_{n}\}$ be a finite-alphabet finite-order Markov process, with an underlying state process $\{W_{n}\}$, whose Markov transition matrix is $P$. Then both $\{X_{n}\}$ and $\{W_{n}\}$ are AMS. Let $\Gamma$ denote the contingent stationary support of $\{W_{n}\}$, then $\{W_{n}\}$ (and $\{X_{n}\}$) are ergodic iff $P$ is irreducible on $\Gamma$. Furthermore, when this is the case, any other initial distribution of the Markov chain that leads to the same contingent stationary support induces the same stationary mean for $\{W_{n}\}$ (and hence also the same stationary mean for $\{X_{n}\}$), and so the corresponding processes are ergodic. In particular, if $\Gamma$ is the full state space, or equivalently, $P$ is irreducible, then these stationary process measures are unique.
\end{lemma}

Now consider a finite state channel defined in Gallager's form \eqref{eq:FSC-cond-prob-App}. Assume the source process $\{X_{n}\}_{n>0}$ is Markov of order $k>0$ and is independent of the initial state $S_{1}$ of the FSC, then the joint process $\{(X_{n},Y_{n},S_{n+1})\}_{n>0}$ is also Markov of order $k$. When $\{X_{n}\}$ is i.i.d. (i.e., $k=0$), $\{(X_{n},Y_{n},S_{n+1})\}_{n>0}$ is simply Markov (i.e., of order-1). (See \cite{Mao-thesis} for the details.) Hence by the lemma above, we have:

\begin{lemma}\label{lem:FSC-Markov-input}
If the source $\{X_{n}\}$ of an FSC is an order-$k$ Markov process with $k\geq0$, then $\{(X_{n},Y_{n},S_{n+1})\}$ is a Markov process of order $\max\{k,1\}$. If the underlying Markov chain for the latter is irreducible on the contingent stationary support, then $\{(X_{n},Y_{n},S_{n+1})\}$ is AMS and ergodic.
\end{lemma}

In our energy harvesting channels we often encounter FSC's that satisfy
\begin{equation}\label{eq:FSC-separable}
p(y_{n}s_{n+1}\cn x_{n}s_{n}) = p(y_{n}\cn x_{n}s_{n})p(s_{n+1}\cn x_{n}s_{n}),
\end{equation}
for which we will show that if the input-state process is AMS ergodic, then so is the full joint process (see Lemma~\ref{lem:ergodicity-FSC-DMC-Y} in Appendix~\ref{sec:joint-marginal}). Thus for such channels we have:

\begin{corollary}\label{cor:FSC-separable-Markov-input}
If the source $\{X_{n}\}$ of an FSC satisfying \eqref{eq:FSC-separable} is an order-$k$ Markov process with $k\geq0$, then $\{(X_{n},S_{n+1})\}$ is a Markov process of order $\max\{k,1\}$. If the underlying Markov chain for the latter is irreducible on the contingent stationary support, then $\{(X_{n},Y_{n},S_{n+1})\}$ is AMS and ergodic.
\end{corollary}

\section{The Shannon-McMillan-Breiman Theorem}\label{sec:SMB}

For a finite alphabet random process $\{X_{n}\}$ whose probability measure is denoted by $p$, we are interested in the convergence of the sample entropy $-\frac{1}{n}\log p(X^{n})$ to the entropy rate
\begin{equation}\label{eq:entropy-rate-def}
H(\X) \triangleq \lim_{n\to\infty}\frac{1}{n}H(X^{n})
\end{equation}
whenever the limit exists. In information theory, this property is called the \emph{asymptotic equipartition property} (AEP) \cite{Cover-Thomas-ElemIT}. When the process is i.i.d., AEP is easily proved using law of large numbers. When $\{X_{n}\}$ is stationary and ergodic, the Shannon-McMillan-Breiman (SMB) theorem for stationary processes \cite{Cover-Thomas-ElemIT} also gives the AEP; in particular, the sample entropy converges to the entropy rate with probability 1. Yet this result is still not general enough for our application in the energy harvesting systems, since the joint input-output process produced by the surrogate channel is often not stationary, but AMS instead. Hence we require an SMB theorem for AMS processes, which is also called the entropy ergodic theorem in \cite{Gray-EIT}.

\begin{theorem}(Shannon-McMillan-Breiman / Entropy Ergodic Theorem \cite{Gray-EIT})\label{thm:SMB}
Let $\{X_{n}\}$ be a finite alphabet random process with an AMS ergodic process distribution $p$, whose stationary mean is denoted by $\op$. Then the entropy rate \eqref{eq:entropy-rate-def} exists and
\[ \lim_{n\to\infty}-\frac{1}{n}\log p(X^{n}) = H(\X), \]
where the convergence is both $p$-a.e. and in $L^{1}$-norm. Furthermore, the value of $H(\X)$ is the same as $H_{\op}(\X)$, the entropy rate defined under the stationary measure $\op$.
\end{theorem}

\section{A Specific Result: Joint and Marginal Processes}
\label{sec:joint-marginal}

In this section we discuss the stationarity and ergodicity of a joint process and its marginals. In the settings of this paper we usually have a joint process, say $\{V_{n},S_{n},Y_{n}\}$, and want to apply the SMB theorem on its various marginal processes, e.g., $\{V_{n},Y_{n}\}$ or $\{Y_{n}\}$. It is enough to show the required AMS and ergodic properties for the joint process $\{V_{n},S_{n},Y_{n}\}$, since from their respective definitions we can easily see that these properties are inherited by the marginal processes from the joint one.

We also have some remarks for the other direction. Consider a general channel $[A,\nu,B']$ whose input and output symbols are $X_{n}$ and $Y'_{n}$, respectively. Let ${[A\times B',\eta,B]}$ be another channel, whose input symbols are the pairs $(X_{n},Y'_{n})$ and output symbols are $Y_{n}$. Assume $\eta$ is a stationary memoryless channel, then it is stationary and strongly mixing and so Adler's theorem applies. In particular, if a source $[A,\mu]$ gives an AMS ergodic hookup $\mu\nu$, then by Theorem~\ref{thm:Adler}, connecting $\mu\nu$ to $\eta$ gives an AMS ergodic hookup $(\mu\nu)\eta$. In other words, the joint process $\{(X_{n},Y'_{n},Y_{n})\}_{n\in\bI}$ is also AMS and ergodic.

For the application in our energy harvesting channels, consider a special class of FSC models whose transition probability satisfies
\begin{equation}\label{eq:FSC-DMC-Y}
p(y_{n}s_{n+1}\cn x_{n}s_{n}) = p(y_{n}\cn x_{n}s_{n})p(s_{n+1}\cn x_{n}s_{n}).
\end{equation}
We can view $p(s_{n+1}\cn x_{n}s_{n})$ as the transition probability of a smaller finite state channel $\nu$, with input symbols $X_{n}$ and output symbols $Y'_{n}=S_{n}$. Furthermore, $Y_{n}$ can be viewed as the output of another DMC $\eta$, whose input symbols are the pairs $(X_{n},Y'_{n})$ with transition probability
\[ p(y_{n}\cn x_{n}y'_{n}) = p(y_{n}\cn x_{n}s_{n}). \]
Applying the argument from the previous paragraph to the channels $\nu$ and $\eta$, we have the lemma below. Consequently, to show the full joint process $\{(X_{n},S_{n},Y_{n})\}_{n>0}$ is AMS and ergodic we only need to consider the smaller finite state channel $p(s_{n+1}\cn x_{n}s_{n})$.

\begin{lemma}\label{lem:ergodicity-FSC-DMC-Y}
For the FSC model \eqref{eq:FSC-DMC-Y} let $\{X_{n}\}_{n>0}$ be an input process that yields an AMS ergodic joint input-state process $\{(X_{n},S_{n})\}_{n>0}$, then the joint input-state-output process $\{(X_{n},S_{n},Y_{n})\}_{n>0}$ is also AMS ergodic.
\end{lemma}

\section*{Acknowledgment}

The authors would like to thank Pascal Vontobel and Guangyue Han for the helpful discussions on the stochastic algorithms for the optimization of achievable rates.

\ifCLASSOPTIONcaptionsoff
  \newpage
\fi


\begin{thebibliography}{10}
\providecommand{\url}[1]{#1}
\csname url@samestyle\endcsname
\providecommand{\newblock}{\relax}
\providecommand{\bibinfo}[2]{#2}
\providecommand{\BIBentrySTDinterwordspacing}{\spaceskip=0pt\relax}
\providecommand{\BIBentryALTinterwordstretchfactor}{4}
\providecommand{\BIBentryALTinterwordspacing}{\spaceskip=\fontdimen2\font plus
\BIBentryALTinterwordstretchfactor\fontdimen3\font minus
  \fontdimen4\font\relax}
\providecommand{\BIBforeignlanguage}[2]{{%
\expandafter\ifx\csname l@#1\endcsname\relax
\typeout{** WARNING: IEEEtran.bst: No hyphenation pattern has been}%
\typeout{** loaded for the language `#1'. Using the pattern for}%
\typeout{** the default language instead.}%
\else
\language=\csname l@#1\endcsname
\fi
#2}}
\providecommand{\BIBdecl}{\relax}
\BIBdecl

\bibitem{Mao-Hassibi-EnHarv}
W.~Mao and B.~Hassibi, ``On the capacity of a communication system with energy
  harvesting and a limited battery,'' in \emph{Proc. of 2013 IEEE International
  Symposium on Information Theory}, Istanbul, Turkey, Jul. 2013.

\bibitem{Mao-Hassibi-EnHarvLinBdCoding}
------, ``New capacity upper bounds and coding aspects for some channels with
  causal {CSIT},'' in \emph{Proc. of 2015 IEEE International Symposium on
  Information Theory}, Hong Kong, China, Jun. 2015.

\bibitem{Mao-Hassibi-CausCSIEnHav}
------, ``Capacity bounds for certain channels with states and the energy
  harvesting channel,'' in \emph{Proc. of the 2014 Information Theory
  Workshop}, Hobart, Australia, Nov. 2014.

\bibitem{Ozel-Yang-Ulukus}
O.~Ozel, J.~Yang, and S.~Ulukus, ``Optimal broadcast scheduling for an energy
  harvesting rechargeable transmitter with a finite capacity battery,''
  \emph{{IEEE} Trans. Wireless Commun.}, vol.~11, no.~6, pp. 2193--2203, Jun
  2012.

\bibitem{Yang-Ulukus}
J.~Yang and S.~Ulukus, ``Optimal packet scheduling in an energy harvesting
  communication system,'' \emph{{IEEE} Trans. Commun.}, vol.~60, no.~1, pp.
  220--230, January 2012.

\bibitem{Ozel-Ulukus-AWGN}
O.~Ozel and S.~Ulukus, ``Achieving {AWGN} capacity under stochastic energy
  harvesting,'' \emph{{IEEE} Trans. Inf. Theory}, vol.~58, no.~10, pp.
  6471--6483, October 2012.

\bibitem{Ozel-Ulukus-0battery}
------, ``{AWGN} channel under time-varying amplitude constraints with causal
  information at the transmitter,'' in \emph{Proc. of the 45th Asilomar
  Conference on Signals, Systems and Computers}, Pacific Grove, CA, Nov. 2011.

\bibitem{Tutuncuoglu-Yener-EnHarvPolicy}
K.~Tutuncuoglu and A.~Yener, ``Optimum transmission policies for battery
  limited energy harvesting nodes,'' \emph{{IEEE} Trans. Wireless Commun.},
  vol.~11, no.~3, pp. 1180--1189, Mar. 2012.

\bibitem{Tutuncuoglu-EnHarvTimingCh}
K.~Tutuncuoglu, O.~Ozel, A.~Yener, and S.~Ulukus, ``Binary energy harvesting
  channel with finite energy storage,'' in \emph{Proc. of 2013 IEEE
  International Symposium on Information Theory}, Istanbul, Turkey, Jul. 2013.

\bibitem{Ozel-EH-CSIR-Discrete}
O.~Ozel, K.~Tutuncuoglu, S.~Ulukus, and A.~Yener, ``Capacity of the discrete
  memoryless energy harvesting channel with side information,'' in \emph{Proc.
  of 2014 IEEE International Symposium on Information Theory}, Honolulu, HI,
  Jun. 2014.

\bibitem{Permuter-FSC-FB}
H.~Permuter, T.~Weissman, and A.~J. Goldsmith, ``Finite state channels with
  time-invariant deterministic feedback,'' \emph{{IEEE} Trans. Inf. Theory},
  vol.~55, no.~2, pp. 644--662, Feb. 2009.

\bibitem{Chen-Berger-FSC-FB}
J.~Chen and T.~Berger, ``The capacity of finite-state {Markov} channels with
  feedback,'' \emph{{IEEE} Trans. Inf. Theory}, vol.~51, no.~3, pp. 780--798,
  Mar. 2005.

\bibitem{Dong-Ozgur-EH-AWGN-Bounds}
Y.~Dong and A.~{\"O}zg{\"u}r, ``Approximate capacity of energy harvesting
  communication with finite battery,'' in \emph{Proc. of 2014 IEEE
  International Symposium on Information Theory}, Honolulu, HI, Jun. 2014.

\bibitem{Shaviv-Minh-Ozgur}
\BIBentryALTinterwordspacing
D.~Shaviv, P.~Nguyen, and A.~{\"O}zg{\"u}r. (2015, Jun.) Capacity of the energy
  harvesting channel with a finite battery. [Online]. Available:
  \url{http://arxiv.org/abs/1506.02024}
\BIBentrySTDinterwordspacing

\bibitem{Shaviv-Ozgur-Permuter}
D.~Shaviv, A.~{\"O}zg{\"u}r, and H.~Permuter, ``Can feedback increase the
  capacity of the energy harvesting channel?'' in \emph{Proc. of the 2015
  Information Theory Workshop}, Jerusalem, Apr.May 2015.

\bibitem{Verdu-Han-Capacity}
S.~Verd{\'u} and T.~S. Han, ``A general formula for channel capacity,''
  \emph{{IEEE} Trans. Inf. Theory}, vol.~40, no.~4, pp. 1147--1157, Jul. 1994.

\bibitem{Vontobel-GBAA}
P.~O. Vontobel, A.~Kav{\v c}i{\'c}, D.~M. Arnold, and H.-A. Loeliger, ``A
  generalization of the {Blahut-Arimoto} algorithm to finite-state channels,''
  \emph{{IEEE} Trans. Inf. Theory}, vol.~54, no.~5, pp. 1887--1918, May 2008.

\bibitem{Gallager-IT_Reliable}
R.~G. Gallager, \emph{Information Theory and Reliable Communication}.\hskip 1em
  plus 0.5em minus 0.4em\relax New York: John Wiley \& Sons, 1968.

\bibitem{Tatikonda-Mitter-Feedback}
S.~Tatikonda and S.~Mitter, ``The capacity of channels with feedback,''
  \emph{{IEEE} Trans. Inf. Theory}, vol.~55, no.~1, pp. 323--349, Jan. 2009.

\bibitem{Shannon-CSIT}
C.~Shannon, ``Channels with side information at the transmitter,'' \emph{IBM
  Journal of Research and Development}, vol.~2, no.~4, pp. 289--293, Oct. 1958.

\bibitem{Caire-Shamai-CSI}
G.~Caire and S.~{Shamai (Shitz)}, ``On the capacity of some channels with
  channel state information,'' \emph{{IEEE} Trans. Inf. Theory}, vol.~45,
  no.~6, pp. 2007--2019, Sep. 1999.

\bibitem{Cover-Thomas-ElemIT}
T.~M. Cover and J.~A. Thomas, \emph{Elements of Information Theory},
  2nd~ed.\hskip 1em plus 0.5em minus 0.4em\relax Hoboken, N.J:
  Wiley-Interscience, 2006.

\bibitem{Mao-thesis}
\BIBentryALTinterwordspacing
W.~Mao, ``Information-theoretic studies and capacity bounds: Group network
  codes and energy harvesting communication systems,'' Ph.D. dissertation,
  California Institute of Technology, 2015. [Online]. Available:
  \url{http://thesis.library.caltech.edu/8834/}
\BIBentrySTDinterwordspacing

\bibitem{Chen-Permuter-Weissman-FSC-Bds}
J.~Chen, H.~Permuter, and T.~Weissman, ``Tighter bounds on the capacity of
  finite-state channels via {Markov} set-chains,'' \emph{{IEEE} Trans. Inf.
  Theory}, vol.~56, no.~8, pp. 3660--3691, Aug. 2010.

\bibitem{Kieffer-Rahe-AMS-Mrkv-Ch}
J.~C. Kieffer and M.~Rahe, ``Markov channels are asymptotically mean
  stationary,'' \emph{Siam Journal of Mathematical Analysis}, vol.~12, no.~3,
  pp. 293--305, 1981.

\bibitem{Gray-Dunham-Gobbi}
R.~M. Gray, M.~O. Dunham, and R.~L. Gobbi, ``Ergodicity of {M}arkov channels,''
  \emph{{IEEE} Trans. Inf. Theory}, vol.~33, no.~5, pp. 656--664, Sep. 1987.

\bibitem{Walters-Ergodic-Theory}
P.~Walters, \emph{An Introduction to Ergodic Theory}, ser. Graduate texts in
  mathematics.\hskip 1em plus 0.5em minus 0.4em\relax New York:
  Springer-Verlag, 1982, vol.~79.

\bibitem{BCJR-algorithm}
L.~R. Bahl, J.~Cocke, F.~Jelinek, and J.~Raviv, ``Optimal decoding of linear
  codes for minimizing symbol error rate,'' \emph{{IEEE} Trans. Inf. Theory},
  vol.~20, no.~2, pp. 284--287, Mar. 1974.

\bibitem{Kschischang-Frey-Loeliger-SumProductAlg}
F.~Kschischang, B.~Frey, and H.-A. Loeliger, ``Factor graphs and the
  sum-product algorithm,'' \emph{{IEEE} Trans. Inf. Theory}, vol.~47, no.~2,
  pp. 498--519, Feb. 2001.

\bibitem{Arnold-Loeliger-info-rate}
D.~Arnold and H.-A. Loeliger, ``On the information rate of binary-input
  channels with memory,'' in \emph{Proc. of 2001 IEEE International Conference
  on Communications}, Helsinki, Finland, Jun. 2001, pp. 2692--2695.

\bibitem{Sharma-Singh-info-rate}
V.~Sharma and S.~K. Singh, ``Entropy and channel capacity in the regenerative
  setup with applications to {Markov} channels,'' in \emph{Proc. of 2001 IEEE
  International Symposium on Information Theory}, Washington, DC, Jun. 2001, p.
  283.

\bibitem{Pfister-et-al-info-rate}
H.~D. Pfister, J.~B. Soriaga, and P.~H. Siegel, ``On the achievable information
  rates of finite-state {ISI} channels,'' in \emph{Proc of 2001 IEEE Global
  Telecommunications Conference (GLOBECOM '01)}, San Antonio, TX, Nov. 2001,
  pp. 2992--2996.

\bibitem{Arnold-Simulation-SMB}
D.~M. Arnold, H.-A. Loeliger, P.~O. Vontobel, A.~Kav{\v c}i{\'c}, and W.~Zeng,
  ``Simulation-based computation of information rates for channels with
  memory,'' \emph{{IEEE} Trans. Inf. Theory}, vol.~52, no.~8, pp. 3498--3508,
  August 2006.

\bibitem{Han-RandomizedCapacity}
G.~Han, ``A randomized approach to the capacity of finite-state channels,'' in
  \emph{Proc. of 2013 IEEE International Symposium on Information Theory},
  Istanbul, Turkey, Jul. 2013.

\bibitem{Blahut-BAA}
R.~E. Blahut, ``Computation of channel capacity and rate-distortion
  functions,'' \emph{{IEEE} Trans. Inf. Theory}, vol.~18, no.~4, pp. 460--473,
  Jul. 1972.

\bibitem{Gray-EIT}
R.~M. Gray, \emph{Entropy and Information Theory}.\hskip 1em plus 0.5em minus
  0.4em\relax New York: Springer-Verlag, 1990.

\bibitem{Tutuncuoglu-EnHarvTimingCh2}
K.~Tutuncuoglu, O.~Ozel, A.~Yener, and S.~Ulukus, ``Improved capacity bounds
  for the binary energy harvesting channel,'' in \emph{Proc. of 2014 IEEE
  International Symposium on Information Theory}, Honolulu, HI, Jul. 2014.

\bibitem{Ozel-EH-CSIR-ESI}
O.~Ozel, K.~Tutuncuoglu, S.~Ulukus, and A.~Yener, ``Capacity of the energy
  harvesting channel with energy arrival information at the receiver,'' in
  \emph{Proc. of the 2014 Information Theory Workshop}, Hobart, Australia, Nov.
  2014.

\bibitem{Naiss-Permuter-EBAA}
I.~Naiss and H.~Permuter, ``Extension of the {Blahut-Arimoto} algorithm for
  maximizing directed information,'' \emph{{IEEE} Trans. Inf. Theory}, vol.~59,
  no.~1, pp. 204--222, Jan. 2013.

\bibitem{Permuter-POST-FB}
H.~Permuter, H.~Asnani, and T.~Weissman, ``Capacity of a {POST} channel with
  and without feedback,'' \emph{{IEEE} Trans. Inf. Theory}, vol.~60, no.~10,
  pp. 6041--6057, Oct. 2014.

\bibitem{Gray-PRE1}
R.~M. Gray, \emph{Probability, Random Processes, and Ergodic Properties},
  1st~ed.\hskip 1em plus 0.5em minus 0.4em\relax Springer, 1987.

\bibitem{Adler-Ergodic-Mixing}
R.~L. Adler, ``Ergodic and mixing properties of infinite memory channels,''
  \emph{Proceedings of the American Mathematical Society}, vol.~12, no.~6, pp.
  924--930, 1961.

\bibitem{Ephraim-Merhav-HMP}
Y.~Ephraim and N.~Merhav, ``Hidden markov processes,'' \emph{{IEEE} Trans. Inf.
  Theory}, vol.~48, no.~6, pp. 1518--1569, Jun. 2002.

\end{thebibliography}
\end{document}